\documentclass{aastex}
\usepackage[onecolumn]{emulateapj5}
\usepackage{amsmath,amssymb}

\newcommand{\rd}{{\rm d}}

\newcommand{\bm}{{\mathbf{m}}}
\newcommand{\bx}{{\mathbf{x}}}

\newcommand{\BE}{{\mathbb{E}}}
\newcommand{\BR}{{\mathbb{R}}}
\newcommand{\CD}{{\cal D}}
\newcommand{\CM}{{\cal M}}
\newcommand{\CS}{{\cal S}}
\newcommand{\CNT}{{{\cal N}_2}}

\newcommand{\hMpc}{{\ifmmode{h^{-1}{\rm Mpc}}\else{$h^{-1}$Mpc}\fi}}
\newcommand{\Beins}{\mbox{1\hspace*{-0.085cm}l}}
\newcommand{\cov}{{\ifmmode{\text{{\it cov}}}\else{ {\it cov} }\fi}}
\newcommand{\cor}{{\ifmmode{\text{{\it cor}}}\else{ {\it cor} }\fi}}
\newcommand{\var}{{\ifmmode{\text{{\it var}}}\else{ {\it var} }\fi}}
\newcommand{\paverage}[1]{\left\langle #1 \right\rangle_{\rm P}}
\newcommand{\phataverage}[1]{\widehat{\left\langle #1 \right\rangle}_{\rm P}}

\newcommand{\aanda}{A\&A}

\renewcommand{\today}{July 26, 2000}

\begin{document}

\title{Luminosity-- and morphology--dependent clustering of galaxies}
\author{Claus Beisbart\altaffilmark{1} and Martin Kerscher\altaffilmark{2,1}}
\altaffiltext{1}{Sektion Physik, Ludwig--Maximilians--Universit{\"a}t, 
Theresienstra{\ss}e 37, 80333 M{\"u}nchen, Germany, 
email: beisbart@theorie.physik.uni-muenchen.de}
\altaffiltext{2}{Department of Physics and Astronomy, 
The Johns Hopkins University, Baltimore, MD 21218,
email: kerscher@pha.jhu.edu}
\shorttitle{Luminosity and morphology dependent clustering}
\shortauthors{Beisbart \& Kerscher} 

\slugcomment{\today\ accepted in the ApJ}

\begin{abstract}
How does the  clustering of galaxies depend on  their inner properties
like morphological  type and luminosity?  We address  this question in
the mathematical  framework of marked point processes  and clarify the
notion of luminosity and  morphological segregation.  A number of test
quantities such  as conditional mark--weighted  two--point correlation
functions   are   introduced.    These   descriptors   allow   for   a
scale--dependent  analysis of  luminosity and  morphology segregation.
Moreover, they  break the degeneracy between  an inhomogeneous fractal
point set  and actual present  luminosity segregation.\\
Using the Southern  Sky Redshift Survey~2 (\citealt{dacosta:southern},
SSRS2) we find both luminosity and morphological segregation at a high
level of significance, confirming claims by previous works using these
data  {}\citep{benoist:biasing,willmer:southern}.   Specifically,  the
average luminosity and the fluctuations  in the luminosity of pairs of
galaxies  are  enhanced out  to  separations  of  15\hMpc.  On  scales
smaller than  3\hMpc\ the  luminosities on galaxy  pairs show  a tight
correlation.
A  comparison  with  the  random--field model  indicates  that  galaxy
luminosities  depend  on the  spatial  distribution and  galaxy-galaxy
interactions. Early--type  galaxies are also  more strongly correlated,
indicating morphological segregation.
The galaxies  in the PSCz  catalog {}\citep{saunders:pscz} do  not show
significant  luminosity  segregation.   This  again  illustrates  that
mainly early--type galaxies contribute to luminosity segregation.
However, based on several  independent investigations we show that the
observed  luminosity   segregation  can   not  be  explained   by  the
morphology--density relation alone.
\end{abstract}

\keywords{methods: statistical -- large-scale structure of universe --
galaxies:    clusters     --    galaxies:    fundamental    parameters
(classification, luminosities) }

\section{Introduction}

The  geometrical  properties  of  the large--scale  structure  in  the
Universe are  a common test  for cosmic structure  formation theories.
However, comparisons between  analytical models and observational data
suffer  from the  fact,  that theoretical  predictions  refer to  mass
correlations  whereas  in  galaxy  catalogs only  luminous  matter  is
observed.   This gap gives  rise to  the bias  problem and  is usually
filled using biasing schemes.  Mostly, these schemes relate properties
of the  density contrast  field to the  distribution of  the galaxies,
thus  combining  descriptors of  a  random  field  with point  process
characteristics. 
Due to the nature of the  {\em dark} matter, only indirect methods are
feasible to  address the  bias problem empirically.   In this  line of
thought, it  seems promising to ask whether  the clustering properties
of galaxies  depend on their  mass, luminosity or  morphological type.
The  idea  behind  this  search  for {\em  luminosity  and  morphology
segregation}  is that  different galaxy  subpopulations may  trace the
dark matter distribution on a different level.\\
Empirical  investigations  concerned  with  this problem  were  mainly
carried out in two directions:
\begin{itemize}
\item The two--point correlation  function was calculated for a series
of volume--limited  subsamples from  galaxy surveys.  A  difference in
the amplitude  of the  two--correlation function between  such samples
was    interpreted    as    an    indication    of    luminosity    or
morphology--segregation.    For  luminosity   segregation   see  e.g.,
{}\citet{ostriker:luminosity,hamilton:evidence,tenreiro:multidimensional,benoist:biasing,willmer:southern}. The
void probability  and cross--correlation  functions have been  used by
{}\citet{maurogordato:void}   and  {}\citet{valotto:dependence}.   For
morphology              segregation              see             e.g.,
{}\citet{tenreiro:morphology-segregation,hermit:morphology-segregation}.
These investigations  are sensitive  to segregation effects  on scales
roughly  between 1  and  10\hMpc.  However,  {}\citet{coleman:fractal}
gave an alternative explanation of  the rising amplitude in terms of a
fractal   galaxy  distribution,   without   any  luminosity--dependent
clustering.
\item  {}\citet{dressler:galaxy} showed that  in clusters  of galaxies
the morphological type of a galaxy is depending on the local (surface)
density; this  is called the morphology--density  relation. For mainly
spherical clusters, where the local  density is closely related to the
radial  distance from  the cluster  center, this  translates  into the
{}\citet{butcher:evolution}  effect. For more  recent accounts  of the
morphology--density                    relation                    see
{}\citet{caon:morphology-segregation,dressler:evolution,andreon:morphologyIII}.
Most  of  these  investigations  focussed on  the  morphology--density
relation {\em inside} clusters, hence on scales smaller than 1.5\hMpc.
But the morphology--density relation can be observed also in groups of
galaxies   {}\citep{postman:morphology-density,maia:groups}   and  for
dwarf galaxies in the field {}\citep{binggeli:abundance}.
\end{itemize}
With the first method,  one compares two--point correlation functions,
whereas with the second, one  considers the relation between the local
number  density  and  the  local  morphology, i.e.,  a  comparison  of
one--point densities. Both methods rely on unweighted descriptors.

The     observations    of     luminosity    segregation     or    the
mor\-phology--density   relation  were  complemented   by  theoretical
considerations.   Motivated by the  offset between  the galaxy--galaxy
and       the       cluster--cluster      correlation       functions,
{}\citet{kaiser:onspatial} and  {}\citet{bardeen:gauss} suggested that
clusters may  be understood as  peaks in the density  field.  Starting
from  a Gaussian random  field they  showed how  the amplitude  of the
correlation  function  increases with  the  threshold  imposed on  the
initial  density field,  i.e., with  the height  of the  peaks  in the
density   field.   This   also   provided  an   explanation  for   the
morphology--density relation  {}\citep{evrard:morphology}.\\
Other authors  developed a conceptual  framework to describe  the bias
(see e.g., {}\citealt{coles:galaxy}, {}\citealt{dekel:stochastic}, and
refs.\ therein).  Within these biasing schemes, characteristics of the
galaxy point  pattern are connected  with descriptions of  the density
field -- often the mass  density contrast and the galaxy over--density
are compared.  The relation is assumed to be (non--) linear and either
deterministic or stochastic {}\citep{dekel:stochastic}.  More involved
biasing  schemes  were  considered  to facilitate  the  extraction  of
reasonable  galaxy  catalogs  from  $N$-body  simulations  (see  e.g.,
{}\citealt{kates:highres},                   {}\citealt{weiss:highres},
{}\citealt{kauffmann:galaxy}).

In this paper,  we introduce a new method to handle  the bias problem. 
Our approach  complements both the more observational  methods and the
analytical  and theoretical  treatments.  We  understand  the galaxies
with their  intrinsic properties  as a realization  of a  marked point
process.  Using conditional weighted  correlation functions, we put an
intermediate step in between the pure point process statistics and the
statistics  of  random fields.  In  our  description stochasticity  is
present  from the  very  beginning.  It  provides  us with  stochastic
models which enable  us to exclude certain families  of models for the
luminosity distribution of galaxies.

\noindent
More precisely, the aim of our  paper is twofold:

On   the   one   hand,   we    want   to   clarify   the   notion   of
luminosity/morphology--dependent clustering by discussing this task in
the    mathematical    framework    of    marked    point    processes
(Sect.~\ref{sect:maths}).  This allows us  to introduce a new class of
indicators       sensitive       to       luminosity       segregation
(Subsect.~\ref{sect:mark-weighted}) and  to discuss models  for marked
point       patterns      (Subsect.~\ref{sect:marked-poisson}      and
Subsect.~\ref{sect:two-species}).
Methods  similar in  spirit  are the  cross--correlation function  and
luminosity--weighted     correlation    functions     considered    by
{}\citet{alimi:cross-correlation},       {}\citet{boerner:correlation},
{}\citet{vallsgabaud:luminosity}, and
{}\citet{tegmark:observational}.
Our methods  allow for  a study of  the interplay between  the spatial
clustering  and  the luminosity  and  morphology  distribution of  the
galaxies,  complementing the  characterization of  the  purely spatial
distribution of the galaxies.

On  the other  hand, we  address the  empirical question,  whether the
luminosities  or  morphological  types  of galaxies  depend  on  their
spatial    distribution    by     analyzing    the    SSRS2    catalog
{}\citep{dacosta:southern} in Sect.~\ref{sect:lum-morph-seg-gal}.  Our
results   show   a   significant   scale--dependent   luminosity   and
morphological  segregation.  To  understand the  data more  closely we
compare our results with the  random field model.  The comparison with
galaxy samples  from the IRAS~1.2Jy  {}\citep{fisher:irasdata} and the
PSCz {}\citep{saunders:pscz} strengthens our conclusions.

In Sect.~\ref{sect:usual} we will discuss the usual way of looking for
luminosity segregation  via the amplitude of  the correlation function
in  the  framework  of  marked  point  processes.   The  criticism  by
{}\citet{coleman:fractal} is reviewed and we show that this degeneracy
between a  fractal spatial distribution and  luminosity segregation is
not  encountered  if  one  uses  the  mark--correlation  functions  we
proposed.  This  strengthens the conclusions of our  empirical work in
Sect.~\ref{sect:lum-morph-seg-gal}.

Investigations inside clusters of galaxies gave clear evidence for the
morphology--density     relation     {}\citep{dressler:galaxy}.     In
Sect.~\ref{sect:morphology-density} we however  show that the observed
luminosity segregation may not be explained by the spatial interaction
of early-- and late--type galaxies {\em alone}. Luminosity segregation
is  already present in  the subsample  consisting only  of early--type
galaxies.

In    Sect.~\ref{sect:summary}   we    summarize   and    provide   an
outlook. Technicalities concerning the estimation of mark--correlation
functions are left to Appendix {}\ref{sect:estimators}.

\section{Marked point distributions}
\label{sect:maths}

Consider a set of  points $X=\{\bx_i\}_{i=1}^{N}$ given by the spatial
coordinates $\bx_i\in\BR^3$  of the galaxies inside  a sample geometry
$\CD$.   Additionally to their  positions in  space we  know intrinsic
properties of the galaxies  like their luminosity, mass, morphological
type etc.   Formally, we  assign to each  point $\bx_i$ a  mark $m_i$,
e.g., the  luminosity of the  galaxy $m_i=L_i$, and obtain  the marked
point  set  $X^M=\{(\bx_i,m_i)\}_{i=1}^{N}$.  We  are  not limited  to
continuous  marks  like  the  luminosity,  also  discrete  marks  like
morphological types  (e.g., {\em spiral}  or {\em elliptical})  can be
used.   The   description  of  the  galaxy  distribution   in  a  {\em
  statistical} way that we will  propose in this article, rests on the
assumption  that the  empirical data  points  may be  considered as  a
realization     of    a     marked    point     process.     Formally,
$X=\{\bx_i\}_{i=1}^{N}$ and $M=\{m_i\}_{i=1}^{N}$ may be thought of as
realizations of  a point process  each, which may be  characterized by
the  usual  point process  statistics.   Physically,  however, we  are
interested  in the interplay  between the  spatial statistics  and the
mark  distribution,   which  is  expressed   in  quantities  combining
information on the space and the mark distribution.

The second--order  theory of marked  point processes was  developed in
detail by {}\citet{stoyan:oncorrelations}  where also a mark--weighted
conditional   correlation   function    was   introduced   (see   also
{}\citealt{stoyan:fractals}).  Some  aspects have been  also discussed
by {}\citet{peebles:lss}.

\subsection{One--point properties}
\label{sect:one-point}

A  point  process  may  be   characterized  by  its  moments.   For  a
homogeneous spatial  point distribution the  first moment is  the mean
number density  $\rho$, which may  be estimated with  $N/|\CD|$, where
$|\CD|$  is the  volume of  the sample  and $N$  the number  of points
inside $\CD$.
Let $\rho_1^M(m)\rd m$ denote the  probability that the value of a mark
lies   within  the  interval   $[m,m+\rd  m]$,   then  the   mean  mark
$\overline{m}$ and the variance of the marks $V$ are given by
\begin{equation}
\overline{m} =\int\rd m\ \rho_1^M(m) m , \quad \text{ and }\quad
V  =\int\rd m\ \rho_1^M(m) (m -\overline{m})^2 ,
\end{equation}
which   may  be   estimated   by 
\[
\frac{1}{N}\sum_{i=1}^{N}m_i   \quad \text{ and }\quad
\frac{1}{N-1}\left(\sum_{i=1}^{N}m_i^2-N\overline{m}^2\right),
\]
respectively.

For  a  homogeneous  marked   point  process,  the  joint  probability
$\rho_1^{SM}(\bx,m) \rd V \rd m$ of finding\footnote{For the sequel we
speak for reasons of simplicity  of ``finding at $\bx$ with mark $m$''
instead of  ``finding in  a volume element  $\rd V$ at  position $\bx$
with mark in the range $[m,m+\rd m]$''.}
a   point  at   position  $\bx$   with   mark  $m$,   splits  into   a
space--independent  mark probability  and the  constant  mean density:
$\rho_1^M(m)\rd m\times\rho\rd V$.   In general, the mark distribution
$\rho_1^M$ is not homogeneous.   Note that this notion of independence
does not rule out luminosity segregation at all and seems a physically
justified assumption, since it simply requires that no region of space
has an a priori specified mark distribution different from that one of
another region.

\subsection{Two--point properties}

The second--order properties of  the spatial distribution of the point
set    $X$    are   fully    specified    by   the    product--density
$\rho_2^S(\bx_1,\bx_2)\rd  V_1\rd  V_2$   giving  the  probability  of
finding a point at $\bx_1$ {\em  and} another point at $\bx_2$.  For a
stationary   and   isotropic   point   distribution   we   have   with
$r=|\bx_1-\bx_2|$
\begin{equation}
\label{eq:correlationdef}
\rho_2^S(\bx_1,\bx_2) = \rho^2 (1+\xi(r))
\end{equation}
with  the two--point correlation  function $\xi(r)$.
Similarly, second--order  properties of the marked point  set $X^M$ are
fully specified by the mark product--density:
\begin{equation}
\rho_2^{SM}((\bx_1,m_1),(\bx_2,m_2))\ \rd V_1 \rd m_1\ \rd V_2 \rd m_2 
\end{equation}
is the joint probability of finding  a galaxy at $\bx_1$ with the mark
$m_1$ and another point at $\bx_2$ with the mark $m_2$.
Hence  the (spatial)  product--density $\rho_2^S(\bx_1,\bx_2)$  is the
marginal density
\begin{equation}
\label{eq:product-as-marginal-space}
\rho_2^S(\bx_1,\bx_2) = \int\rd m_1 \int\rd m_2\ 
\rho_2^{SM}((\bx_1,m_1),(\bx_2,m_2)).
\end{equation}
With  an appropriate  chosen integration  measure  similar definitions
apply for discrete marks.\\
Now consider a finite domain $\CD$. The normalization of $\rho_2^S$ is
given by
\begin{equation}
\label{eq:def-n2}
\CNT 
=\int_\CD\rd x_1^3\int_\CD\rd x_2^3\ \rho_2^{S}(\bx_1,\bx_2) 
= \BE[N(N-1)], 
\end{equation}
with $N$  the number  of points of  one realization inside  $\CD$, and
$\BE$ the mean value over several realizations.\\
Respecting this normalization,  a marginal product density for the
marks can be defined by
\begin{equation}
\label{eq:product-as-marginal-marks}
\rho_2^M(m_1,m_2) = \frac{1}{\CNT}\int_\CD\rd x_1^3 \int_\CD\rd x_2^3\ 
\rho_2^{SM}((\bx_1,m_1),(\bx_2,m_2)).
\end{equation}
$\rho_2^M(m_1,m_2) \rd m_1\rd m_2$  quantifies the probability to find
the marks  $m_1$ and $m_2$  at two given  points in the  distribution. 
Mathematically,  $\rho_2^M(m_1,m_2)$   quantifies  a  real  two--point
property.  Physically, however,  we expect -- at least  in our case --
that intrinsic correlations in mark space are not present, i.e., that
\begin{equation}
\label{eq:no-mark-correlation}
\rho_2^M(m_1,m_2) = \rho_1^M(m_1)\rho_1^M(m_2).
\end{equation}
Otherwise the  probability of  finding a galaxy  with mark $m_i$  in a
fixed sample  would depend on  the other marks regardless  how distant
they  are, a  consequence  which may  seem  reasonable in  biosciences
(epidemiology) but not  in our case of large  galaxy surveys. In other
words,  {\em  spatial} mark  correlations  may  be  present, but  {\em
  globally} the presence  of a mark with value $m$  in the sample does
not prearrange the values of  the other marks.  
Typically,   the   one--point    mark   distribution   $\rho_1^M$   is
inhomogeneous in mark--space. Therefore, one cannot check the relation
{}\eqref{eq:no-mark-correlation}  from one  realization  only; several
independent samples are needed.  In  future redshift surveys it may be
possible to extract  approximately independent subsamples seperated by
a large distance, allowing for such a check.
Throughout      this      paper,      we      will      adopt      the
assumption~\eqref{eq:no-mark-correlation}.

\subsection{Mark correlations depending on the spatial distance}
\label{sect:lum-seg}

In the following we want to  know, whether the clustering in space and
the luminosity distribution are correlated.  We define the conditional
mark density:
\begin{equation}
\CM_2(m_1,m_2|\bx_1,\bx_2) = 
\begin{cases}
\frac{\rho_2^{SM}((\bx_1,m_1),(\bx_2,m_2))}{\rho_2^S(\bx_1,\bx_2)}
& \text{ for } \rho_2(\bx_1,\bx_2)\ne 0,\\
0 & \text{ otherwise }.
\end{cases} 
\end{equation}
For    a     stationary    and    isotropic     point    distribution,
$\CM_2(m_1,m_2|\bx_1,\bx_2)$ is the probability density\footnote{
  The notation $\CM_2(m_1,m_2|r)$ is  somehow imprecise, since it does
  not  remind us  of the  fact that  the marks  refer to  given points
  $\bx_1$ and $\bx_2$.  For simplicity,  we do not use a more accurate
  notation like $\CM_2(m_1(\bx_1),m_2(\bx_2)|r)$.  }
of  finding the  marks  $m_1$ and  $m_2$  at two  galaxies located  at
$\bx_1$ and  $\bx_2$, respectively, under the  condition that galaxies
at  these positions are  present in  the data.  For the  following, we
assume that  this quantity is only  a function of  the galaxy distance
$r=|\bx_1-\bx_2|$:  $\CM_2(m_1,m_2|r)$.  This  assumption  expresses a
sort of  homogeneity and isotropy,  however, it does not  presuppose a
well--defined mean density and is thus only a weak requirement.

The full mark product--density can be written as 
\begin{equation}
\label{eq:markproduct-factorization}
\rho_2^{SM}((\bx_1,m_1),(\bx_2,m_2)) = 
\CM_2(m_1,m_2|\bx_1,\bx_2)\ \rho_2^S(\bx_1,\bx_2).
\end{equation}
$\CM_2(m_1,m_2|r)$ is  a function depending on three  variables and is
therefore  hard  to  estimate.   With the  mark--weighted  correlation
functions  and  the  discrete  mark--correlation function  we  further
distill       the        information       as       discussed       in
Subsection~\ref{sect:mark-weighted}.\\
If  the  distribution  of  the  marks  is  {\em  independent}  of  the
distribution  of  the points,  the  conditional  mark density  becomes
independent of $r$:
\begin{equation}
\label{eq:markcorr-independence}
\CM_2(m_1,m_2|r) = 
\rho_1^M(m_1)\rho_1^M(m_2),
\end{equation}
Intuitively, this independence may be understood in the following way:
After  having distributed  galaxies in  space, we  choose marks  (as a
realization of a second independent stochastic process) and distribute
them randomly without any regard  to the clustering of the galaxies.\\
Equation  \eqref{eq:markcorr-independence}  is  the  basic  assumption
behind     projection      formulas     like     Limber's     equation
{}\citep{peebles:lss}.
If, on the other hand, $\CM_2(m_1,m_2|r)$ does depend on $r$, we speak
of  e.g., {\em  mark segregation}:  The probability  of  observing two
marks $m_1$ and $m_2$ (e.g.,  luminosities) on the galaxies at $\bx_1$
and $\bx_2$ varies with the separation $r$ of these two galaxies. \\
Note, that for every empirical  dataset of marked points (which we may
think of as realization of a marked point process) we can artificially
construct another  dataset with the  same spatial features  showing no
mark segregation  by redistributing the marks to  the points randomly.
This boostrap resampling strategy for  the marks provides a method for
testing the statistical significance of mark correlations.

\subsection{Spatial correlations depending on the marks}
\label{sect:spatial-on-marks}

There are  complementary definitions of  this sort of  independence or
luminosity segregation.  For example, we can think the other way round
and define a conditional density that there be two galaxies at $\bx_1$
and $\bx_2$, under the condition that their marks be $m_1$ and $m_2$:
\begin{equation}
\label{eq:def-S2}
\CS_2(\bx_1,\bx_2|m_1,m_2) = 
\begin{cases}
\frac{\rho_2^{SM}((\bx_1,m_1),(\bx_2,m_2))/\CNT}{\rho_2^M(m_1,m_2)}
& \text{ for } \CNT \rho_2^M (m_1,m_2)\ne 0,\\
0 & \text{ otherwise },
\end{cases} 
\end{equation}
with $\CNT$ given in  Eq.~\eqref{eq:def-n2}.  If the conditional space
correlation    is    independent   of    $m_1$    and   $m_2$,    then
$\CS_2(\bx_1,\bx_2|m_1,m_2)=\rho_2^{S}(\bx_1,\bx_2)/\CNT$.    In   the
case of luminosity  segregation, on the other hand,  the values of the
marks influence the spatial clustering.
Using $\CS_2$ we will discuss  the usual way of looking for luminosity
segregation in Subsect.~\ref{sect:usual-luminosity-seg}.

\subsection{$n$--point properties}

For  completeness  we   mention  that  $n$--point--properties  may  be
discussed in the same way. Basic quantities are the $n$--point product
densities   $\rho^{SM}_n  ((\bx_1,m_1),\ldots,(\bx_n,m_n))$   and  the
conditional  densities $\CM_n(m_1,\ldots,m_n|\bx_1,\ldots,\bx_n)$.  At
this  level   the  issue  may  be  re--discussed,   whether  the  mark
distribution depends on the spatial clustering.

Robust   statistics  for   the  clustering   of  galaxies   in  space,
incorporating  higher--order  correlations,  are  the  $J$--func\-tion
({}\citealt{vanlieshout:j},               {}\citealt{kerscher:regular},
{}\citealt{kerscher:global})    and    the    Minkowski    functionals
(\citealt{mecke:robust},        for         a        review        see
{}\citealt{kerscher:statistical}).    A   first   extension   of   the
$J$--functions  to  discretely  marked  point  sets  is  discussed  by
{}\citet{vanlieshout:indices}.  The application to galaxy catalogs and
the   generalization   for  continuous   marks   is  currently   under
investigation.

\section{Mark--weighted conditional correlation functions and 
models for marked point distributions}

Since the joint space  and mark product--density $\rho_2^{SM}$ and the
conditional mark  density $\CM_2$ depend on three  variables at least,
they  are  not  easy  to  handle.  Therefore,  we  discuss  quantities
accessible both  to straight--forward interpretation  and to numerical
estimation.    Particularly,   we   investigate   the   mark--weighted
conditional densities.

\subsection{Mark--weighted conditional correlation functions}
\label{sect:mark-weighted}

For a non--negative  weighting  function $f(m_1,m_2)$  we  define  the
average over pairs with separation $r$:
\begin{equation}
\label{eq:def-kappa-f}
\paverage{f}(r) = \int\rd m_1 \int\rd m_2\ f(m_1,m_2)\ \CM_2(m_1,m_2|r).
\end{equation}
$\paverage{f}(r)$ is the expectation  value of the weighting function $f$
(depending  only on the  marks), under  the condition  that we  find a
galaxy--pair with separation $r$ in the data.
With this definition we  separate the mark correlation properties from
the    spatial    clustering     properties    of    the    underlying
point--distribution, as can be seen directly from
\begin{equation}
\label{eq:kappa-f-separate}
\paverage{f}(r)  = 
\frac{\int\rd m_1 \int\rd m_2\ f(m_1,m_2)\ 
\rho_2^{SM}((\bx_1,m_1),(\bx_2,m_2))}{\rho^S_2(r)} 
\end{equation}
for $\rho^S_2(r)\ne0$.   We are  free to choose  appropriate weighting
functions adopted to our problem.   In the following we discuss common
choices from the literature and introduce some new ones.

\subsubsection{Continuous marks}
\label{sect:continuous-marks}

Using  several  positive  weighting  functions we  construct  statistical
indicators to  investigate the mark correlation properties  of a point
set (see  also {}\citealt{stoyan:fractals} and {}\citealt{schlather:mark},
we assume that the marks are positive numbers):
\begin{enumerate}
\item
At first we consider the mean mark:
\begin{equation}
k_{m}(r) = \frac{\paverage{m_1+m_2}(r)}{2\ \overline{m}} .
\end{equation}
$k_{m}$  equal unity  indicates the  absence of  mark  segregation.  A
preferred clustering  of marks e.g.,  $m>\overline{m}$ at a  scale $r$
can be concluded from $k_{m}(r)>1$.
\item
Closely  related is  Stoyan's  $k_{mm}$--function\footnote{Also called
(normalized) mark--correlation  function, see however  the comments by
\citet{schlather:mark}.}  {}\citep{stoyan:fractals}:
\begin{equation}
k_{mm}(r) = \frac{\paverage{m_1 m_2}(r)}{\overline{m}^2} .
\end{equation}
With $k_{mm}$ we  investigate the square of the  geometric mean of the
marks  on  points  at  a  distance  of  $r$.  Therefore,  a  preferred
clustering of marks at a  scale $r$ can be inferred from $k_{mm}(r)>1$
similar to  $k_{m}$.  Note that if the  mark is the mass  of a galaxy,
$k_{mm}$  may   serve  as  an  estimator  for   the  conditional  mass
correlations    $\BE[\varrho(0)\varrho(\bx)]/\rho^2_2(0,\bx)$,   where
$\varrho(\bx)$  is  the  mass--density  at  position  $\bx$,  thus  it
quantifies the ratio between galaxy and mass correlations.
\item
The mark variogram {}\citep{waelder:variograms} is defined by
\begin{equation}
\label{eq:def-gamma}
\gamma(r) = \paverage{\tfrac{1}{2}(m_1-m_2)^2}(r) 
= \paverage{m_1^2}(r) -  \paverage{m_1 m_2}(r) .
\end{equation}
$\gamma (r)$ equals the variance $V$ of the mark distribution, if mark
segregation is  absent; it  exceeds $V$ at  some scale $r$,  if points
that are about $r$ apart from  each other, tend to have very different
marks.
\item
Another tool  for investigating the variance of  the mark distribution
is the mark covariance function {}\citep{cressie:statistics}
\begin{equation}
\label{eq:def-cov}
\cov(r) 
= \paverage{m_1 m_2}(r) - \paverage{m_1}(r)\paverage{m_2}(r) 
= \paverage{m_1 m_2}(r) - \paverage{m_1}^2(r) .
\end{equation}
Thus,  luminosity  segregation  can  be detected  by  looking  whether
$\cov(r)$ does significantly differ from zero for some $r$.
\item
Both  $\gamma(r)$  and $\cov(r)$  mix  the  two--point and  one--point
fluctuations of  the mark distribution.  To  quantify the fluctuations
of  the mark  at  one point  only,  given there  is  another point  at
distance $r$, we suggest to use
\begin{equation}
\var(r) = \paverage{\left(m_1-\paverage{m_1}(r)\right)^2}(r).
\end{equation}
From  Eq.~\eqref{eq:def-gamma} and  {}\eqref{eq:def-cov}  one directly
obtains
\begin{equation}
\label{eq:var-gamma-cov}
\var(r) = \gamma(r) + \cov(r).
\end{equation}
\item 
Closely  related to  $\cov(r)$  is the  mark--correlation function  of
{}\citet{isham:marked}
\begin{equation}
\cor(r) = \frac{\paverage{m_1 m_2}(r) - \paverage{m_1}^2(r)}
{\paverage{m_1^2}(r) - \paverage{m_1}^2(r)}  
= \frac{\cov(r)}{\var(r)} ,
\end{equation}
the covariance function divided by the fluctuations of the mark.
\end{enumerate}

{}\citet{schlather:mark}  showed that  there  is an  ambiguity in  the
definitions of these mark characteristics at $r$ equal zero, but there
is no  problem for $r>0$.   Since we always  have to use a  finite and
non--zero $r$  to estimate these mark  characteristics, this ambiguity
is a technical point we do not need to consider further.
As   another   characteristic    for   marked   point   distributions,
{}\citet{capobianco:autocovariance}  consider  the  extension  of  the
$k_{mm}$ function on a two--dimensional grid.

\subsubsection{Discrete marks}
\label{sect:discrete-mark-weighted}

To  investigate   the  correlation  properties   between  galaxies  of
different  morphological types  the marks  $m_i$ are  chosen out  of a
finite range of attributes $m_i\in\{t_\alpha\}_{\alpha=1}^A$.  We also
could use other intrinsic  properties, like spectral features etc. of
the  galaxies to  define these  discrete marks.   Similarly,  a finite
binning may be used  for continuous marks.  Consider pairwise disjoint
bins $I_\alpha$  in luminosity  space, then the  mark is chosen  to be
$m_i=t_\alpha$ if the luminosity of the galaxy is $L_i\in I_\alpha$.

For  discrete  marks  the  following symmetric  weight  functions  for
$\alpha,\beta=1,\ldots,A$ are appropriate:
\begin{equation}
f_{t_\alpha t_\beta} (m_1,m_2) = 
\delta_{m_1t_\alpha} \delta_{m_2t_\beta} +
(1-\delta_{\alpha\beta}) \delta_{m_2t_\alpha} \delta_{m_1t_\beta} ,
\end{equation} 
where   the   Kronecker   $\delta_{m_1t_\alpha}$   equals   unity   if
$m_1=t_\alpha$      and     zero     otherwise.       According     to
Eq.~\eqref{eq:def-kappa-f}  we consider  the  (normalized) conditional
cross--correlation functions
\begin{equation}
C_{t_\alpha,t_\beta}(r) = \paverage{f_{t_\alpha t_\beta}}(r).
\end{equation}
Clearly,          $\sum_{\alpha=1}^{A}         \sum_{\beta=\alpha}^{A}
f_{t_\alpha,t_\beta} = 1$ and therefore also
\begin{equation}
\sum_{\alpha=1}^{A}\sum_{\beta=\alpha}^{A} C_{t_\alpha,t_\beta}(r) = 1 
\end{equation}
for all $r$.
If  the  marks  are  independent  on  the  distribution  in  space
one can show that
\begin{equation}
\label{eq:kappa-discrete-independent}
C_{t_\alpha,t_\beta}(r) = 
\frac{2\ \rho_{t_\alpha} 
\rho_{t_\beta}}{\rho^2}
\quad \text{ for } t_\alpha\ne t_\beta, \text{ and }\quad
C_{t_\alpha,t_\alpha}(r)  = 
\frac{\rho_{t_\alpha}^2}{\rho^2},
\end{equation}
with  the number density  $\rho_{t_\alpha}$ of  points with
mark $t_\alpha$.

Summarizing, there is a variety  of test quantities which allows us to
search  for luminosity  segregation  in real  data.   Note that  these
quantities are  applicable to  a single data  set without the  need of
constructing  a  series  of  volume--limited subsamples.   With  these
methods  we are  able to  gain new  insights into  the  luminosity and
morphological       dependent       clustering       of       galaxies
(Sect.~\ref{sect:lum-morph-seg-gal}).     As   we    will    show   in
Subsect.~\ref{sect:robust}, these methods break the degeneracy between
fractal spatial structure and luminosity segregation.

\subsection{Marked Poisson processes} 
\label{sect:marked-poisson}

Before applying  these test quantities  to real data we  explain their
properties  with a  simple  model, where  the  marks are  artificially
constructed from  the spatial pattern.  Other models  are discussed in
Subsect.~\ref{sect:randomfield} and Subsect.~\ref{sect:two-species}.

\begin{figure}
\epsscale{0.33}
\plotone{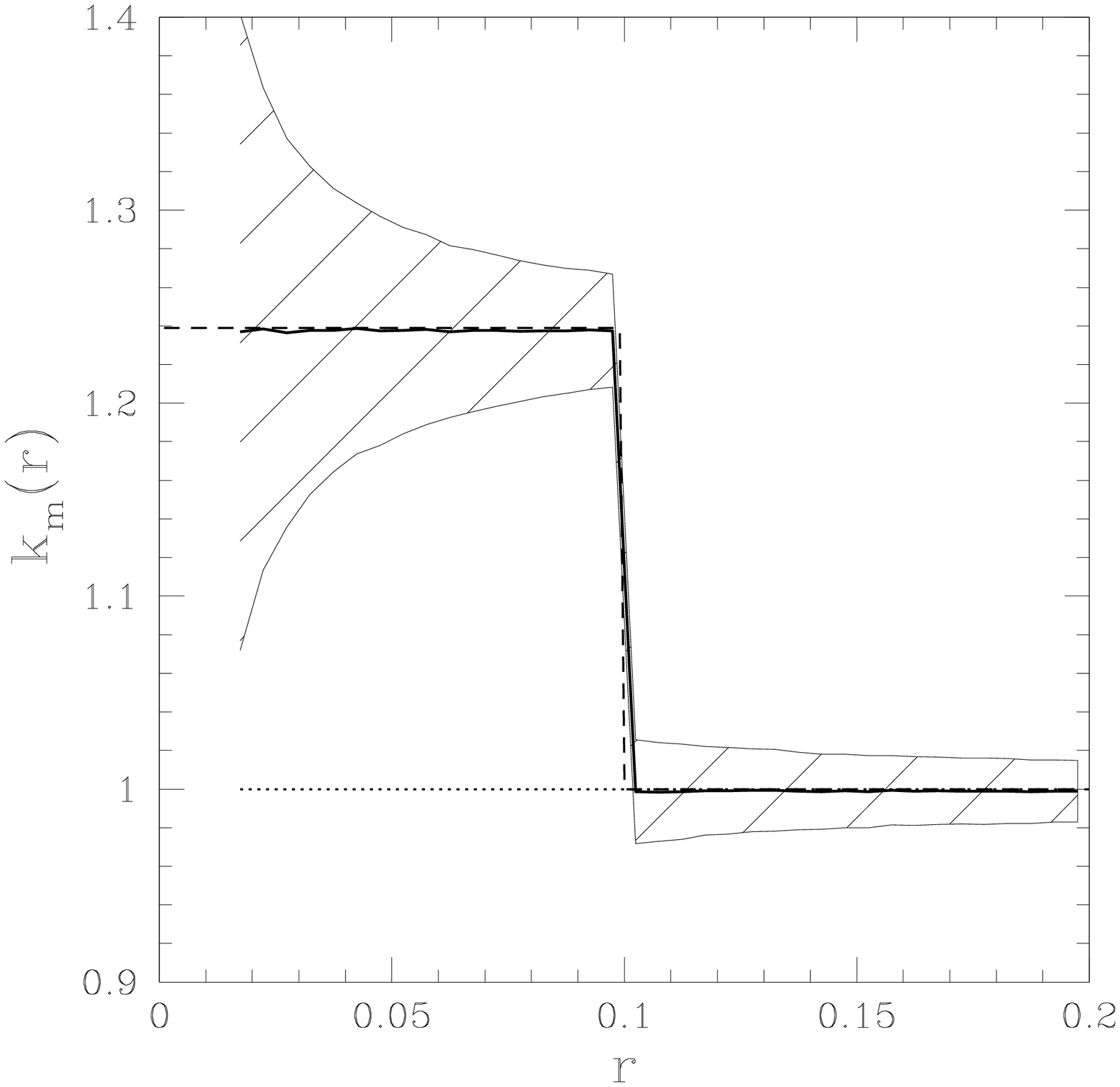} \plotone{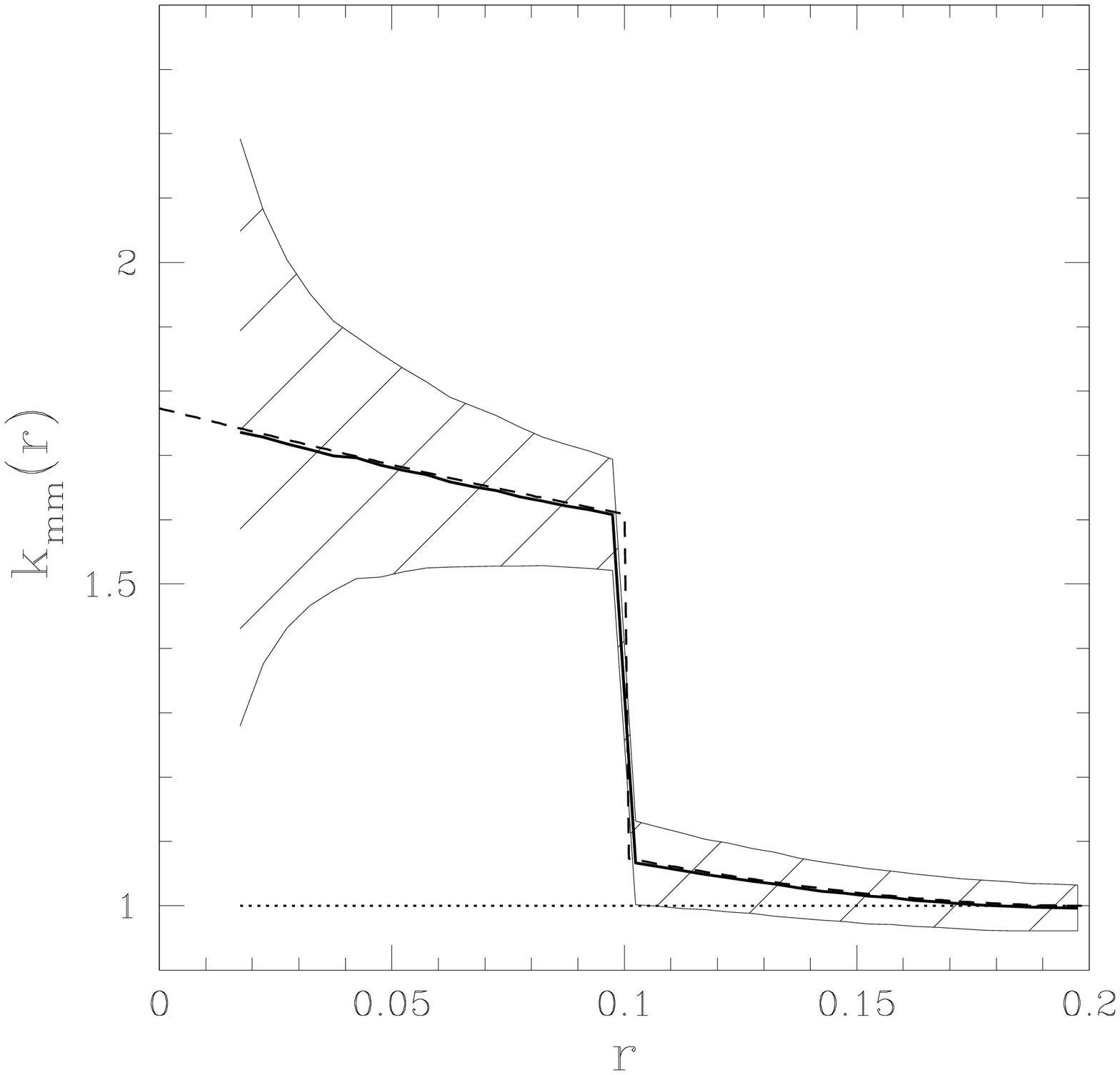}
\plotone{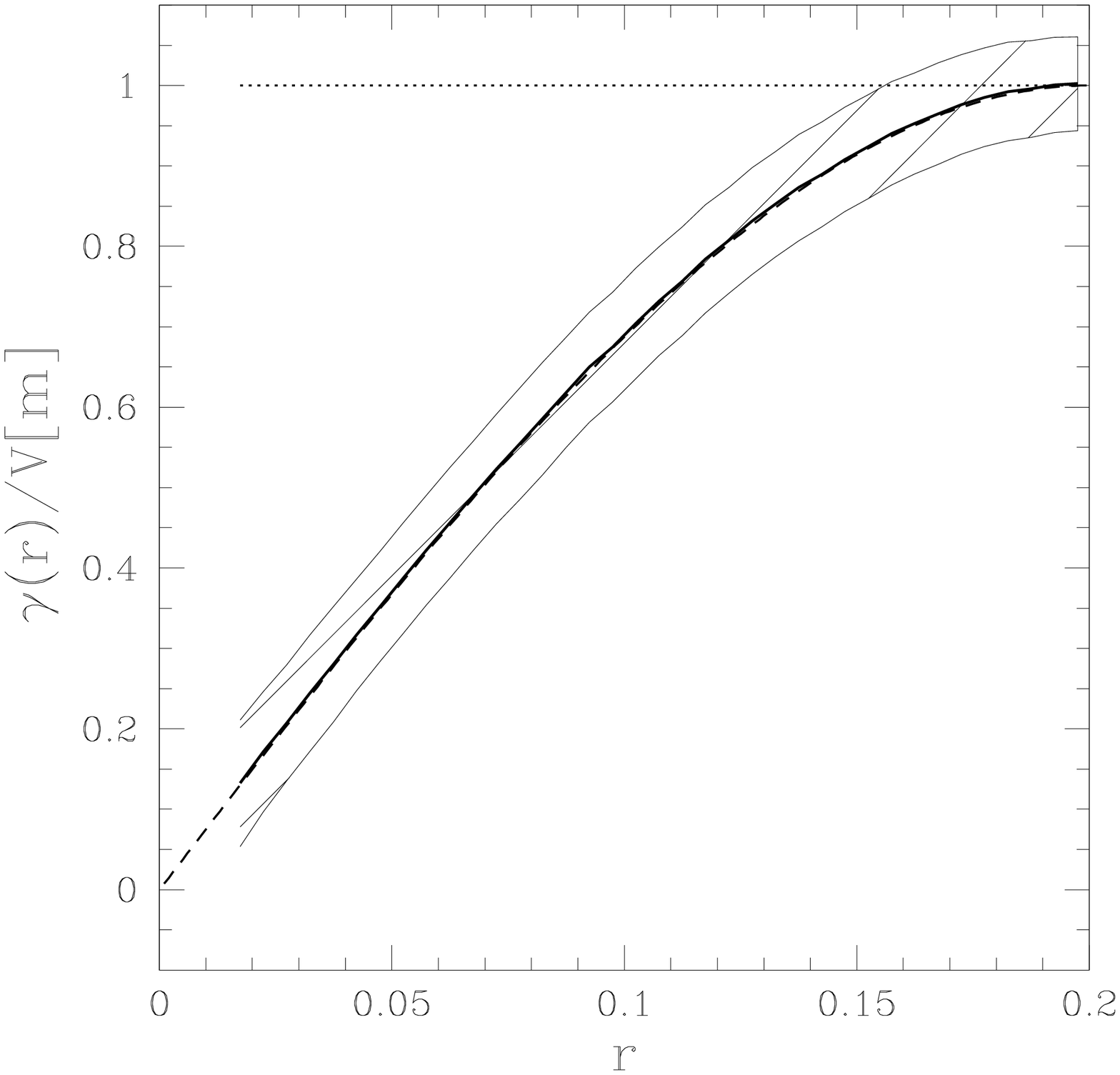} \plotone{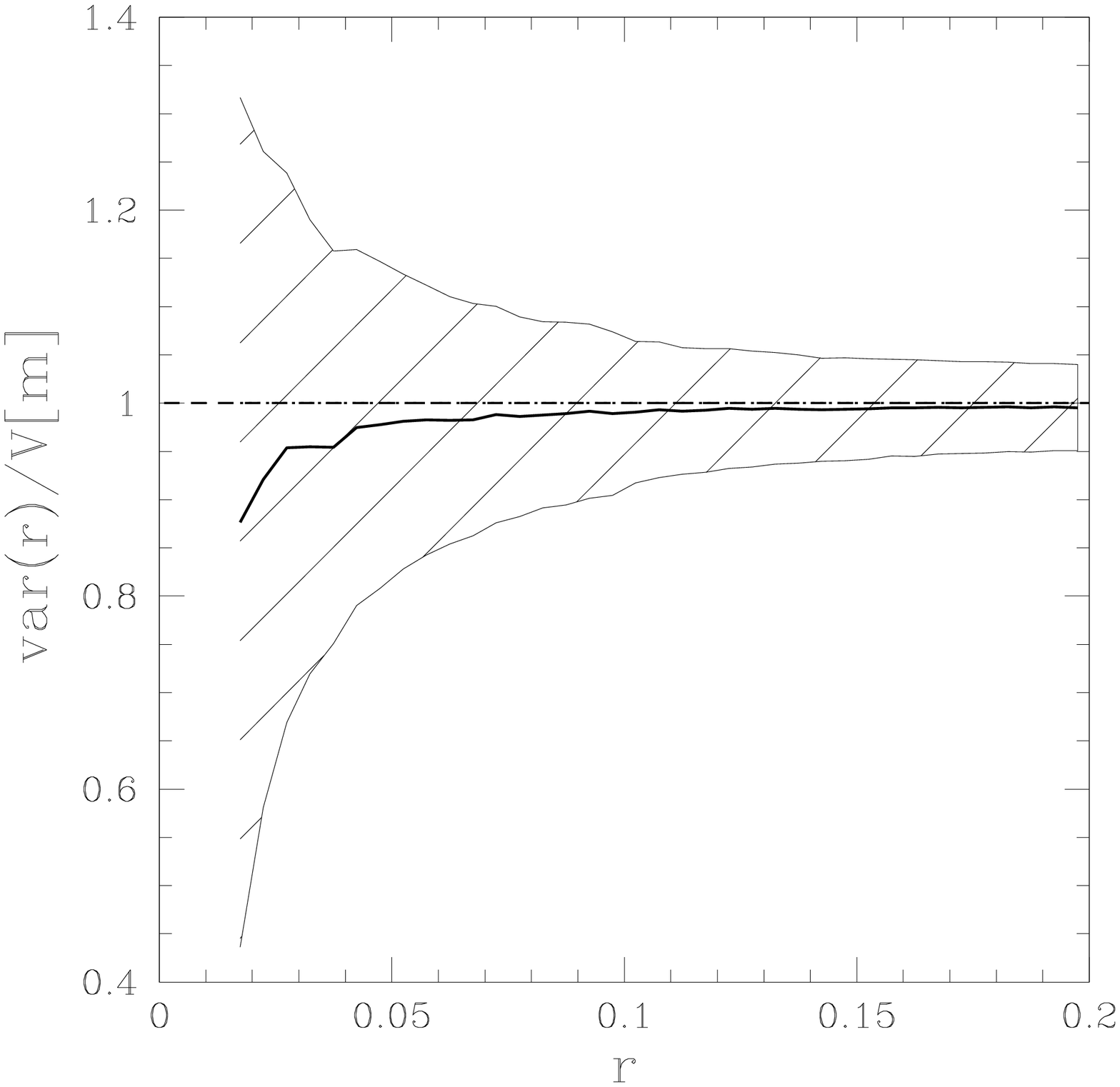}
\plotone{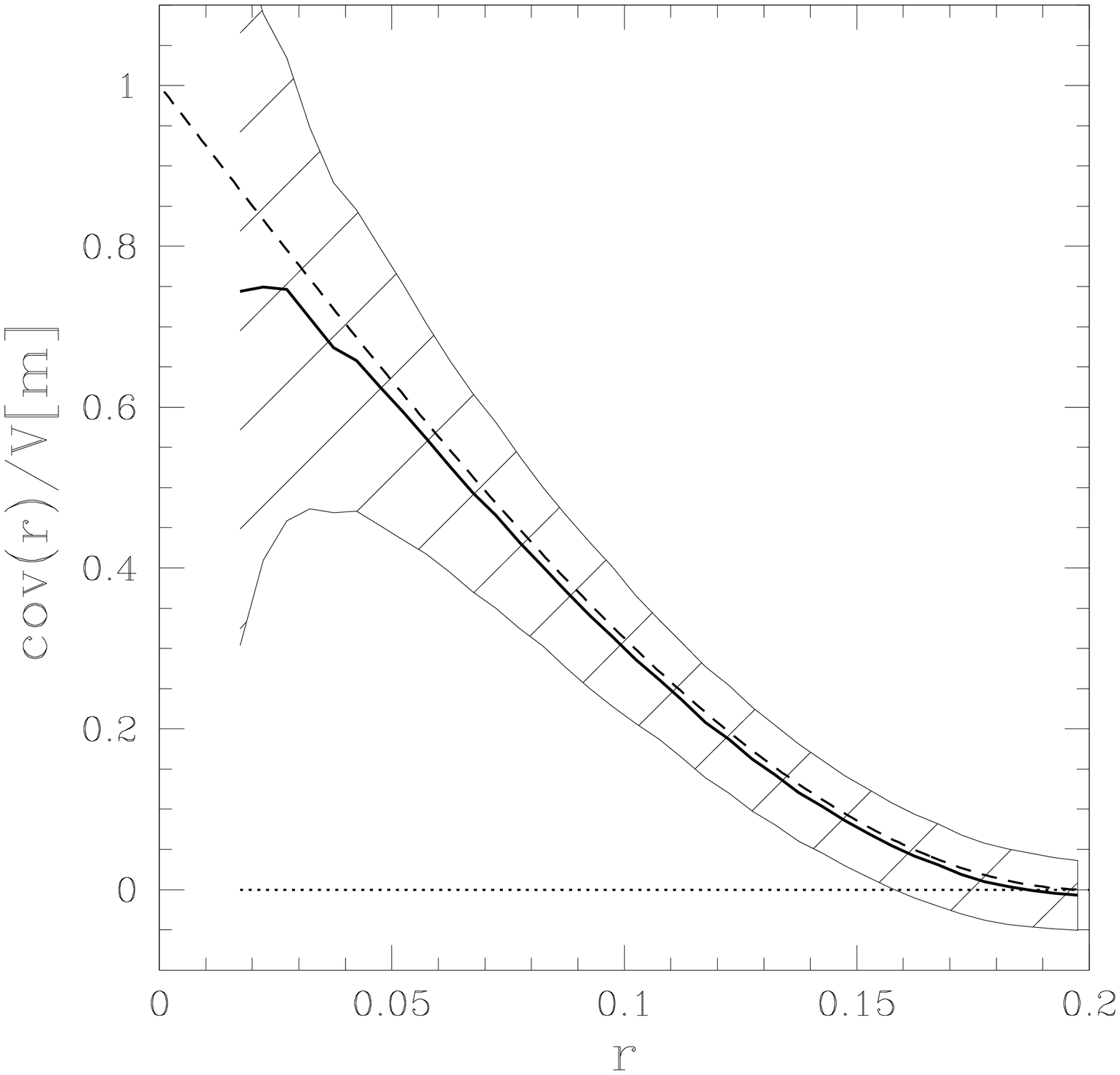} \plotone{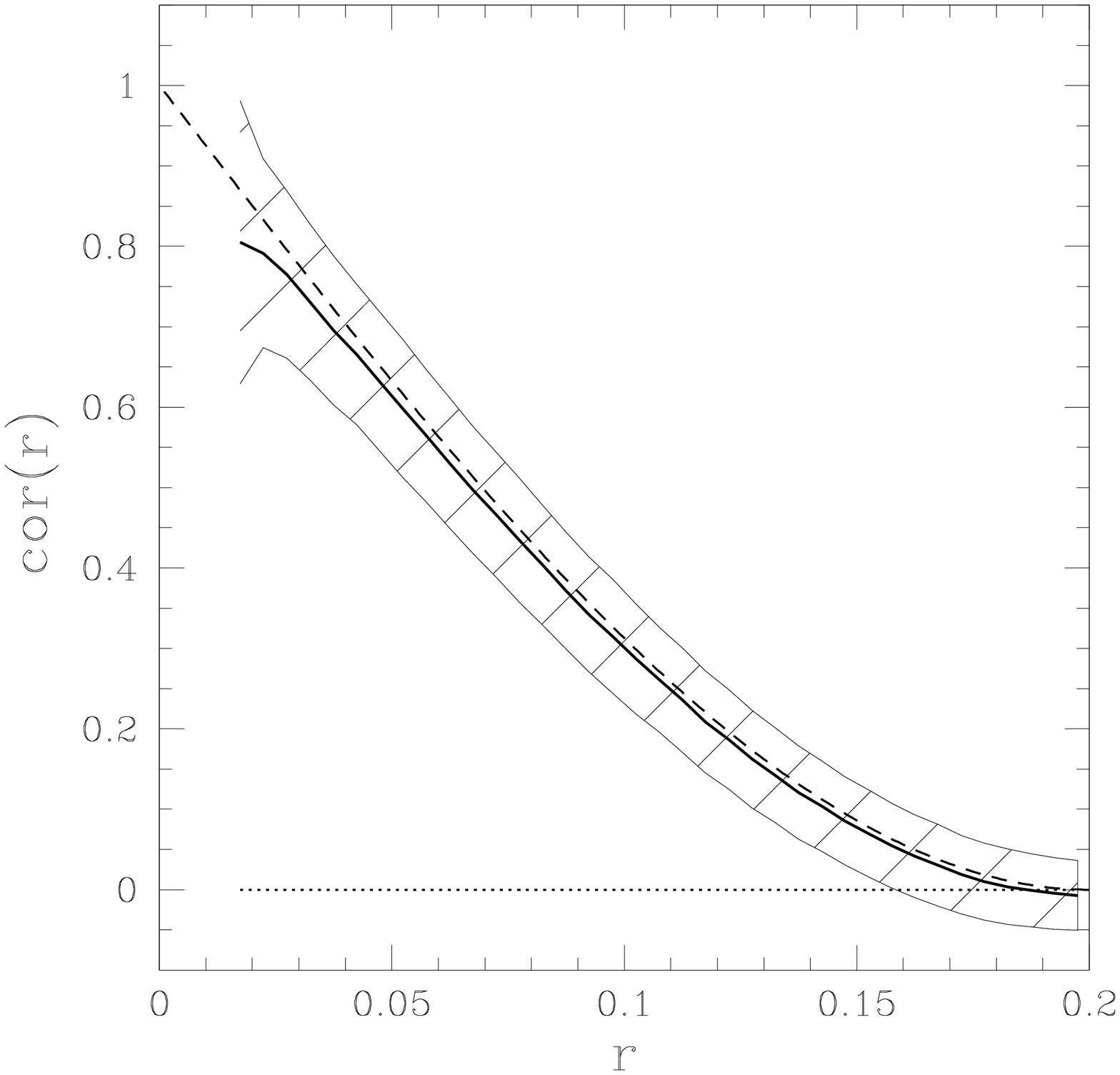}
\figcaption{\label{fig:nnmodel}  The $k_{m}(r)$, the  $k_{mm}(r)$, the
  normalized  $\gamma(r)/V$  and   $\var(r)/V$,  and  the  covariances
  $\cov(r)$  and $\cor(r)$ for  a marked  Poisson process  with number
  density of $1000$ inside the  unit box.  The points were marked with
  $N_i(0.1)$.  The  dashed line is the theoretical  result, the shaded
  area marks the 1$\sigma$  range estimated from 5000 simulations, and
  the dotted line  is the result for randomly  distributed marks.  $r$
  is given in units of the box--length.}
\end{figure}

We start with Poisson--distributed points $\bx_i$, with number density
$\rho$ and assign to each  point the mark $m_i=N_i(R)$, where $N_i(R)$
is the number of other points within a sphere of radius $R$ around the
point $\bx_i$.
Explicit  formulas for  $\gamma(r)$  and $k_{mm}(r)$  were derived  by
{}\citet{waelder:variograms}.   In  Fig.~\ref{fig:nnmodel} we  compare
numerical  simulations with  the theoretical  curves.
Points,  which are  members of  a pair  with small  separation,  are on
average  situated  in over--dense  regions,  have  more neighbors  and
therefore  get higher  marks.   This is  reflected  by $k_{m}(r)$  and
$k_{mm}(r)$  larger  than  unity   on  small  scales  ($k_{m}(r)$  and
$k_{mm}(r)$ indeed show a jump at $r=R$).
Since  nearby  points  get   similar  marks,  the  mark  variogram  is
suppressed  on small  scales,  which  can be  seen  directly from  the
reduced $\gamma(r)$  on small scales.  However,  the mean fluctuations
of the mark at one point  are not influenced by the presence of nearby
other points for a Poisson process, and consequently $\var(r)=V$.
The strong correlation of marks on  small scales can be seen also from
the     the     covariance     $\cov(r)$     and     correlation
$\cor(r)$.  
Empirically,  both $\cov(r)$  and  $\cor(r)$ and  also $k_{m}(r)$  and
$k_{mm}(r)$ exhibit  the same information content. We  also found this
in     our     analysis      of     the     galaxy     catalogs     in
Sect.~\ref{sect:lum-morph-seg-gal}.   Moreover,   $\gamma(r)$  may  be
expressed with $\cov(r)$ and $\var(r)$ (Eq.~\eqref{eq:var-gamma-cov}).
Therefore,  we  will  focus  in  the following  only  on  $k_{mm}(r)$,
$\var(r)$ and $\cov(r)$.

\section{Luminosity and morphological segregation in the galaxy 
distribution}
\label{sect:lum-morph-seg-gal}

Having clarified  the basic notion  of luminosity segregation,  we now
apply the above--defined characteristics  to real data and discuss the
empirical   question  whether   there  is   evidence   for  luminosity
segregation in the large--scale  structure of the galaxy distribution.
We  study  luminosity-- and  morphology--dependent  clustering in  the
Southern Sky  Redshift Survey~2 (SSRS2, {}\citealt{dacosta:southern}).
This survey is  99\% complete with a limiting  magnitude of $m_B=15.5$
within     the     region    $-40^\circ\le\delta\le-2.5^\circ$     and
$b\le-40^\circ$ and the  region $\delta\le0^\circ$ and $b\ge35^\circ$.
We will focus on a volume--limited subsample with 100\hMpc\ depth with
1179 galaxies.  We  obtained the same results looking  at samples with
different   limiting   depths   (see   Sect.~\ref{sect:errors}).    In
Sect.~\ref{sect:iras} we  compare with the results  from IRAS selected
samples.

\subsection{Luminosity as a continuous mark}
\label{sect:ssrs-luminosity-cont}

For  a galaxy  at  a distance  $r_i=|\bx_i|$  from our  galaxy with  a
magnitude $\text{mag}(\bx_i)$ the  luminosity $L_i$ is proportional to
$r_i^2\  10^{-0.4\ \text{mag}(\bx_i)}$.  Since  we look  at normalized
quantities, the absolute scaling of the luminosity is unimportant, and
we  assign to  a  galaxy  at $\bx_i$  the  mark $m_i=r_i^2\  10^{-0.4\
\text{mag}(\bx_i)}$.
To estimate $k_{mm}$, $\var$, and  $\cov$ we show the results obtained
with   the   estimator   without   boundary  corrections,   which   is
distinguished   by  its  simplicity   and  unbiasedness.    The  other
estimators gave fully consistent results.  A systematic examination of
the estimators  further justifying this approach is  given in Appendix
{}\ref{sect:estimators}.  The errorbars for  the case of no luminosity
segregation were estimated by randomly redistributing the marks of the
galaxies, keeping their positions in space fixed.

\begin{figure}
\epsscale{0.33}
\plotone{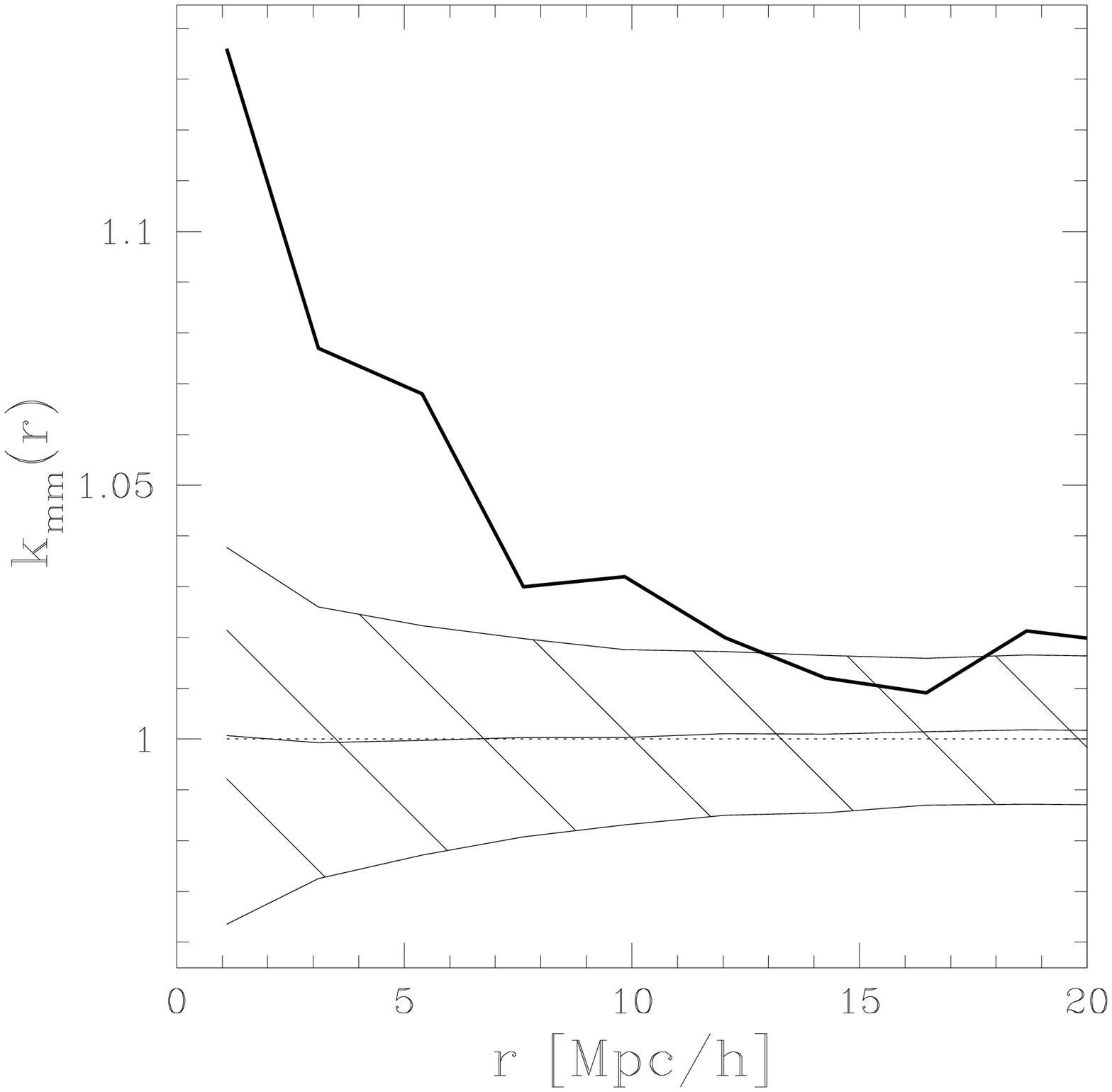}\plotone{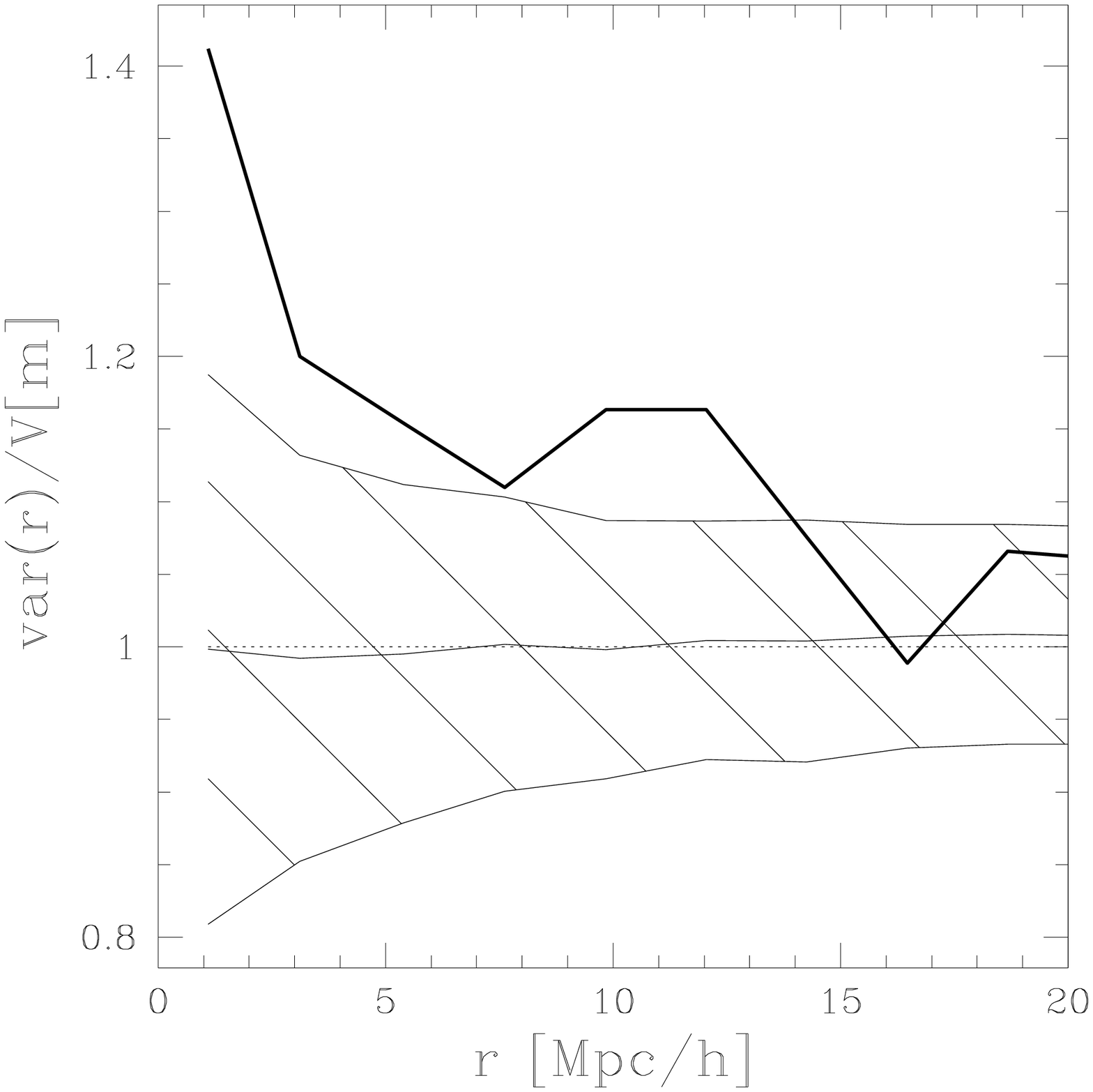}\plotone{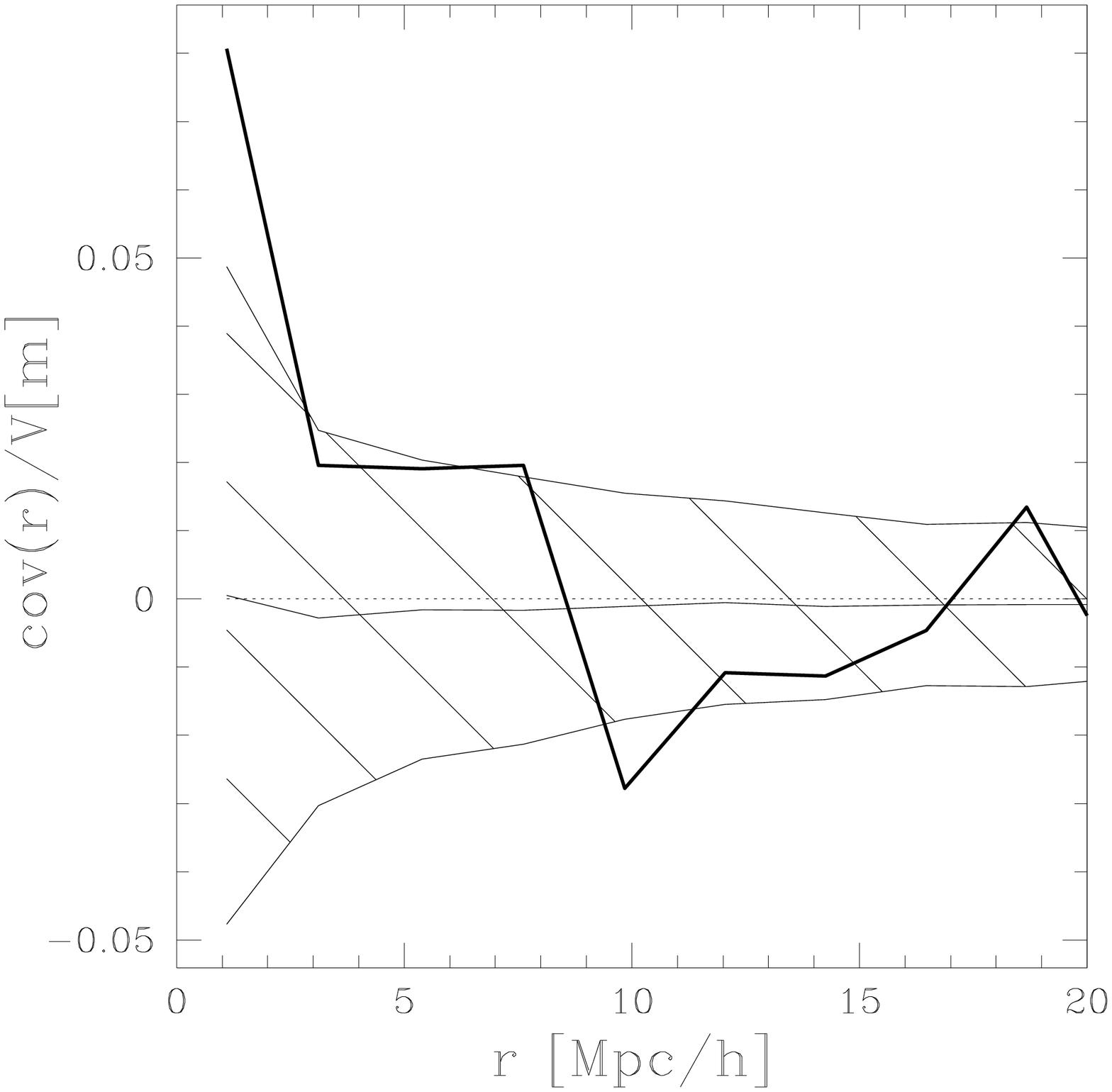}
\caption{\label{fig:ssrs2-markcorr}  The  mark--correlation  functions
for  a  volume--limited  subsample  of  the  SSRS2  with  a  depth  of
$100\hMpc$.   The  shaded  areas  denote  the  1--$\sigma$  error  for
randomized marks estimated from 1000 realizations.}
\end{figure}
Already at  a first glance  Fig.~\ref{fig:ssrs2-markcorr} reveals that
all test quantities show evidence  of luminosity segregation at a high
level of significance, especially $k_{mm}$ and $\var$.  The increasing
$k_{mm}$  towards small  scales  supports the  hypothesis that  bright
galaxies exhibit stronger clustering  than the dim ones ($k_{m}$ shows
the same feature).
The  strong  signal of  $\var$  is  a  result which  escaped  previous
analyses;  the  luminosity  {\em  fluctuations}  of  galaxies  with  a
neighbor  closer  than  $15\hMpc$   are  enhanced,  showing  that  the
luminosity distribution  is broader for these galaxies  in addition to
their higher mean luminosity as detected by $k_{mm}$.
Both  $k_{mm}$ and $\var$ show a signal out
to  15\hMpc,  indicating  that  luminosity  segregation  is  not  only
confined to clusters of galaxies.
The  covariance  $\cov$ measures  the  correlations between  the
luminosities on {\em both} galaxies.   It shows only weak evidence for
luminosity  segregation on  large scales,  however, on  scales smaller
than  3\hMpc,  the  $\cov>0$  indicates  an  excess  correlation
between the luminosities of two galaxies: Close pairs of galaxies tend
to assume similar luminosities.
At $r\sim 10\hMpc$, $k_{mm}$ and especially $\var$ show a second peak,
indicating that  the average  luminosity of the  galaxy pairs  and the
fluctuations of  the luminosity on  each galaxy are  enhanced.  $\cov$
shows  a  negative minimum  corresponding  to  an increased  diversity
between the luminosities of the  two galaxies. Clearly this is at most
a  two--$\sigma$  result,  however   these  features  also  appear  in
volume--limited samples with different depths.

\subsection{The random field model} 
\label{sect:randomfield}

To understand  the data  in more detail  we compare with  a particular
model  for  marked  point   processes  which  shows  mark  segregation
{}\citep{waelder:variograms}.   In the  {\em random  field  model} the
marks $m_i$ are  assigned to the points $\bx_i$  of a (unmarked) point
process   using   an   {\em   independent}  random   field   $u(\bx)$:
$m_i=u(\bx_i)$.  This  is a basic model in  geo--statistics (see e.g.,
{}\citealt{cressie:statistics}).  If the  point process and the random
field are homogeneous,  so is the marked point  process.  In this case
one obtains for $r>0$ \citep{waelder:variograms}
\begin{equation}
k_{mm}(r) = \frac{1}{\overline{m}^2} \BE[u(0)u(r)] ,
\end{equation}
and 
\begin{equation}
\gamma(r) = \frac{1}{2} \BE \left[ (u(0)-u(r))^2\right] .
\end{equation}
Here, $\BE$  is the  average over several  realizations of  the random
field,  thus $\BE[u(0)u(r)]$  is the  covariance of  the  field. Using
well--known     properties     of     random     field     covariances
{}\citep{adler:randomfields,waelder:variograms},  a  relation for  the
random field model can be derived:
\begin{equation}
\label{eq:random-field-constraint-1}
\gamma(r)  = \BE \left[u(0)^2\right] - 
\overline{m}^2 k_{mm}(r) 
= V + \overline{m}^2  - \overline{m}^2 k_{mm}(r).
\end{equation}
This  enables  us  to test  whether  a  marked  point process  may  be
understood in terms  of the random field model.  --  Note, that in the
random field model the marks  are given by an underlying random field,
which is not affected by the spatial distribution of the points.  This
does not cover the general case,  where the marks on the points may be
influenced by  spatial interactions  of the points,  as in  the marked
Poisson  process  in  Sect.  \ref{sect:marked-poisson}.   Indeed,  the
relation  {}\eqref{eq:random-field-constraint-1} is not  fulfilled for
this model as inferred directly from Fig.~\ref{fig:nnmodel}.

\begin{figure}
\epsscale{0.33}
\plotone{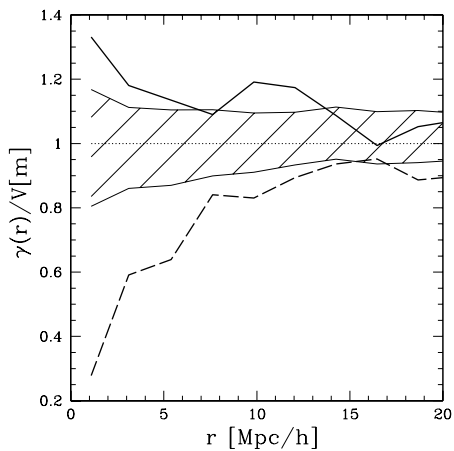}
\caption{\label{fig:ssrs2-randomfield-test}    Both   the   normalized
variogram    $\gamma(r)/V$   (solid    line)   and    the   normalized
$\gamma^{\text{rf}}(r)/V$  (dashed   line),  calculated  according  to
Eq.~{}\eqref{eq:random-field-constraint-1} are  shown. The shaded area
is marking  the 1--$\sigma$  region for $\gamma(r)/V$  with randomized
marks. }
\end{figure}
From Fig.~\ref{fig:ssrs2-randomfield-test} we  see that for the galaxy
distribution   the    estimated   variogram   $\gamma(r)$    and   the
$\gamma^{\text{rf}}(r)$,                calculated                from
Eq.~\eqref{eq:random-field-constraint-1}, show  the opposite behavior. 
Hence, the  luminosity segregation observed in this  galaxy sample can
not be described by a random field model. Therefore, the luminosity of
a galaxy  does not trace  an independent luminosity field,  but rather
depends  on the  spatial interactions  with other  galaxies.   Such an
interaction  is expected  physically  in clusters  of galaxies,  where
galaxies  merge. Beyond  cluster  scales this  ``interaction'' may  be
caused  by a common  origin in  the same  large--scale feature  of the
density distribution.

\subsection{Luminosity classes as discrete marks}
\label{sect:ssrsii-lum-discrete}

\begin{figure}
\epsscale{0.33}
\plotone{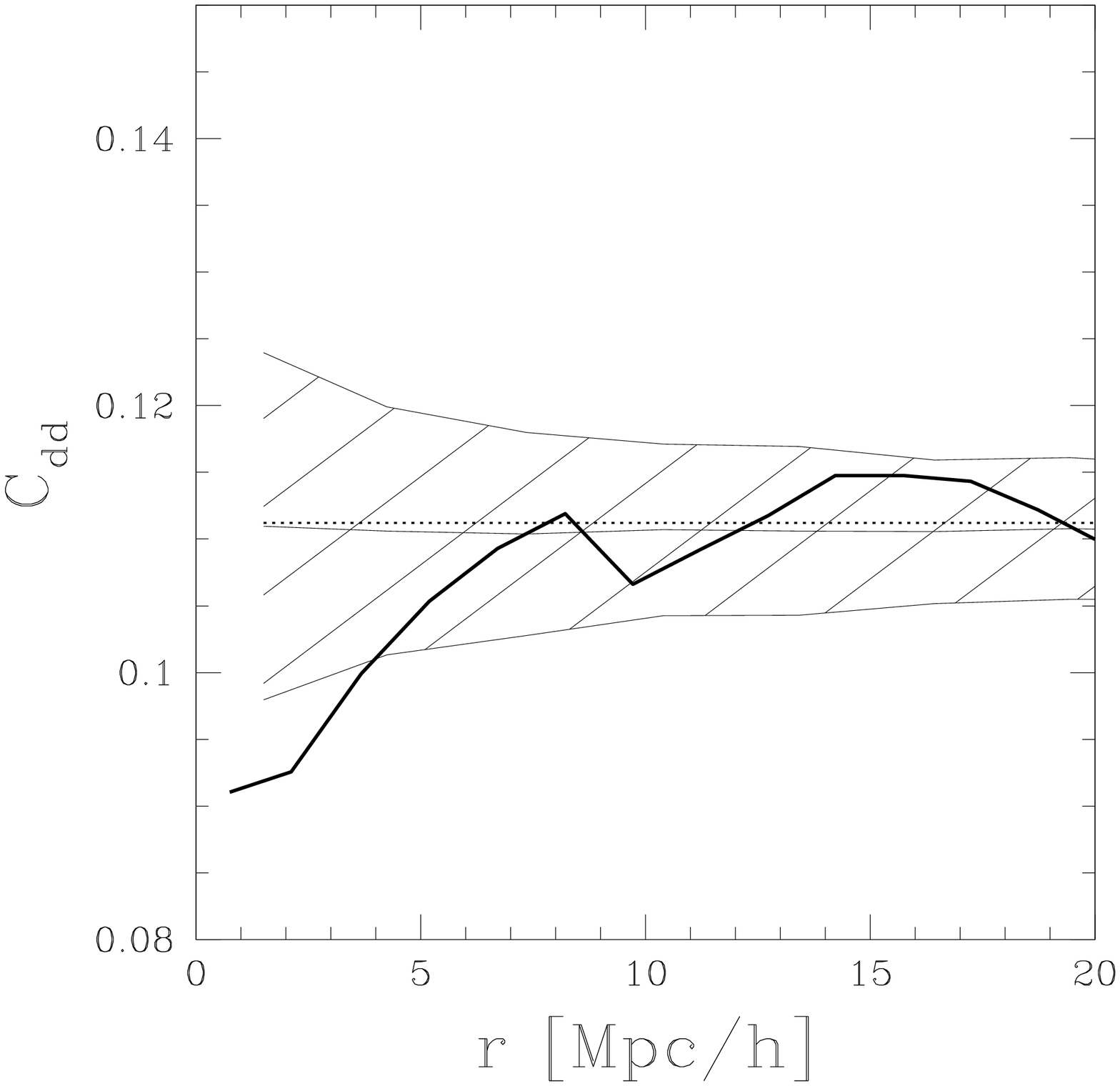}\plotone{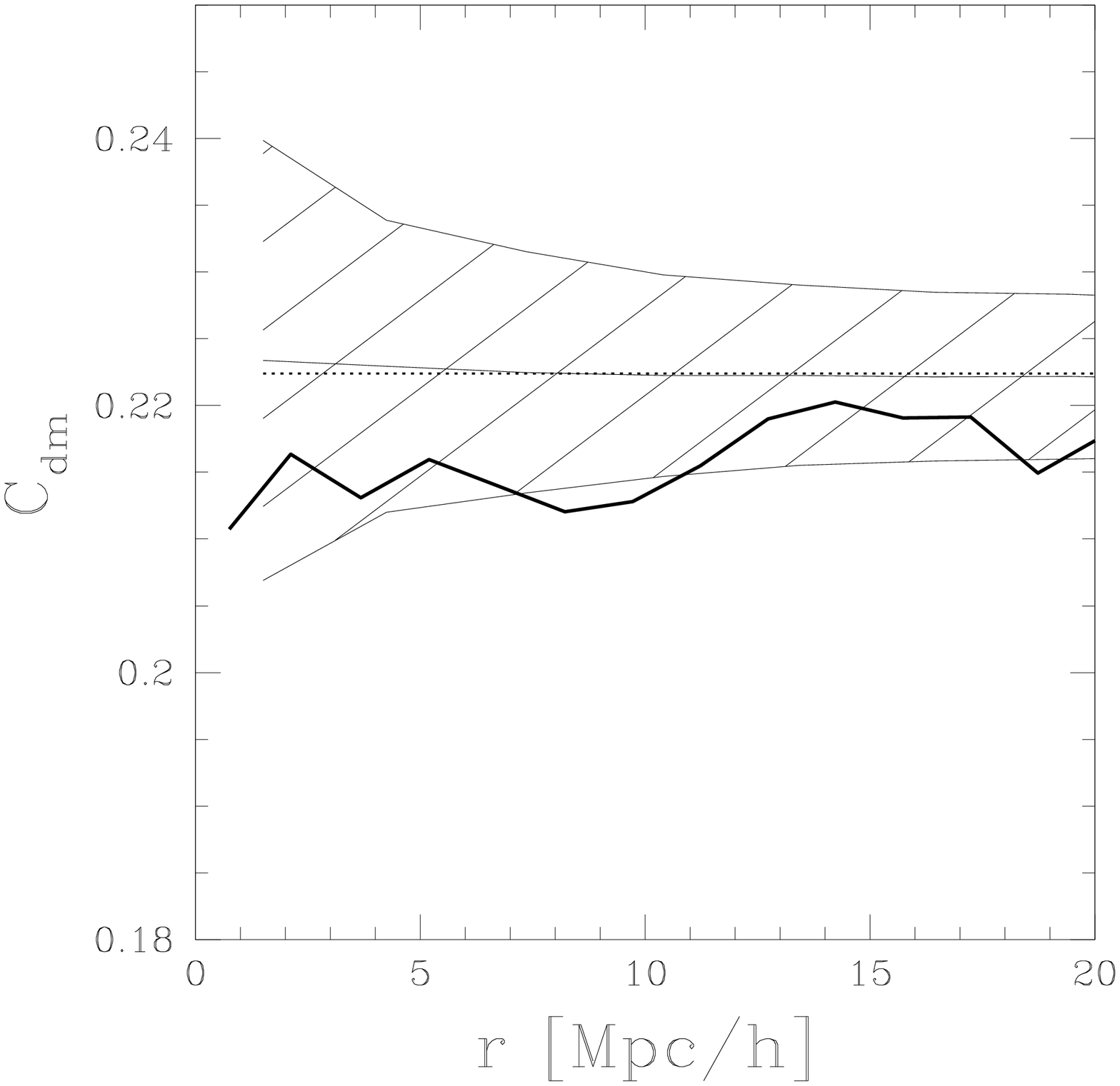}
\plotone{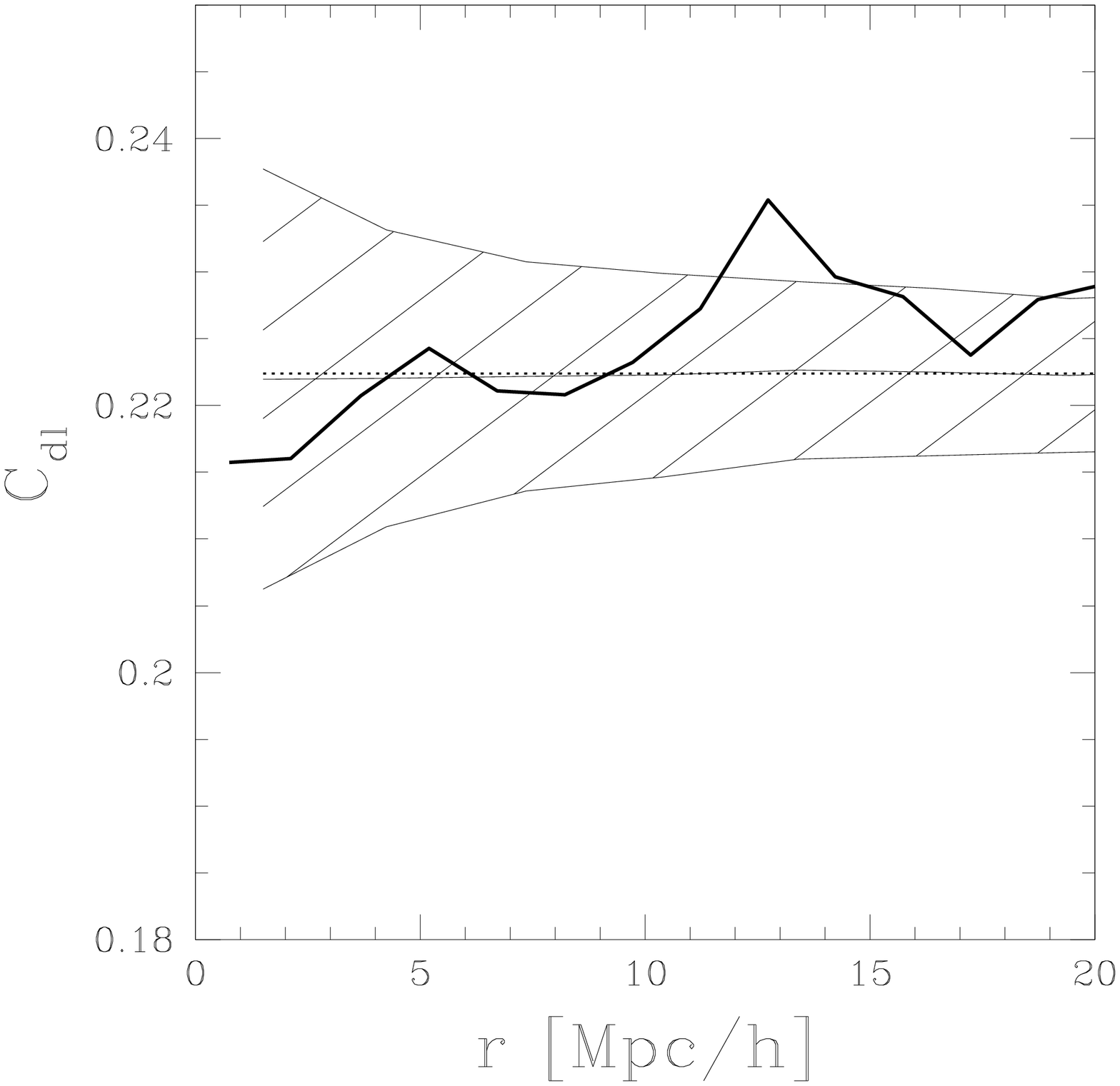}\plotone{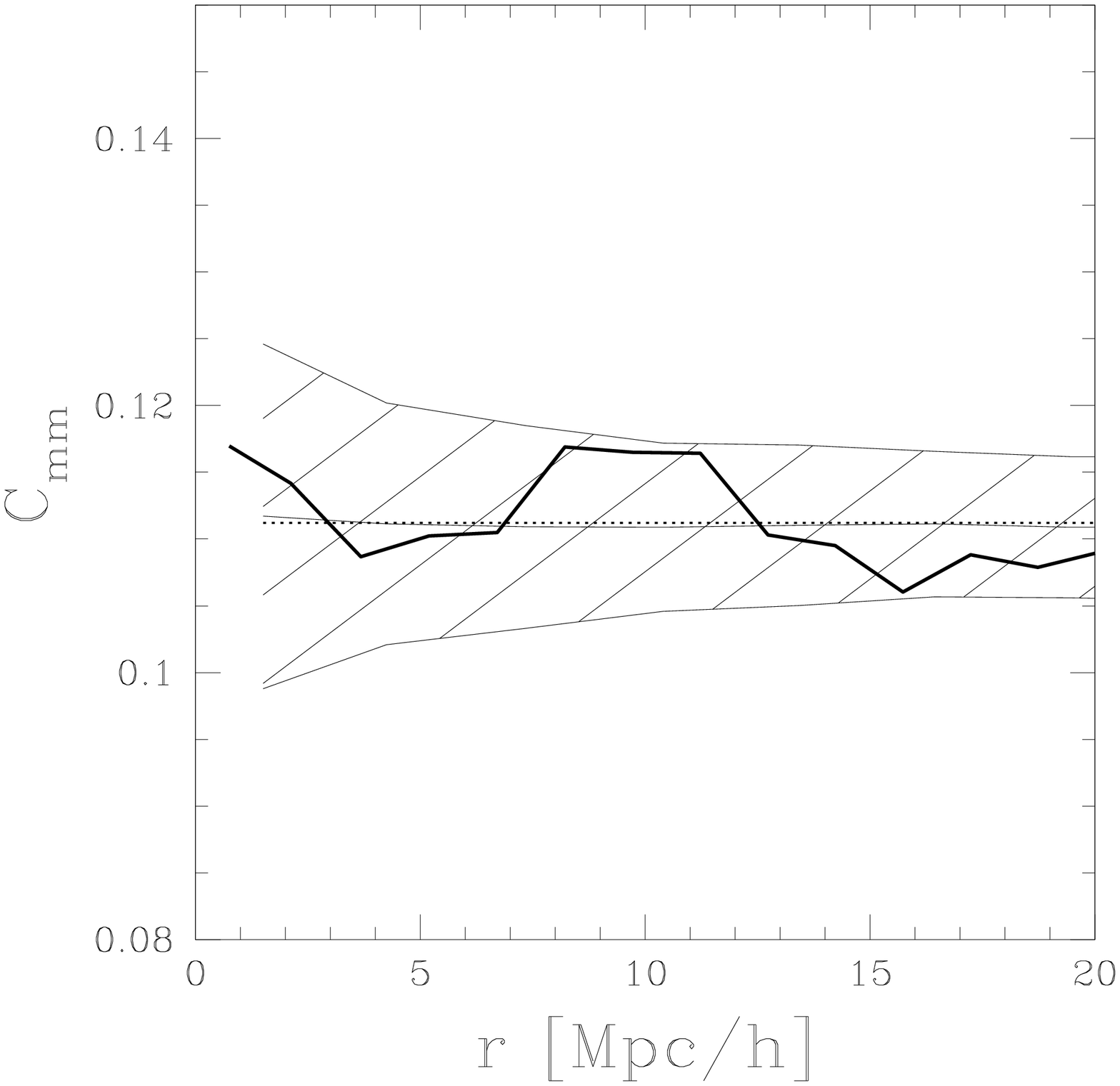}
\plotone{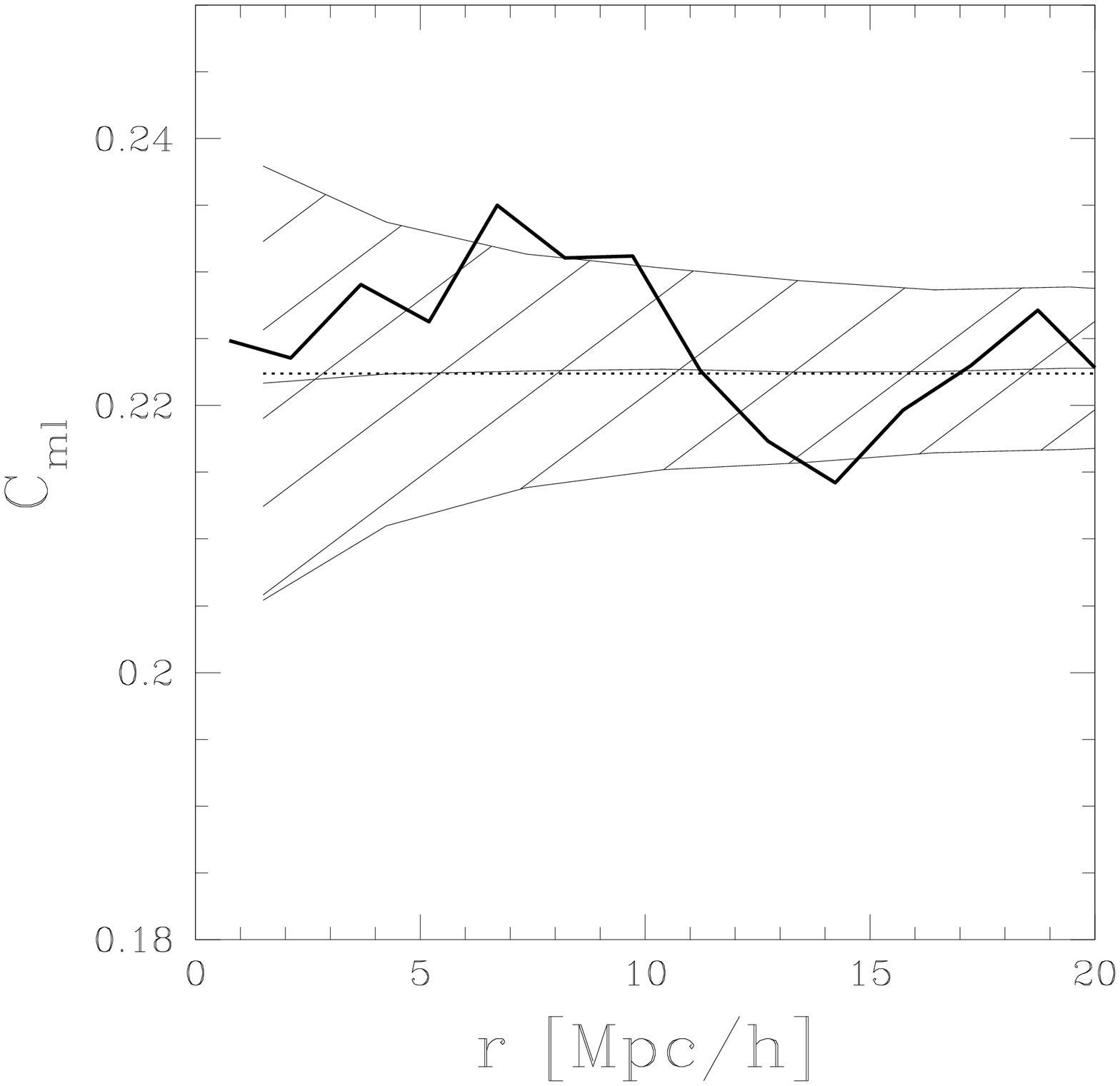}\plotone{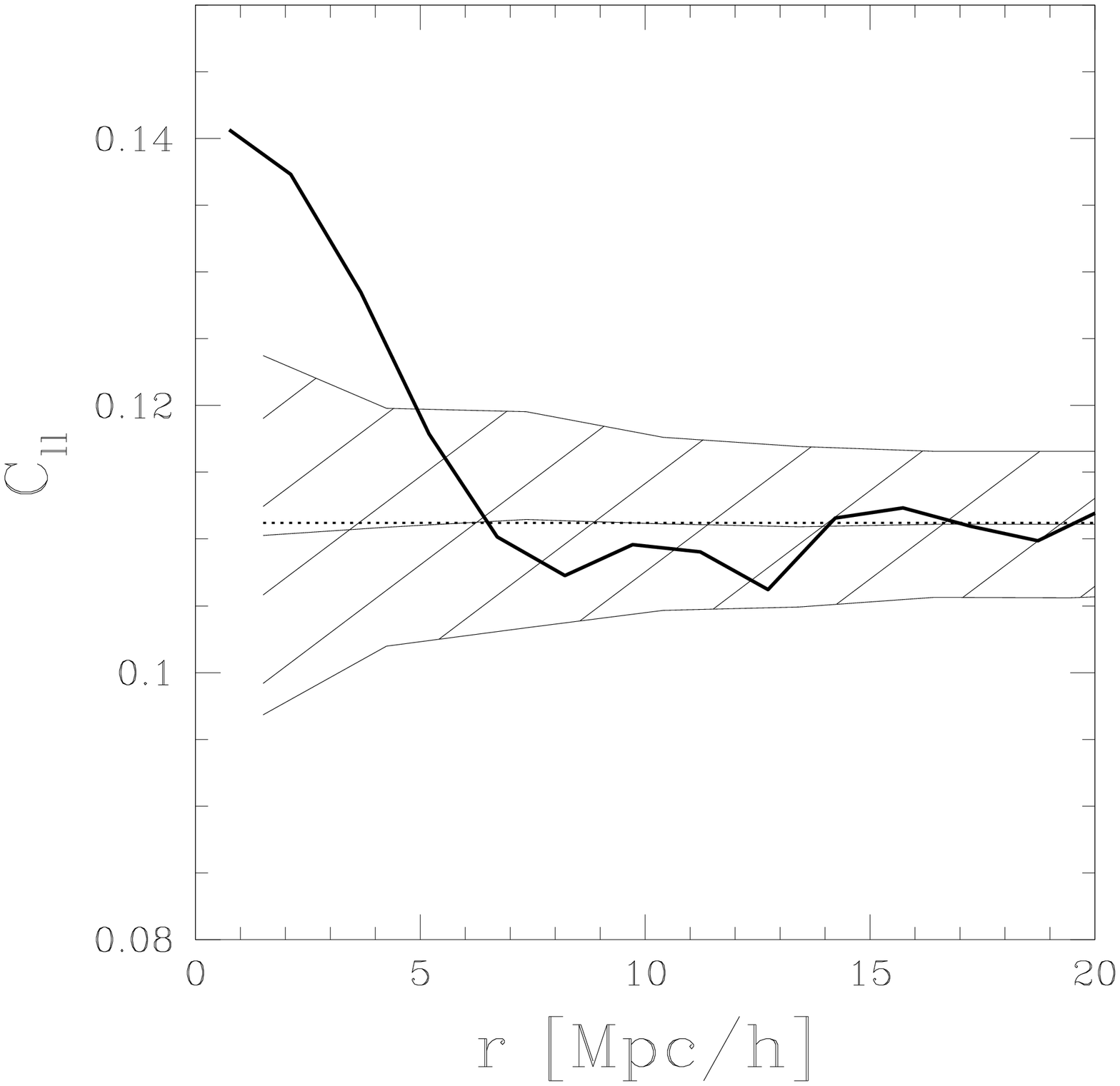}
\caption{\label{fig:luminosity-crosscorr}        The       conditional
cross--correlation functions of dim  ($d$), medium ($m$), and luminous
($l$)  galaxies in  the volume--limited  sample with  100\hMpc\ depth.
The shaded  1--$\sigma$ region  was determined from  1000 realizations
with randomized marks.}
\end{figure}

Now  we split  the volume--limited  sample with  100\hMpc\  depth into
three distinct  subsamples with  393 galaxies each.   These subsamples
consist of  luminous, medium and  dim galaxies, labeled with  $l$, $m$
and $d$ respectively.
The  conditional   cross--correlation  functions  $C_{dd}$,  $C_{dm}$,
$C_{dl}$,    $C_{mm}$,    $C_{ml}$,     $C_{ll}$    are    shown    in
Fig.~\ref{fig:luminosity-crosscorr},      estimated      from      the
volume--limited  sample  with  100\hMpc\  depth  using  the  estimator
employing no boundary correction.
They  show  that our  above  interpretation  of  $k_{mm}(r)$ based  on
Fig.~\ref{fig:ssrs2-markcorr}  points  into  the right  direction.  At
scales up to $5\hMpc$, the  bright galaxies cluster more strongly than
the other ones, this effect is at the expense of the dim galaxies, the
galaxies  with medium  luminosities  do not  contribute to  luminosity
segregation.
However, an  analysis based on  luminosity classes cannot  explain the
strong peak  of $\var$  and $\cov$ at  small scales since  both embody
{\em  fluctuations} of  the marks. 
Note,  that this  partition in  luminous, medium  and dim  galaxies is
arbitrary and  neither physically justified nor  suggested directly by
the data.  We  also divided the sample into  two luminosity classes of
equal size.  Here the  cross--correlations are all compatible with the
randomized results and no  luminosity segregation seems to be present.
This  emphasizes  the  discriminative  power of  the  continuous  mark
correlation  functions: The  conditional  cross--correlation functions
for the binned marks may be blind to luminosity segregation.  But with
a  carefully   adapted  binning  they  are  able   to  strengthen  the
conclusions obtained with the continuous mark--correlations functions.

\subsection{Morphological types as discrete marks}
\label{sect:ssrs2-morphology}

Using  the morphological type  of a  galaxy as  a mark  we investigate
morphological  segregation  using  the conditional  cross--correlation
functions defined in Sect.~\ref{sect:discrete-mark-weighted}.

\begin{figure}
\epsscale{0.33}
\plotone{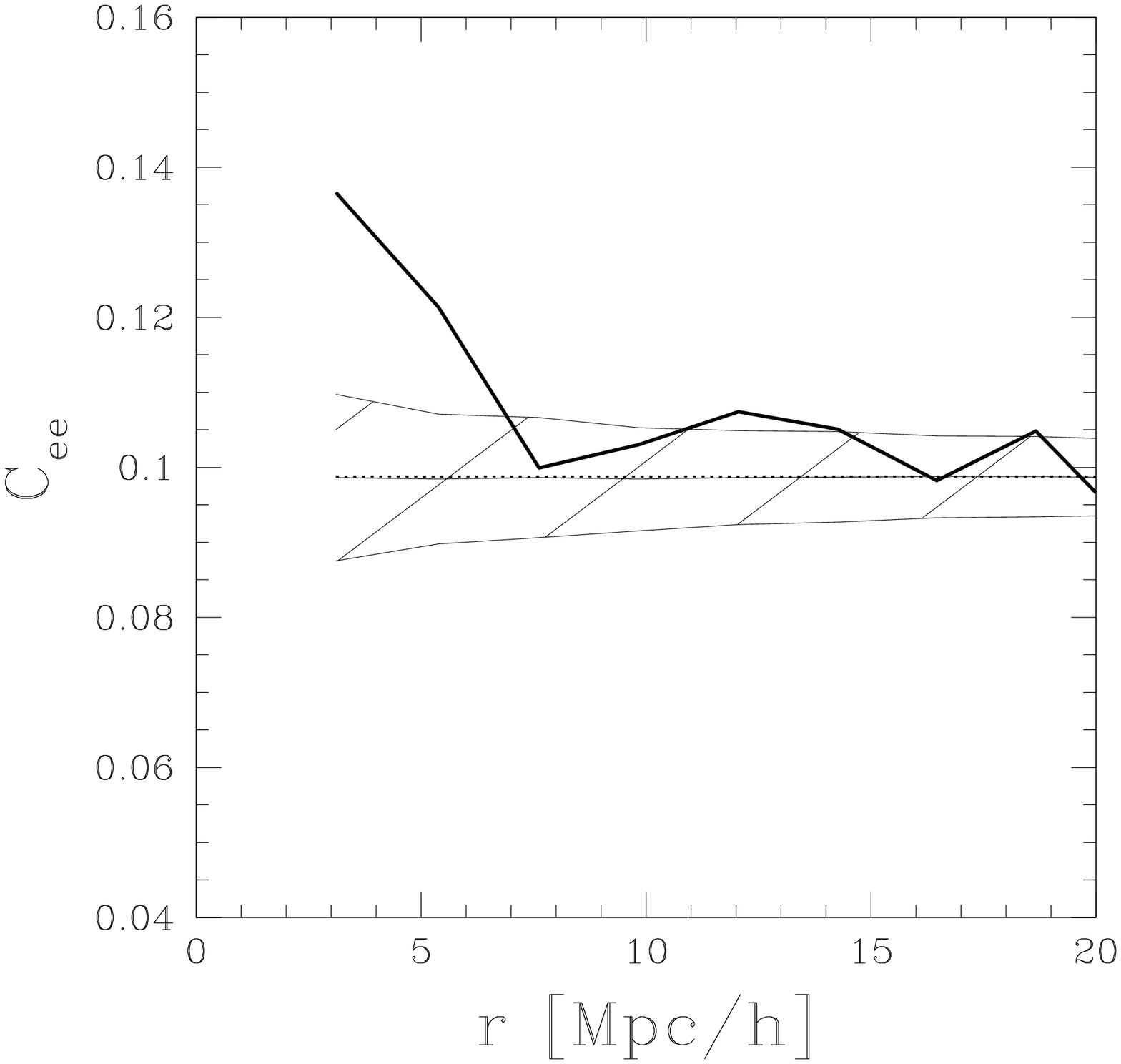}\plotone{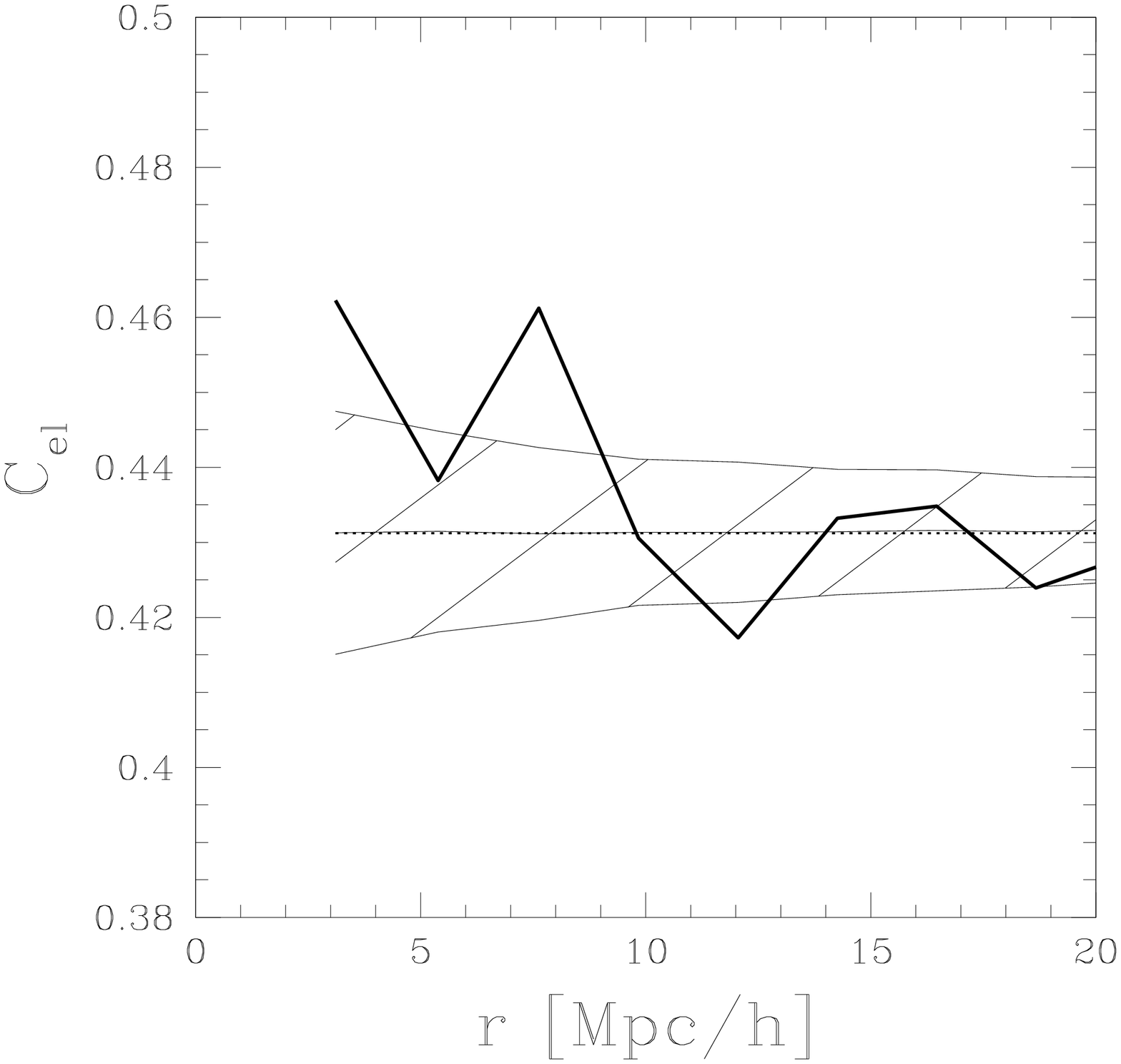}\plotone{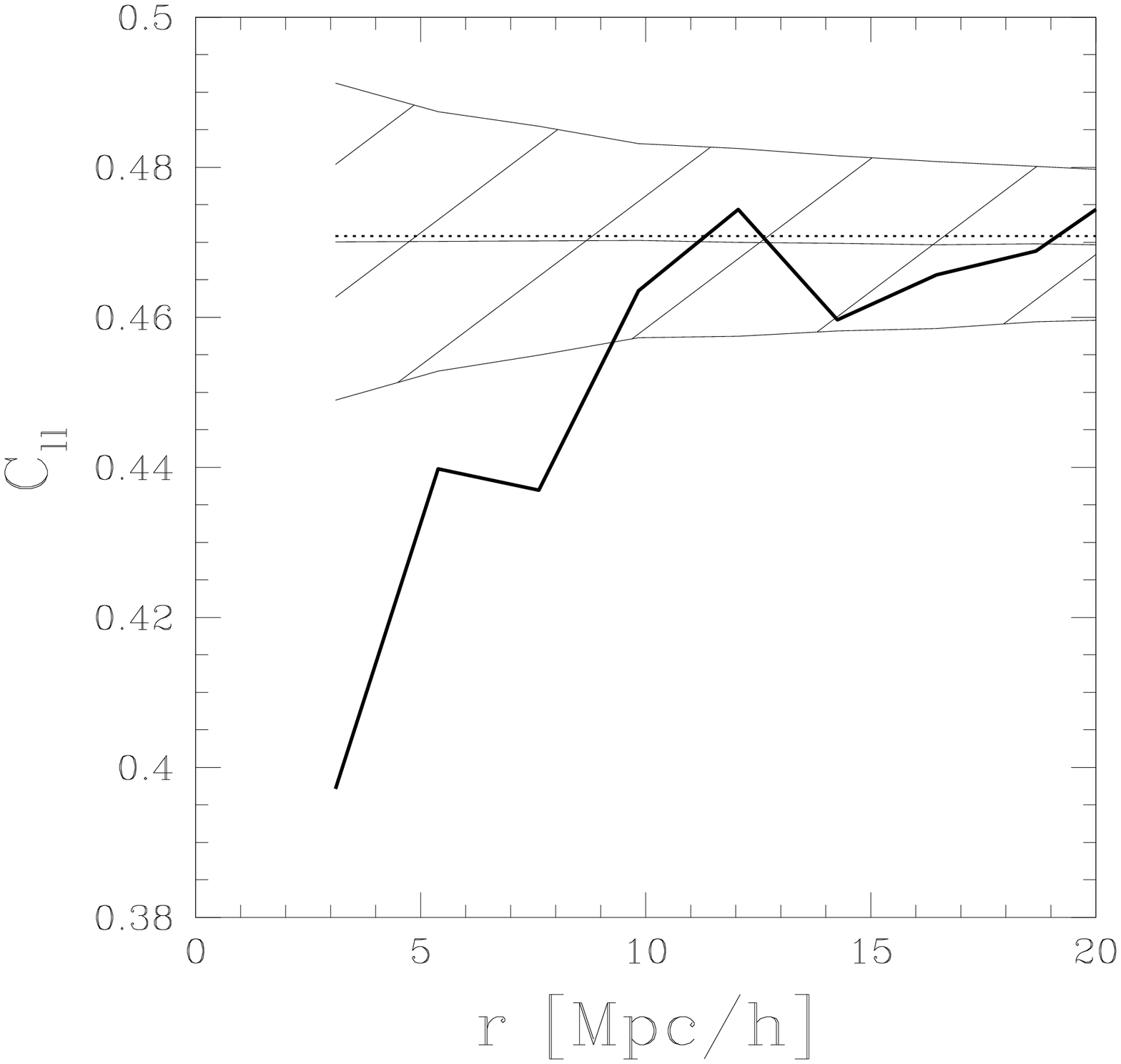}
\caption{\label{fig:morhpology-crosscorr}
The conditional  cross--correlation functions of early  ($e$) and late
($l$)  type  galaxies in  the  volume--limited  sample with  100\hMpc\
depth.  The  1--$\sigma$ region was determined  from 1000 realizations
with randomized marks.}
\end{figure}

The  morphological  classification  of  the galaxies  in  the  {SSRS2}
catalog was compiled from  different sources.  So, only wide classes
will give reliable results {}\citep{dacosta:southern}.
We  compare the clustering  properties of  two classes,  consisting of
spiral,  irregular  and peculiar  galaxies,  labeled  with $l$  (late
type),  and  elliptical and  lenticular  galaxies,  labeled with  $e$
(early type).  We discard the small fraction of unclassified galaxies.
In       Fig.~\ref{fig:morhpology-crosscorr}      the      conditional
cross--correlation functions $C_{ee}(r)$, $C_{el}(r)$, $C_{ll}(r)$ are
shown, estimated from the volume--limited sample with 100\hMpc\ depth,
using no boundary correction.

The  results  demonstrate,  that  the  clustering  properties  of  the
{SSRS2}--galaxies  depend  on  morphology.   Although  the  late--type
galaxies   predominate  the   catalog,  especially   the  small--scale
clustering is disproportionally due to pairs of early--type galaxies.
In  Subsect.~\ref{sect:ssrsii-lum-discrete} we  saw that  the luminous
galaxies  tend  to cluster  stronger.   At  this  point, the  question
arises, whether  the morphology segregation is  a possible explanation
of  the luminosity  segregation or  vice  versa. We  will discuss  the
connection  between   both  sorts  of   mark  segregation  in   Sect.  
\ref{sect:two-species}.

\subsection{Error estimates}
\label{sect:errors}

\begin{figure}
\epsscale{0.33}
\plotone{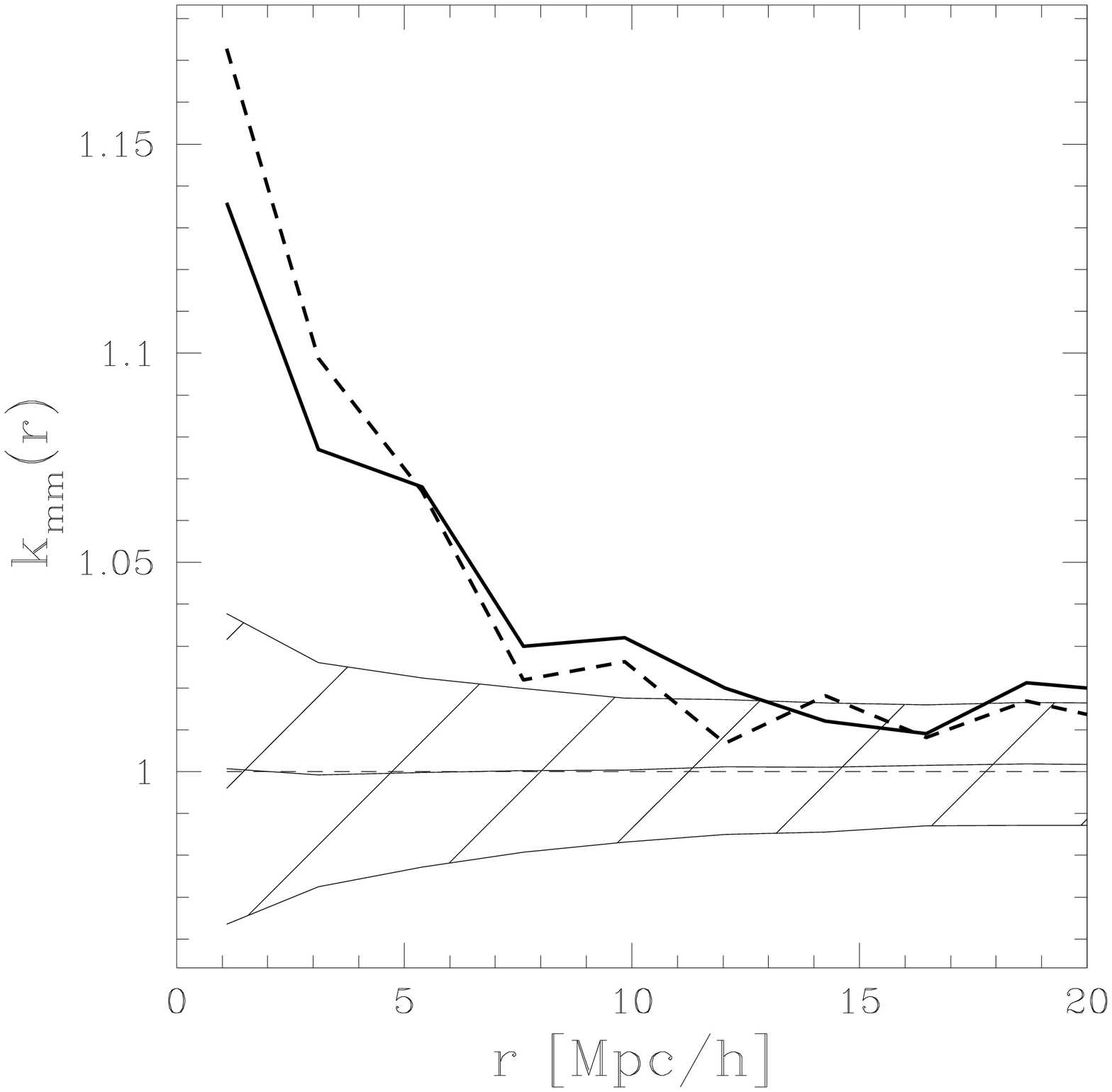}\plotone{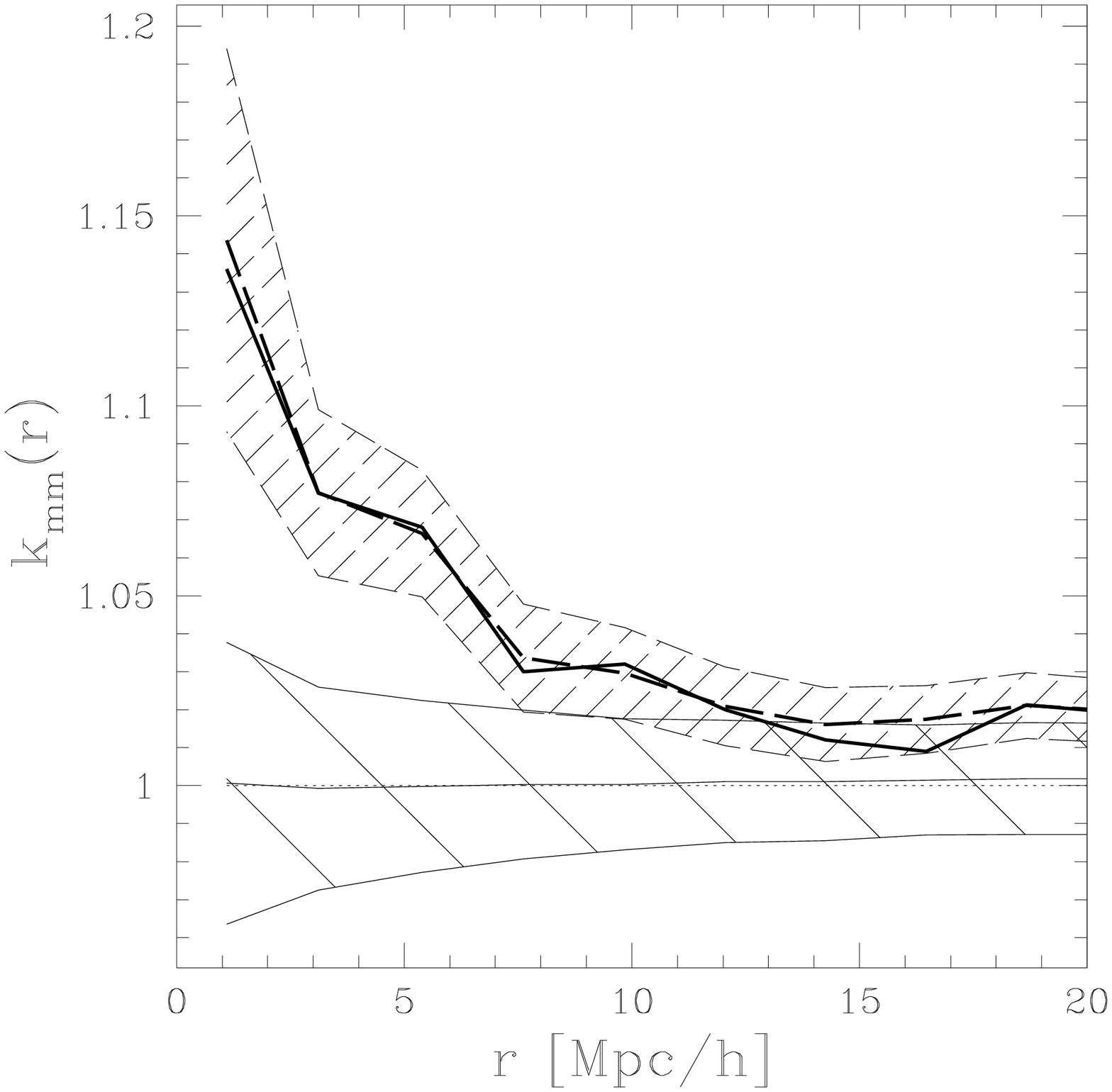}
\caption{\label{fig:markcorr-tests}
In the left  figure the $k_{mm}(r)$ function is  shown calculated from
the  measured magnitudes  using the  Euclidean distances  (solid line)
compared  to the $k_{mm}(r)$  function calculated  from $K$--corrected
magnitudes using the luminosity distances.
In the right plot again the ``pure'' $k_{mm}(r)$ function (solid line)
is  compared to  the $k_{mm}(r)$  function calculated  with randomized
peculiar velocities (shaded area  with dashed lines).  The shaded area
with solid lines corresponds to the results with randomized marks. }
\end{figure}

In the preceding sections we  have shown results for a volume--limited
sample  with  100\hMpc\  depth.   We also  considered  volume--limited
samples with  a limiting depth  of 60\hMpc, 80\hMpc\ and  120\hMpc, all
giving similar results.
Moreover, the  results do  not change if  we use  luminosity distances
instead of  Euclidean and  apply a type--dependent  $K$--correction as
used by {}\citet{benoist:biasing} (see Fig.~\ref{fig:markcorr-tests}).

Systematic  errors  may occur,  since  we  performed  our analysis  in
redshift space, i.e., we estimate the luminosity $L$ of a galaxy using
its  redshift  $z$:  $L\propto  z^2  10^{-0.4\text{mag}}$.   Therefore
peculiar velocities not only change the spatial correlations, but also
the values  of the  marks may be  biased in  a systematic way.   It is
difficult to correct for such  an effect, since in--fall and streaming
motions lead to correlated peculiar velocities.
To estimate  the order of  magnitude of this  error we randomly  add a
line--of--sight peculiar velocity to  each galaxy following a Gaussian
distribution with zero mean and a width of $300$km/s in agreement with
the value for  the pairwise velocity dispersion in  the SSRS2 given by
{}\citet{marzke:pairwise}.   In   randomizing  the  radial  velocities
independently we  overestimate this  error since correlated  pairs are
eventually torn  apart~\footnote{In adding random  peculiar velocities
we also  account for  possible errors within  the measurements  of the
redshifts,  which  are in  fact  much  smaller  than the  imprints  of
peculiar velocities.}.  Repeating this  procedure several times we can
show that the mean values of $k_{mm}$, $\var$ and $\cov$ do not change
compared  to   the  results  in   Fig.~\ref{fig:ssrs2-markcorr}.   The
additional fluctuations introduced by  this procedure are smaller than
the statistical errors quantified by  randomizing the marks, as can be
seen in Fig.~\ref{fig:markcorr-tests}.   Both $k_{mm}$ and $\var$ show
a  signal outside  the one--$\sigma$  range of  this  luminosity error
combined  with  the statistical  errors,  whereas  $\cov$ is  becoming
marginally consistent.

Note,   that   in   volume--limited   samples  a   special   sort   of
Malmquist--bias may influence the luminosities:
The luminosities are estimated using  the flux and the redshift as the
distance  indicator.    Hence  the  distance  is   influenced  by  the
individual  peculiar velocity  of  the galaxy.   Consider  a shell  at
distance $r$.   For geometrical reasons  more galaxies from  the outer
side get scattered into the shell than galaxies get scattered outside.
Hence,  in the  mean more  galaxies are  assigned too  small distances
resulting in underestimated luminosities.
Considering only galaxies with a distance smaller than 90\hMpc\ in the
volume--limited sample with 100\hMpc\ depth we obtain nearly identical
results for the mark--correlation  functions.  Therefore, this sort of
bias does not affect our analysis.

\subsection{IRAS selected galaxies}
\label{sect:iras}

Up  to now  we investigated  luminosity segregation  in  the optically
selected  SSRS2  catalog, with  the  luminosities  estimated from  the
B--magnitude. To see how our  results depend on the selection criteria
imposed on  the catalog  we look at  the mark  correlations determined
from the infrared  selected IRAS 1.2~Jy and PSCz  galaxy catalogs (for
details see {}\citealt{fisher:irasdata}, {}\citealt{saunders:pscz}).

\begin{figure}
\epsscale{0.33}
\plotone{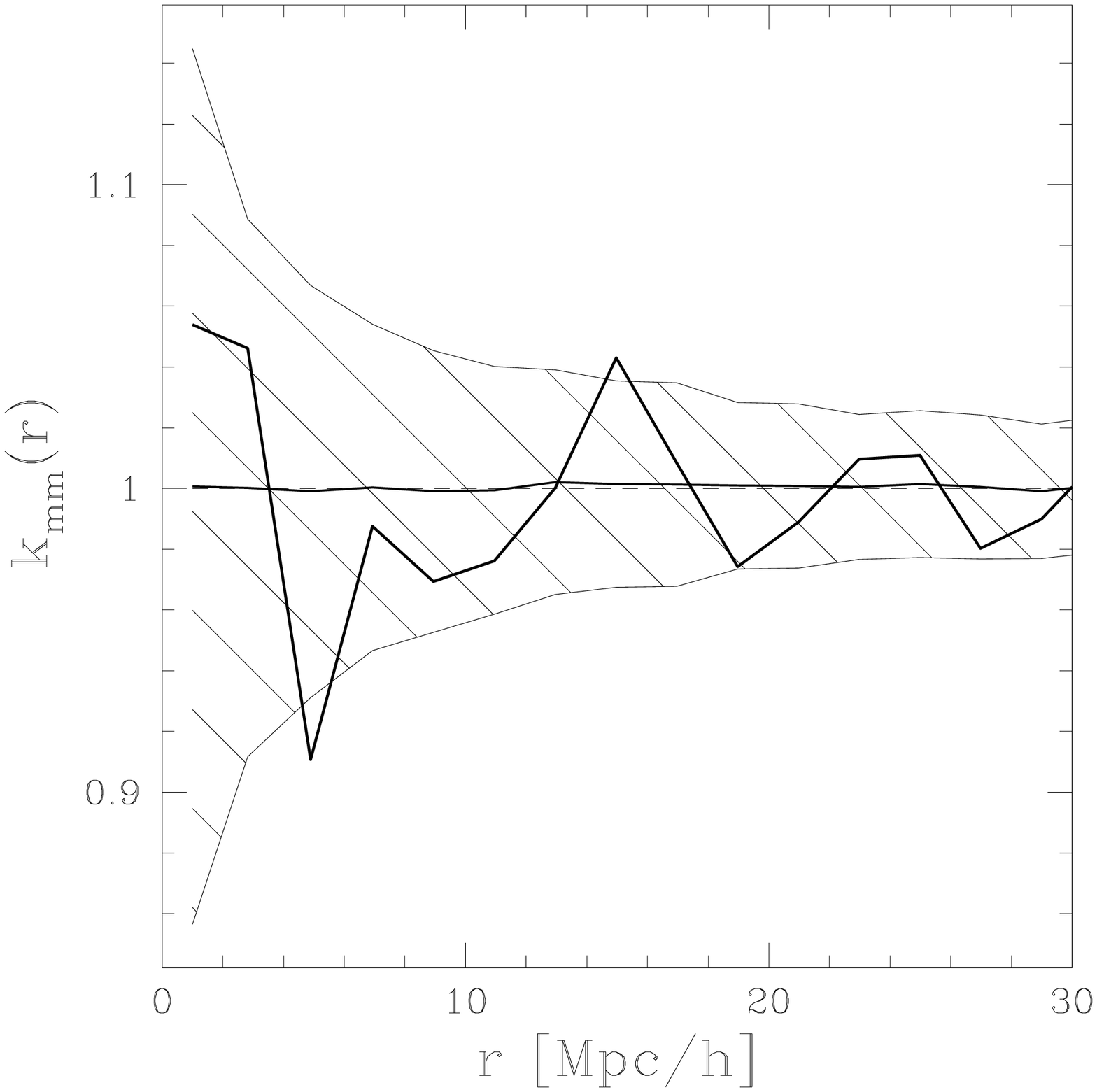}\plotone{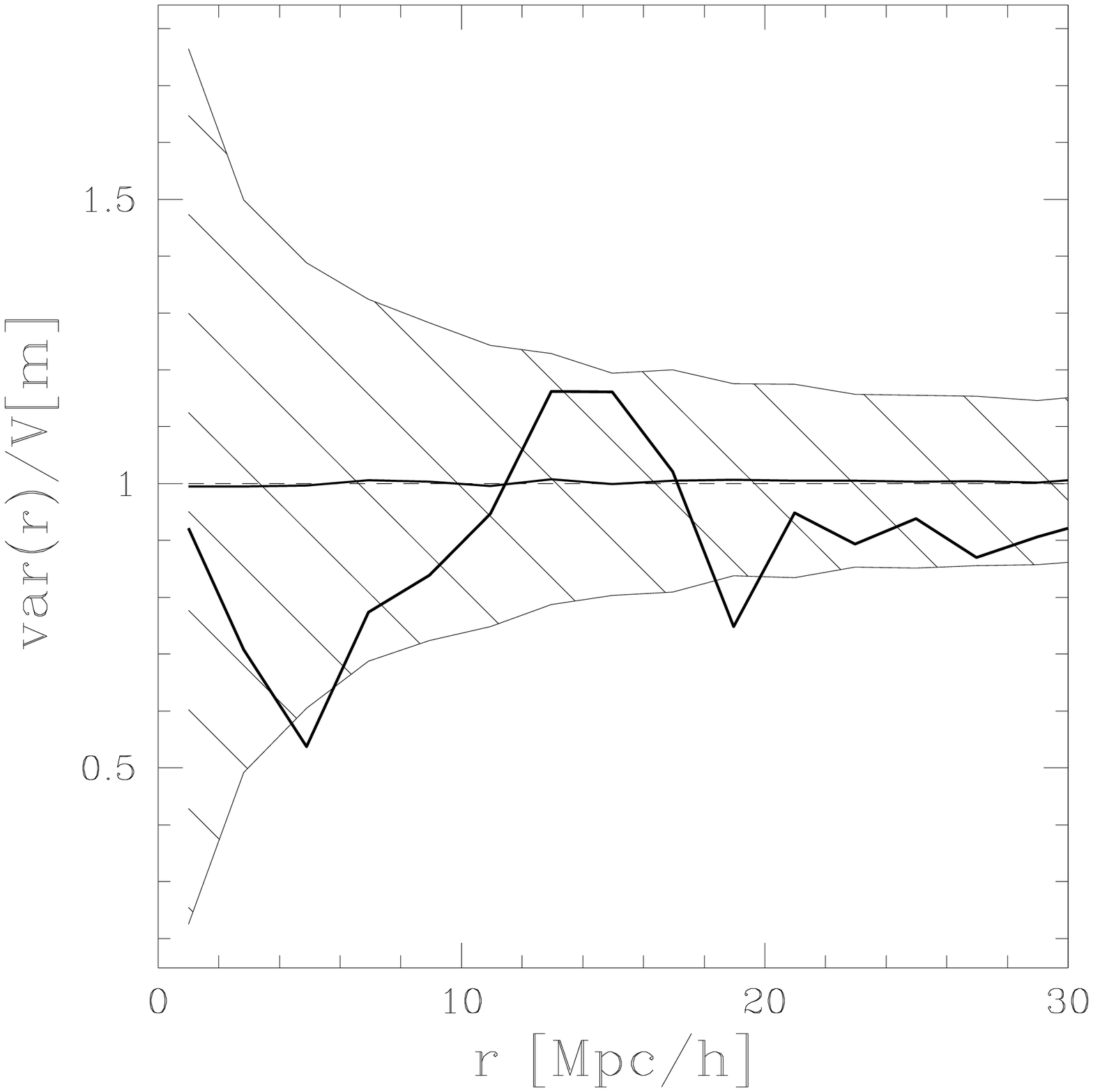}\plotone{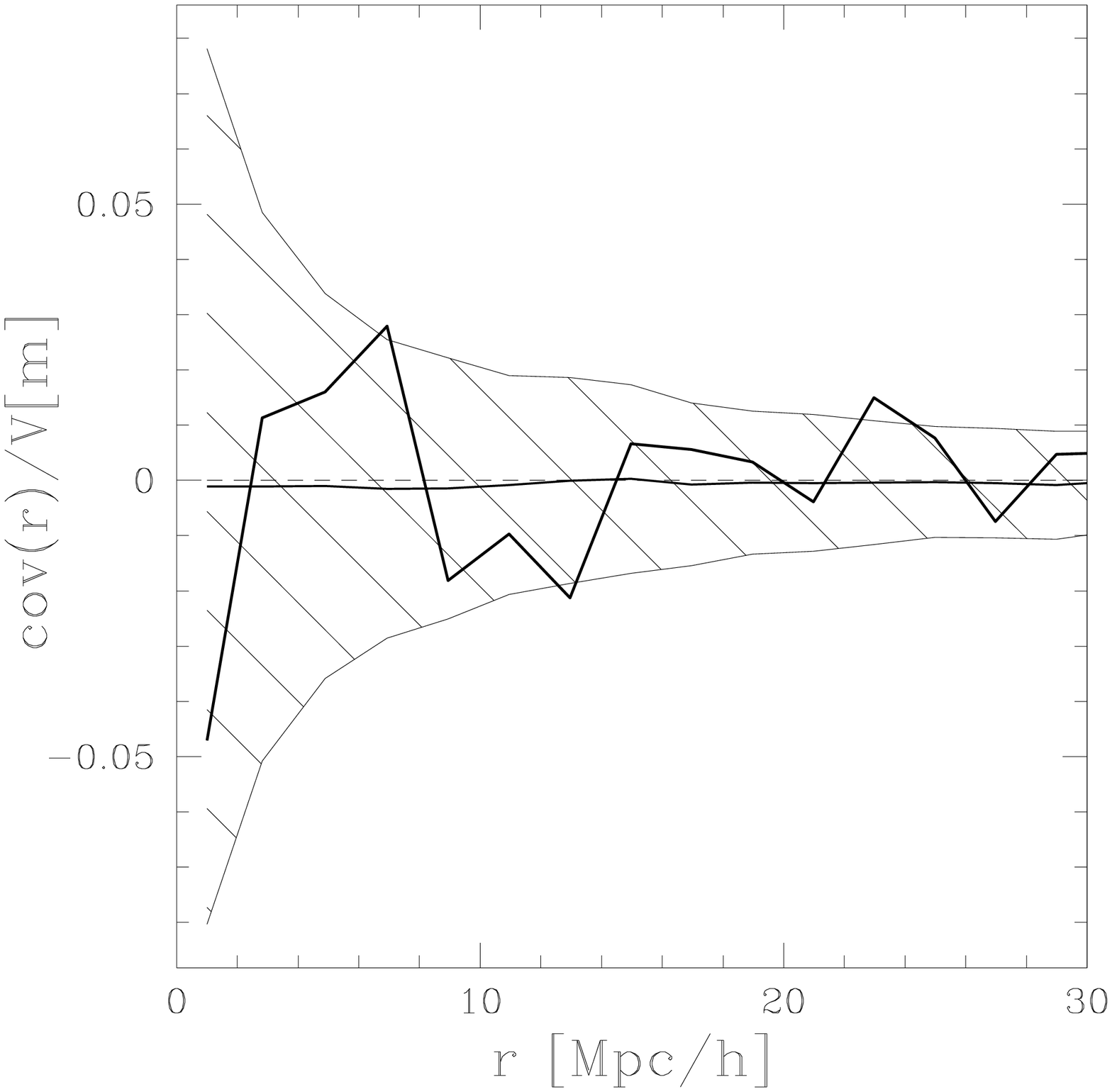}
\caption{\label{fig:pscz-markcorr} The mark--correlation functions for
a  volume--limited subsample  of  the  PSCz catalog  with  a depth  of
$100\hMpc$.   The  shaded  areas  denote  the  1--$\sigma$  error  for
randomized marks estimated from 1000 realizations.}
\end{figure}

We analyze  2259 galaxies  in the volume--limited  sample of  the PSCz
galaxy  catalog with a  depth of  100\hMpc\ inside  the mask  given by
{}\citet{saunders:pscz}.                  Similarly                 to
Sect.~\ref{sect:lum-morph-seg-gal} we  use $m_i=r_i^2\ f(\bx_i)$  as a
continuous mark,  proportional to  the luminosity of  the galaxy  at a
distance  of  $r_i=|\bx_i|$  with   an  observed  flux  $f(\bx_i)$  at
60~microns.
From  Fig.~\ref{fig:pscz-markcorr}  we  conclude that  no  significant
luminosity segregation is present in the PSCz galaxy catalog. The same
result holds  for volume--limited  samples with different  depths, and
for volume--limited  samples extracted  from the IRAS  1.2~Jy catalog.
This confirms  the results by {}\citet{bouchet:moments}  from the IRAS
1.2~Jy   and   especially   the   investigation   of   the   PSCz   by
{}\citet{szapudi:correlationspscz}   who  used   a   variant  of   the
conditional          cross--correlations         discussed         in
Subsect.~\ref{sect:discrete-mark-weighted}.
Similarly, only a weak dependence  on spectral features was reported by
{}\citet{mann:warmcool} for the QDOT catalog.
In Sects.~\ref{sect:separately} and  {}\ref{sect:wayround} we will see
that  the  luminous early--type  galaxies  play  a  dominant role  for
luminosity  and  morphology segregation.   This  is  supported by  the
negative results  from these IRAS selected  samples, since early--type
galaxies  are  significantly  underrepresented  in  infrared--selected
galaxy samples.

There is however an  interesting feature in the deeper volume--limited
samples from the PSCz.   Both $k_{mm}(r)$ and $\var(r)$ are consistent
with a random mark distribution, but the covariance $\cov(r)$ shows an
almost   three--$\sigma$  peak   near  $r=20\hMpc$.    This  increased
covariance at 20\hMpc\ is currently beyond an explanation, however the
feature is also visible  in volume--limited samples with 200\hMpc\ and
300\hMpc\  depth,  and  stable  against distance  cuts  and  different
binning.

\section{Luminosity segregation via amplitudes}
\label{sect:usual}

Previous investigations  detecting luminosity segregation  have used a
sequence  of  volume--limited  samples  and compared  the  correlation
amplitude of  the two--point correlation function  $\xi(r)$ (see e.g.,
{}\citealt{willmer:southern}).                                       In
Subsect.~\ref{sect:usual-luminosity-seg}  we  show  how  this  can  be
incorporated   into   the   more   general   formalism   provided   in
Sect.~\ref{sect:maths}.  In Subsect.~\ref{sect:faking} we reassess the
arguments given  by {}\citet{coleman:fractal} showing that  there is a
degeneracy   between  a   scale--invariant   point  distribution   and
luminosity  segregation  {\em  if}   the  analysis  is  based  on  the
amplitudes   of  $\xi$.   The   mark  characteristics   introduced  in
Subsect.~\ref{sect:mark-weighted} do not  suffer from this artifact as
shown in Subsect.~\ref{sect:robust}.  This strengthens our conclusions
in Sect.~\ref{sect:lum-morph-seg-gal} that  there is indeed luminosity
and morphological segregation.

\subsection{Luminosity segregation from a series of volume--limited samples}
\label{sect:usual-luminosity-seg}

Consider  a flux--limited  sample  with limiting  flux $f_{\rm  lim}$.
Every  galaxy at a  distance $|\bx_i|$  with observed  flux $L_i/(4\pi
|\bx_i|^2)$ larger  than some limiting flux $f_{\rm  lim}$ is included
in the sample.
We  construct  volume--limited subsamples  by  introducing a  limiting
depth  $R$  and  a  limiting  luminosity $L_{\rm  lim}$  with  $L_{\rm
lim}/(4\pi R^2)  = f_{\rm  lim}$ and by  admitting only  galaxies with
$|\bx|<R$   and   $L>L_{\rm   lim}$.    In  such   a   volume--limited
sample\footnote{
  In  general, we  have more  freedom in  constructing volume--limited
  samples: varying $R$ and $L_{\rm lim}$ independently, as long as the
  constraint:   $R^2<\frac{L_{\rm  lim}}{4\pi   f_{\rm   lim}  }$   is
  respected.  Holding $L_{\rm  lim}$ fixed, we can vary  $R$ and look,
  whether  the  statistical properties,  e.g.,  the  amplitude of  the
  correlation  function $\xi$,  differs between  these  samples.  This
  would allow to test  for fractal spatial structures independent from
  luminosity segregation.}
the  observed number  density  $\rho_{1,R}^{S}(\bx)=\rho_{1,R}^{S}$ is
spatially constant
\begin{equation}
\rho_{1,R}^{S}
= \rho\ \int_{L_{\rm lim}}^\infty\rd L\ \rho_{1}^{M}(L) ,
\end{equation}
if the underlying galaxy pattern is homogeneous.

For  two--point  properties we  can  proceed  similarly.  The  spatial
two--point density  in the volume--limited samples  for $|\bx_1|<R$ and
$|\bx_2|<R$ is
\begin{equation}
\rho_{2,R}^{S}(\bx_1,\bx_2) =
\int_{L_{\rm lim}}^\infty \rd L_1 \int_{L_{\rm lim}}^\infty \rd L_2\
\rho_{2}^{SM} ((\bx_1,L_1),(\bx_2,L_2)) .
\end{equation}
Using the definition~\eqref{eq:def-S2}  of the conditional probability
density  $\CS_2$ and the  assumption~\eqref{eq:no-mark-correlation} we
get
\begin{equation}
\rho_{2,R}^{S}(\bx_1,\bx_2) = 
\CNT \int_{L_{\rm lim}}^\infty \rd L_1 \int_{L_{\rm lim}}^\infty \rd L_2\
\CS_{2} (\bx_1,\bx_2|L_1,L_2)\ \rho_1^{M}(L_1)\rho_1^{M}(L_2) . 
\end{equation}
With   $r=|\bx_1-\bx_2|$,    the   two--point   correlation   function
$\xi_{R}(r)$ in a volume--limited sample is then
\begin{equation}
\label{eq:xiobserved} 
\xi_{R}(r) + 1  = 
\frac{\CNT}{\left(\rho\ \int_{L_{\rm lim}}^\infty\rd L\ 
\rho_1^{M}(L)\right)^2} \
\int_{L_{\rm lim}}^\infty \rd L_1 \int_{L_{\rm lim}}^\infty \rd L_2\
\CS_{2}(\bx_1,\bx_2|L_1,L_2) \rho_1^{M}(L_1)\rho_1^{M}(L_2) .
\end{equation}
If       no       luminosity       segregation       is       present,
$\CS_2(\bx_1,\bx_2|L_1,L_2)=\rho_2^{S}(\bx_1,\bx_2)/\CNT$            and
therefore:
\begin{equation}
\xi_{R}(r) = \xi(r) .
\end{equation}
If, on the  other hand, the clustering of the  galaxies does depend on
the  luminosities, the  two--point correlation  function  is different
between volume--limited  samples of  varying depths, and  also differs
from the two--point correlation function of all galaxies.

\begin{figure}
\epsscale{0.33}
\plotone{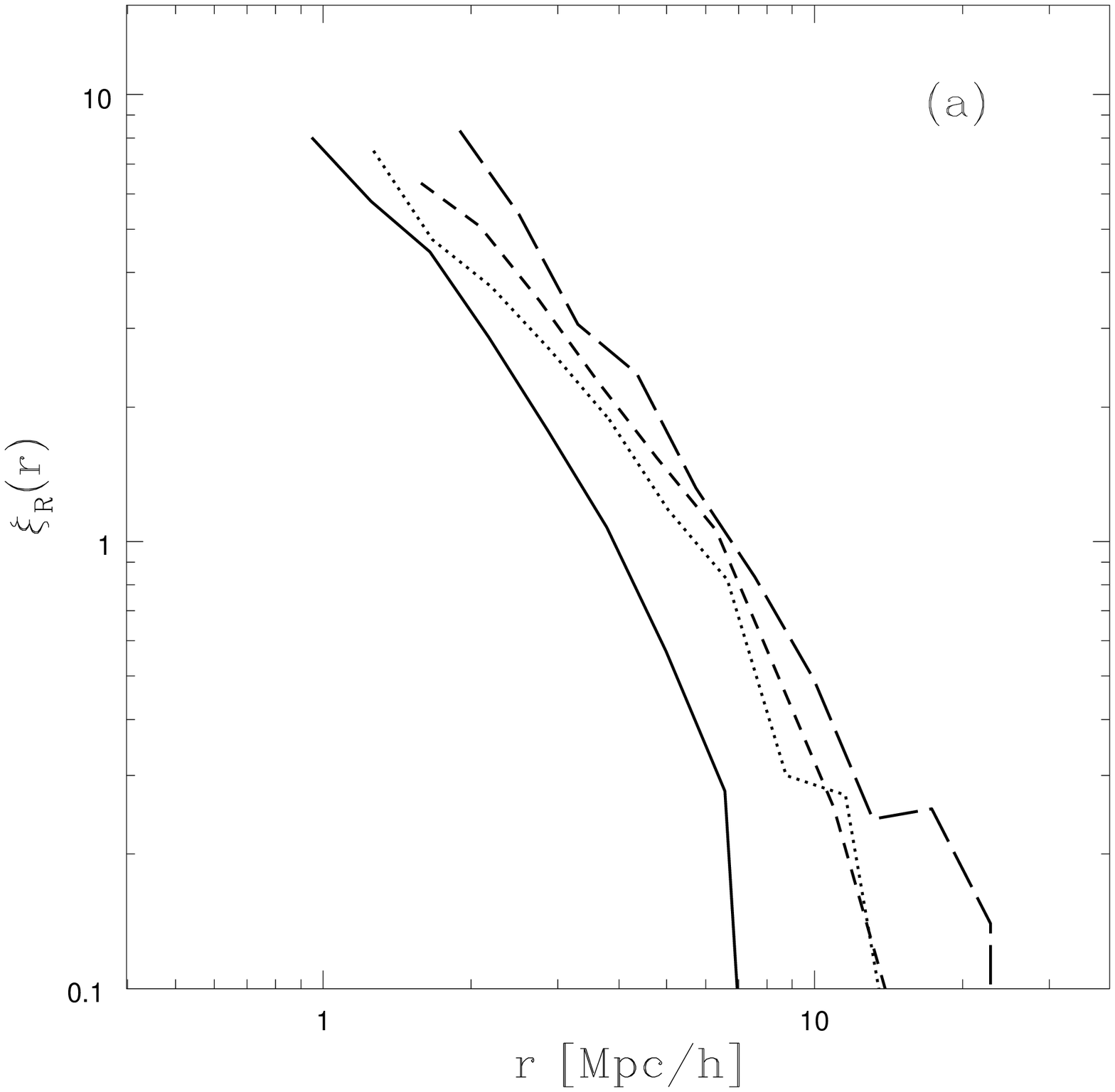}\plotone{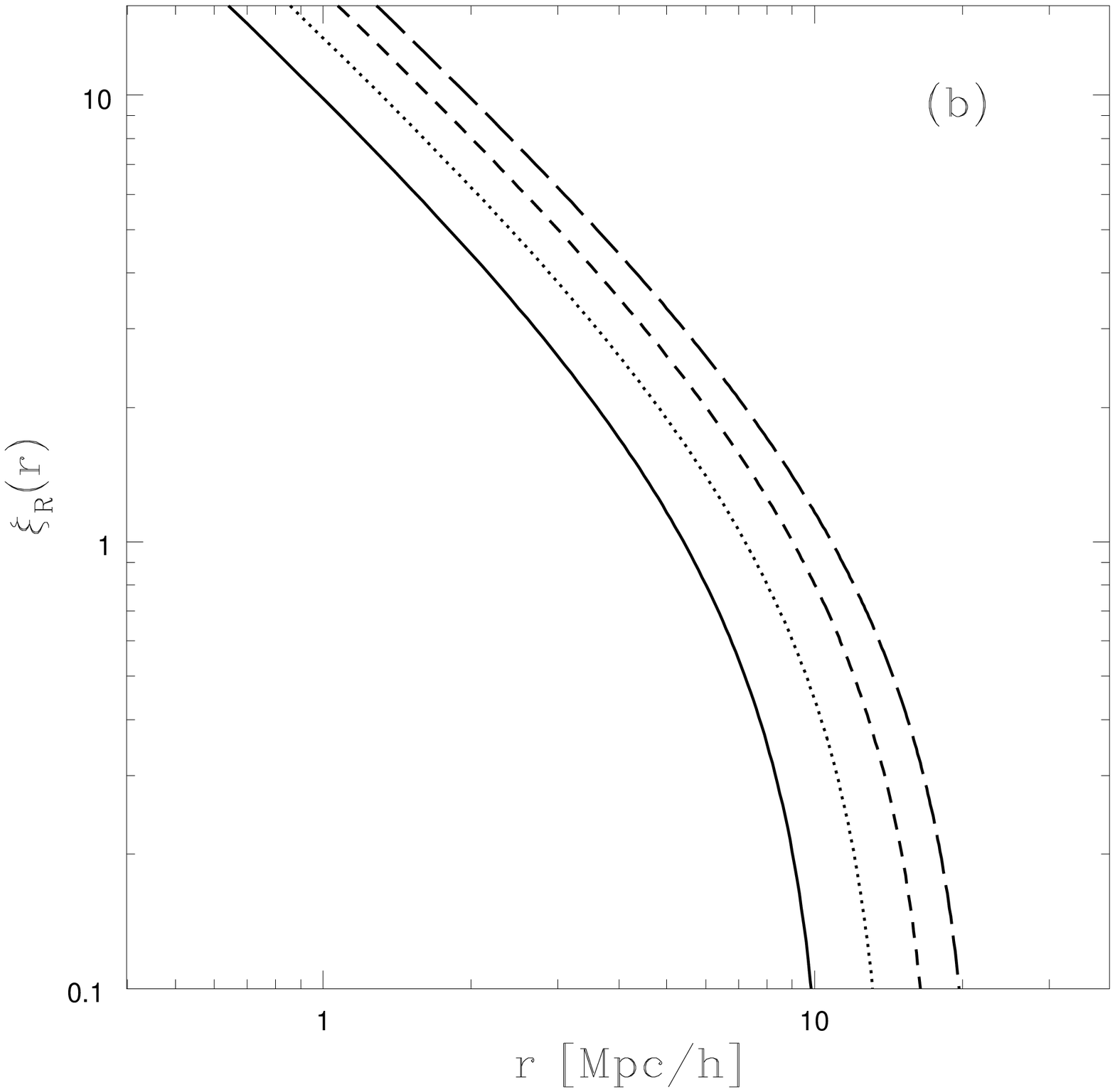}
\caption{\label{fig:xi2-luminosity-seg} In  plot $(a)$ the two--point
  correlation  functions $\xi_{R}(r)$  for volume--limited  samples of
  the  SSRS2 with  $R=60\hMpc$ (solid),  80\hMpc\  (dotted), 100\hMpc\ 
  (short dashed) and 120\hMpc\ (long dashed) depth are shown.  In plot
  $(b)$  the  two--point  correlation  functions  $\xi_{R}(r)$  for  a
  fractal  with  fractal dimension  $D=2$  are  shown,  with the  same
  increasing depths $R$ mimicking luminosity segregation.}
\end{figure}

As  an illustration  we calculate  $\xi_{R}(r)$ from  volume--limited
samples of the SSRS2 with increasing limiting depth $R$.
Our results in Fig.~\ref{fig:xi2-luminosity-seg} completely agree with
the results  reported by {}\citet{willmer:southern},  showing a higher
amplitude of $\xi_{R}(r)$ for the deeper volume--limited samples.  See
also the  comprehensive investigations of  {}\citet{cappi:fractal} and
{}\citet{benoist:biasinghigher}.  We  used several estimators  for the
two--point correlation function {}\citep{kerscher:twopoint}, including
the  minus estimator  shown  in Fig.~\ref{fig:xi2-luminosity-seg}  and
found  that this  behavior  of  the amplitude  is  independent of  the
estimator.

\subsection{Faking luminosity segregation}
\label{sect:faking}

In     this    section     we    illustrate     the     argument    by
{}\citet{coleman:fractal}  who  showed  that  there  is  a  degeneracy
between  luminosity segregation  determined with  the  standard method
(Subsect.~\ref{sect:usual-luminosity-seg})   and   a  fractal   galaxy
distribution.  Indeed, a general inhomogeneous galaxy distribution can
fake a sort  of ``luminosity segregation''.  Here, we  use a ``fractal
point set'' as a simple,  yet analytically tractable model for general
inhomogeneous point distributions.

The argument is based on the  scaling behavior of the number of points
inside a sample $N(R)\propto R^D$ for  a fractal point set in a sample
with linear  extent $R$, where  $D$ is the  (correlation--) dimension.
For a  fractal point set  the two--point correlation  function behaves
like
\begin{equation}
\xi_{R}(r) +1 \propto R^D\ r^{D-3},
\label{eq:fractalxi}
\end{equation}
with an amplitude  of $\xi_{R}$ depending on the  extent of the sample
(for  details see  {}\citealt{labini:scale}).  We  illustrate  this in
Fig.~\ref{fig:xi2-luminosity-seg}  showing  that fractal  correlations
according  to formula  \eqref{eq:fractalxi}  can mimic  a behavior  of
$\xi_{R}$ as observed in the galaxy data.\\
To summarize,  the behavior  of $\xi$ in  a series  of volume--limited
samples can  be explained  either by a  fractal point  distribution or
luminosity segregation  or both. So $\xi$  does not seem to  be a good
method to assess  one of both claims. Note, that Pietronero's argument
is based on the assumption that no luminosity segregation is present.

\subsection{Robustness of mark--correlation functions}
\label{sect:robust}

In the  preceding section we have  seen that to  search for luminosity
segregation employing the amplitude  of $\xi_R$ may be uncertain.  Now
we    show    that   the    mark    characteristics   introduced    in
Subsect.~\ref{sect:mark-weighted} do not suffer from this degeneracy.

All the quantities we used to investigate luminosity and morphological
segregation were  defined using  the average $\paverage{f}(r)$  over a
weight function  $f$.  With $\paverage{f}(r)$ we look  at the averages
of  some  mark--dependent  weight  function  $f(m_1,m_2)$,  under  the
condition that the points holding  the marks are separated by $r$.  We
do not investigate the spatial  distribution of the points.  As can be
seen   directly  from   Eq.~\eqref{eq:kappa-f-separate}   the  spatial
two--point correlations are ``divided out''.
Hence,  quantities like $\paverage{f}(r)$  are not  only well--defined
for homogeneous  point distributions,  but also give  reliable results
for inhomogeneous point distributions like fractals.

\begin{figure}
\epsscale{0.33}
\plotone{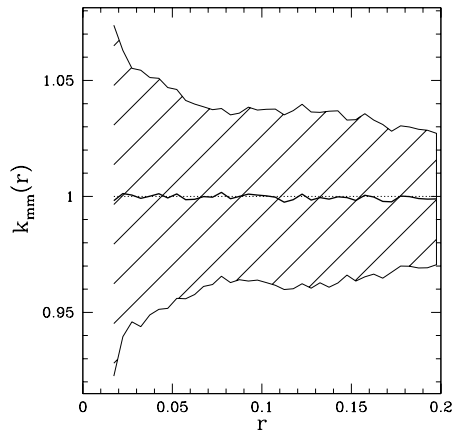}
\caption{  \label{fig:poifrac}  The $k_{mm}(r)$  for  a fractal  point
distribution with  random marks and  number density $1000$  inside the
unit box.  The shaded area  marks the 1--$\sigma$ range estimated from
1000 simulations, and the  dotted line is the theoretical expectation.
$r$ is given in units of the box--length.}
\end{figure}

To illustrate this  we use a ``fractal point  set'' kindly provided by
Alessandro Amici. This fractal  is a three--dimensional realization of
the  random--$\beta$ model  with  a  fractal dimension  of  two. On  a
randomly selected set of points  from this fractal we distribute marks
chosen  randomly out  of  $[0,1]$.  This  resembles a  volume--limited
sample   with   no    luminosity   segregation.    We   estimate   the
mark--correlation  functions  using  the  estimator  without  boundary
corrections.  The function $k_{mm}(r)$ shown in Fig.~\ref{fig:poifrac}
gives the correct  result that no mark correlation  is present. Hence,
our methods  give stable results  even on such an  inhomogeneous point
distribution.  This is  also the case for all  other functions and for
any estimator considered (Appendix ~\ref{sect:estimators}).
Therefore, our results obtained from the SSRS2 galaxy survey discussed
in  Sect.~\ref{sect:lum-morph-seg-gal}  can not  be  explained with  a
scale--invariant spatial distribution, showing no luminosity, alone.

There is  also a more technical  advantage of our  method: to estimate
the  correlation   function  $\xi(r)$  one  has   to  employ  boundary
corrections.    Quantities   like   $k_{mm}$  only   use   conditional
probabilities and  may be estimated without  boundary corrections (see
Appendix ~\ref{sect:estimators}), reducing the estimators' variance.

\section{The morphology--density relation}
\label{sect:morphology-density}

The  morphology--density  relation  states  that inside  clusters,  in
regions with  a high (surface)  density of galaxies, the  abundance of
early--type galaxies  is enhanced whereas the  abundance of late--type
galaxies is reduced {}\citep{dressler:galaxy}.
This relation is very well established, and therefore it seems natural
to ask whether the observed luminosity segregation can be explained by
the morphology--density relation alone.   In this section we present a
number of  reasons why this is  {\em not} the case. 

As a first  test we discarded all galaxies in  a spherical region with
1.5\hMpc\   and  also   3\hMpc\   radius  around   the  APM   clusters
{}\citep{dalton:apmcluster}, and conduct a analysis similar to the one
in   Subsect.~\ref{sect:ssrs-luminosity-cont}    restricted   to   the
intersection  of the  SSRS2 and  the  APM cluster  catalog.  The  mark
correlation functions  did not show any significant  change.  This may
not  be decisive,  since only  a few  of the  (rich) APM  clusters are
included; however,  it supports our view that  the observed luminosity
segregation is not caused by clusters of galaxies alone.

But in the  spirit of the morphology--density relation,  one could try
to explain  the observed luminosity segregation in  the following way:
the  two  populations of  galaxies,  the  early--  and the  late--type
galaxies, cluster in a different  way (which is, e.g., manifest in the
morphology--density  relationship  and   in  the  observed  morphology
segregation Sect.~\ref{sect:ssrs2-morphology}).  If these classes show
different average luminosities,  the morphology--density relation will
generate the luminosity segregation.  This is the main idea behind the
two--species  model discussed  below.  A  first indication,  that this
kind of  model is  not able to  explain luminosity  segregation, comes
from the  observation, that both  early-- and the  late--type galaxies
show very similar  luminosity distributions within the volume--limited
sample we considered.

We conduct  additional tests of this  idea, which allow  for a further
understanding   of  the  luminosity   segregation:  we   consider  the
two--species model in  Subsect.~\ref{sect:two-species} in more detail,
and we  investigate the early-- and late--type  galaxies separately in
Subsect.~\ref{sect:separately};  moreover,   we  look  for  morphology
segregation   in   dim   and   luminous   subsamples   separately   in
Subsect.~\ref{sect:wayround}.

\subsection{The two--species model}
\label{sect:two-species}

\begin{figure}
\epsscale{0.33}
\plotone{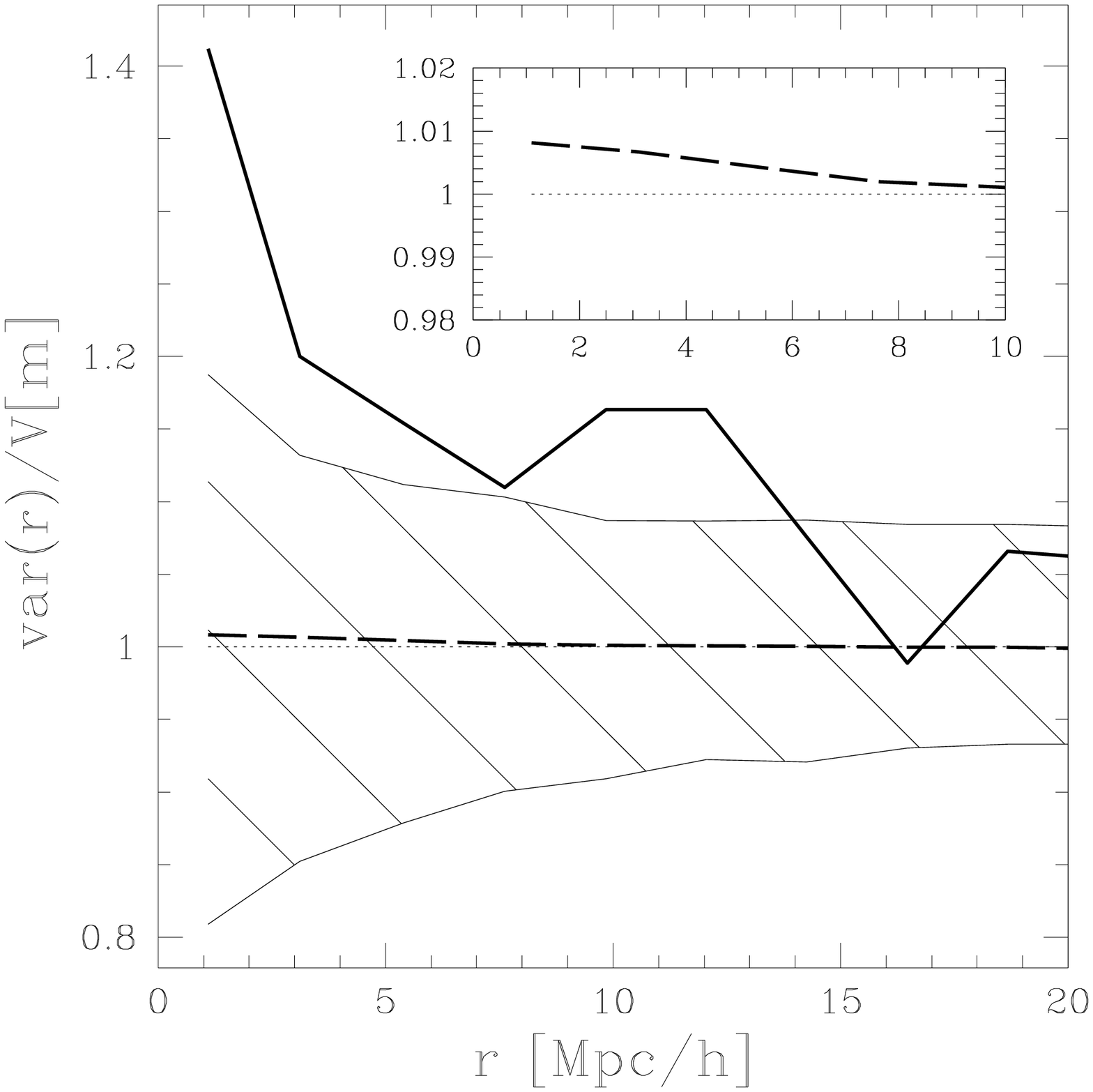}
\caption{\label{fig:ssrs2-markcorr-twospecies}  The  mark--correlation
function $\var(r)$ for a volume--limited subsample of the SSRS2 with a
depth of  $100\hMpc$ (solid line) compared  with the mark--correlation
function calculated  from the  two--species model (dashed  line).  The
shaded  area  denotes  the  1--$\sigma$  error  for  randomized  marks
estimated from 1000  realizations.  The inset compares the  case of no
luminosity  segregation  (dotted  line)  and  the  prediction  of  the
two--species model (dashed line)}
\end{figure}

As already outlined  above, in the two--species model  we consider two
subpopulations of  galaxies, with  different spatial clustering  and a
different  mark  distribution. Within  each  class  there  is no  mark
segregation. Thus  this model explains in  a very simple  way how mark
correlations arise  from the spatial  interplay of the two  classes of
galaxies.  The  subclasses  will   be  formed  by  early--  $(e)$  and
late--type $(l)$ galaxies.\\
Let $\rho_l$,  $\overline{m}_l$, $V_l$ denote the  number density, the
mean mark,  and the  variance of  the marks of  galaxies of  type $l$,
respectively,  and similarly  for subclass  $e$.  The  one--point mark
distributions are denoted  by $\rho_{1,e}^M(m)$ and $\rho_{1,l}^M(m)$.
The  spatial  (cross--)   correlations  are  given  by  $\xi_{ee}(r)$,
$\xi_{ll}(r)$,  and $\xi_{el}(r)$  (symmetrically defined  in  $e$ and
$l$, i.e., $(\xi_{el}+\xi_{le})/2$).
We use  the morphological type and  the luminosity as  components of a
compound   mark  $\bm=\{t,m\}$,   where   $t\in\{e,l\}$  denotes   the
morphological  type and  $m$ is  the  luminosity of  the galaxy.   The
two--point properties within the two--species model are then given by:
\begin{multline}
\rho_2^{SM}\left((\bx_1,\{t_1,m_1\}),(\bx_2,\{t_2,m_2\})\right)= \\
\delta_{t_1 e}\delta_{t_2 e}\ \rho_{e}^2\
\rho_{1,e}^M(m_1)\rho_{1,e}^M(m_2)\ (1+\xi_{ee}(r)) 
+  \delta_{t_1 l}\delta_{t_2 l}\ \rho_{l}^2\
\rho_{1,l}^M(m_1)\rho_{1,l}^M(m_2)\ (1+\xi_{ll}(r)) \\
+ \left(\delta_{t_1 e}\delta_{t_2 l}\ \rho_{1,e}^M(m_1)\rho_{1,l}^M(m_2) 
+ \delta_{t_1 l}\delta_{t_2 e}\ \rho_{1,l}^M(m_1)\rho_{1,e}^M(m_2)\right) 
\ \rho_{e}\rho_{l} (1+\xi_{el}(r)).
\end{multline}
With    $q_l=\rho_l/(\rho_l+\rho_e)$,   $q_e=1-q_l$,    the   combined
two--point correlation is function
\begin{equation}
1+\xi(r) = 
q_e^2(1+\xi_{ee}(r))+ q_l^2(1+\xi_{ll}(r)) + 2q_eq_l(1+\xi_{el}(r)) ,
\end{equation}
and using the definitions in Sect.~\ref{sect:continuous-marks} one may
calculate  the  luminosity  correlation  functions for  this  specific
model.  We  measured $q_e$,  $\xi_{ee}$, as well  as $\overline{m}_e$,
$V_e$ etc.\  in the volume--limited  sample with 100\hMpc\  depth from
the SSRS2.  Using these quantities we calculated the mark--correlation
functions       for       the       two--species      model.        In
Fig.~\ref{fig:ssrs2-markcorr-twospecies}   we   compare   the   $\var$
function from  the two--species model with the  actual observed values
(similar results  are obtained  for $k_{mm}$ and  $\cov$).  Obviously,
the  two--species   model  is  not   able  to  explain   the  observed
luminosity--correlations.   This  shows  that  the  spatial  interplay
between   different   morphological  types,   as   suggested  by   the
morphology--density  relation, is  only  in part  responsible for  the
observed  luminosity  segregation.   A  necessary ingredient  is  that
luminosity segregation is already present  in one of the subclasses at
least (see the next section).

The    results     for    the    two--species     model    shown    in
Fig.~\ref{fig:ssrs2-markcorr-twospecies}         were         obtained
selfconsistently from  the empirically determined  parameters as given
by the division of the sample into early-- and late--type galaxies.
We may go further and treat the two--species model as a toy model with
scale--invariant     (cross--)     correlation    functions     (e.g.,
$\xi_{ee}\propto  r^{-\gamma}$) and  free  parameters $\overline{m}_e$
etc. to fit the data.  However, we find that an acceptable qualitative
description of  the observed luminosity  segregation in terms  of this
model is only satisfied when  the parameters of the two--species model
are highly unrealistic.

\subsection{Early--and late--type galaxies separately}
\label{sect:separately}

\begin{figure}
\epsscale{0.33}
\plotone{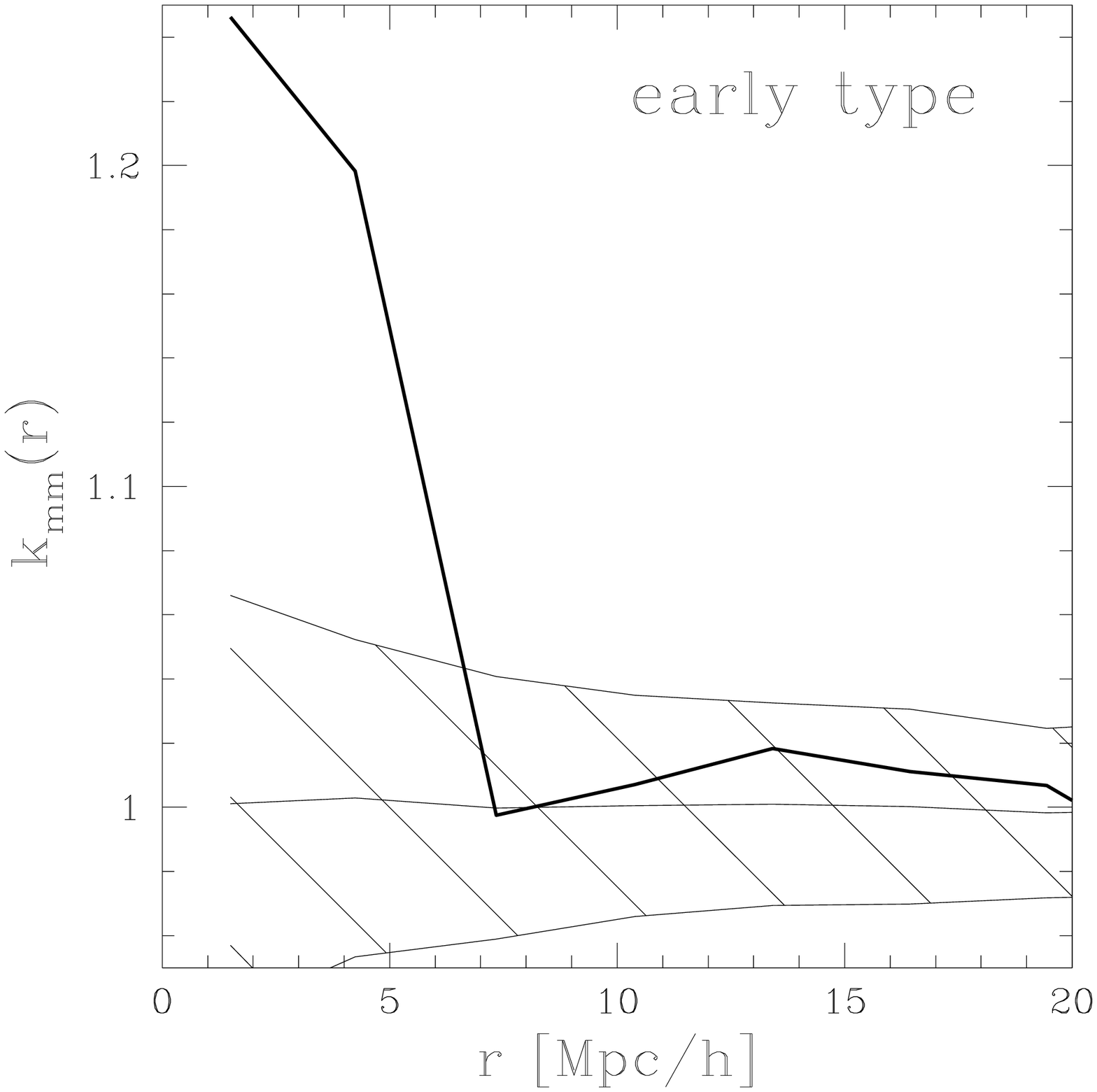}\plotone{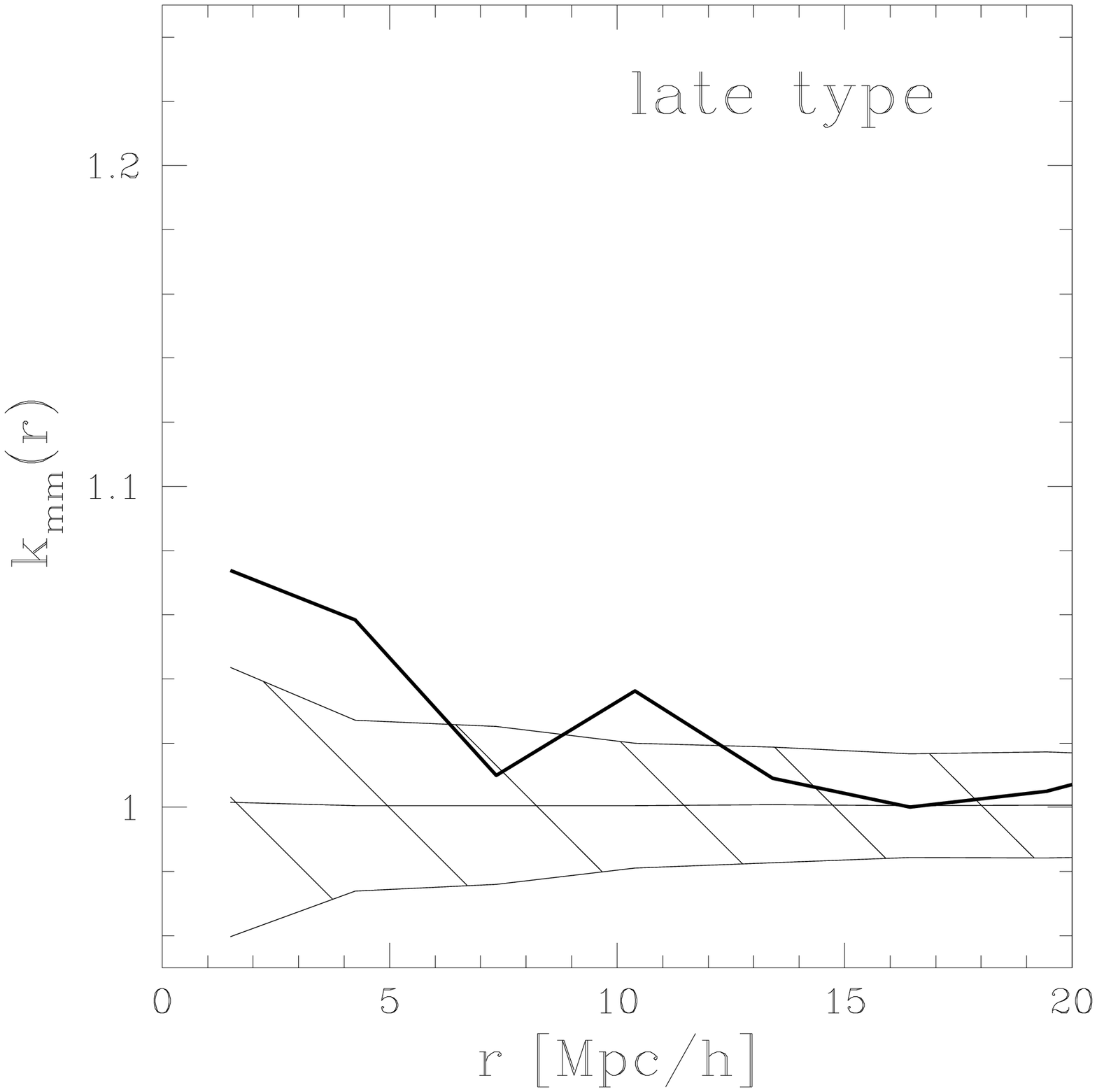}
\plotone{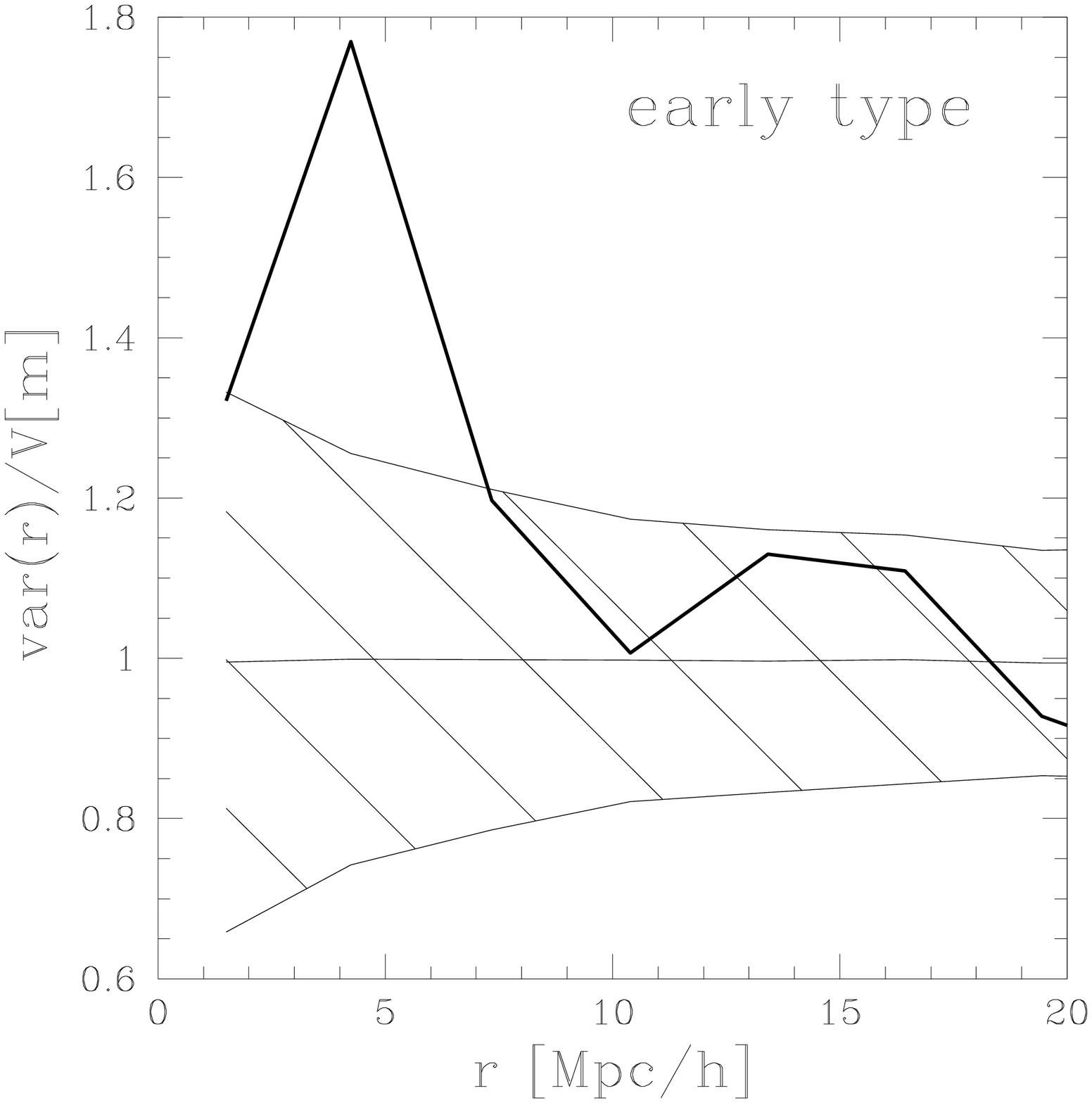}\plotone{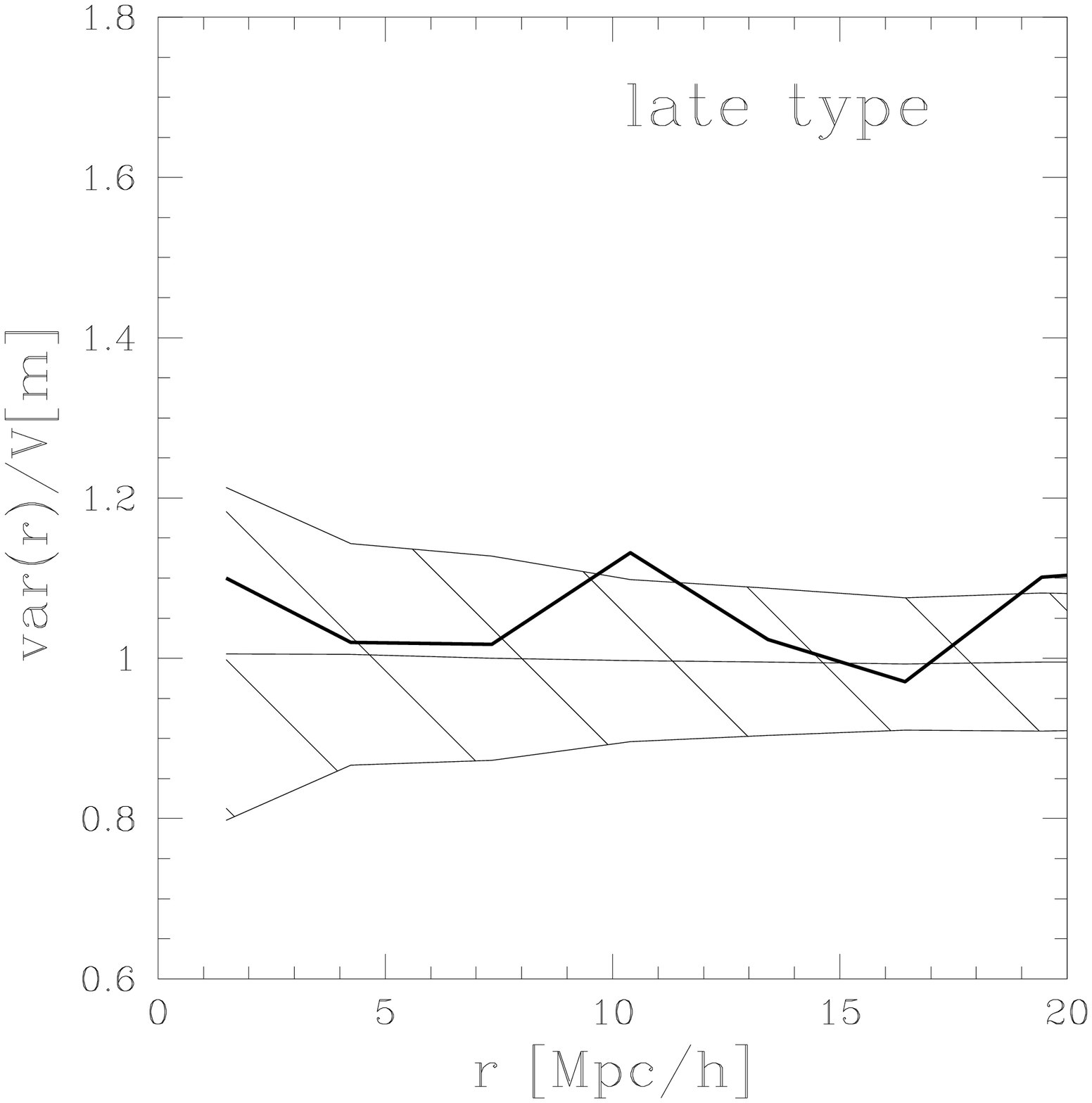}
\plotone{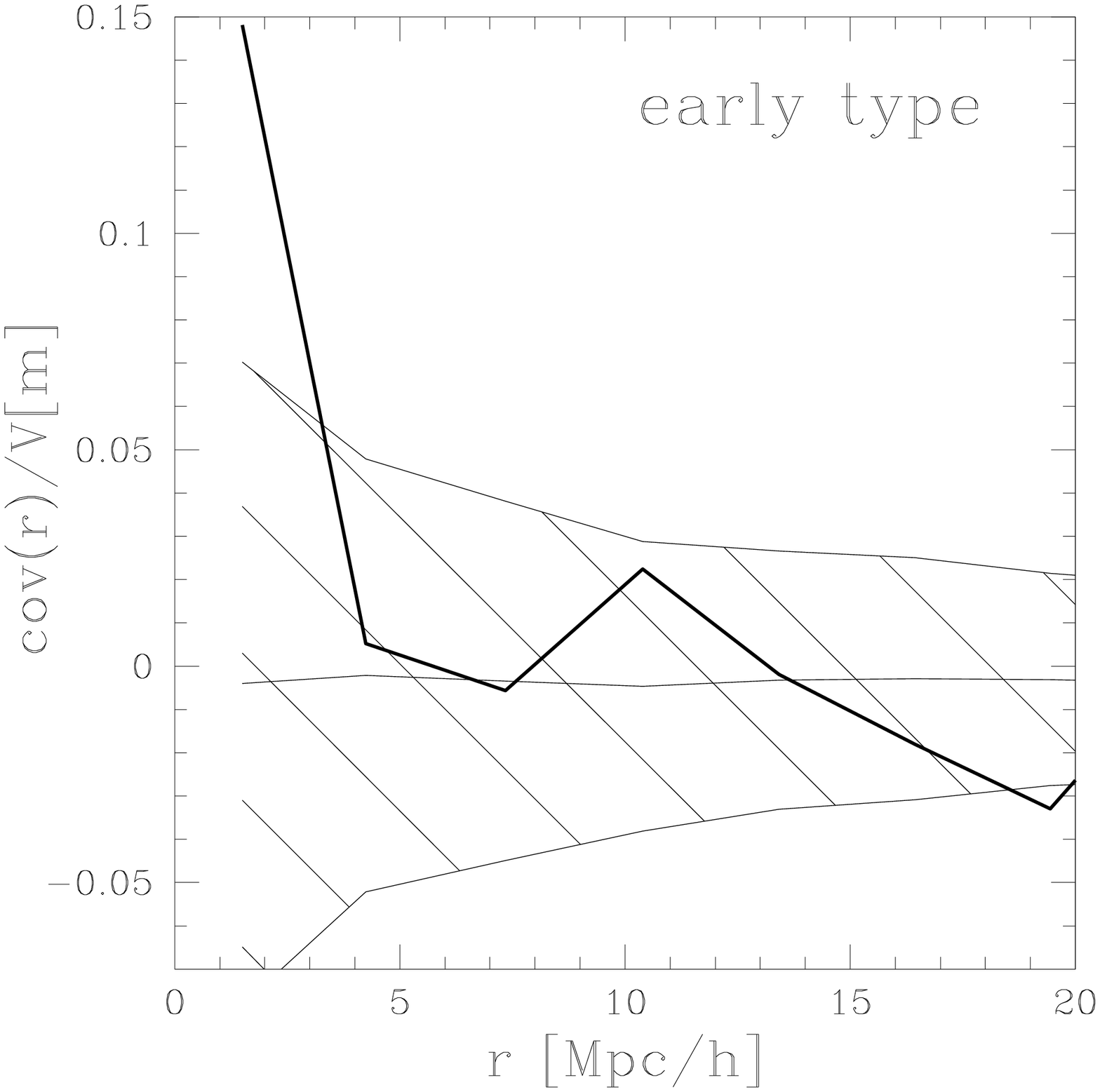}\plotone{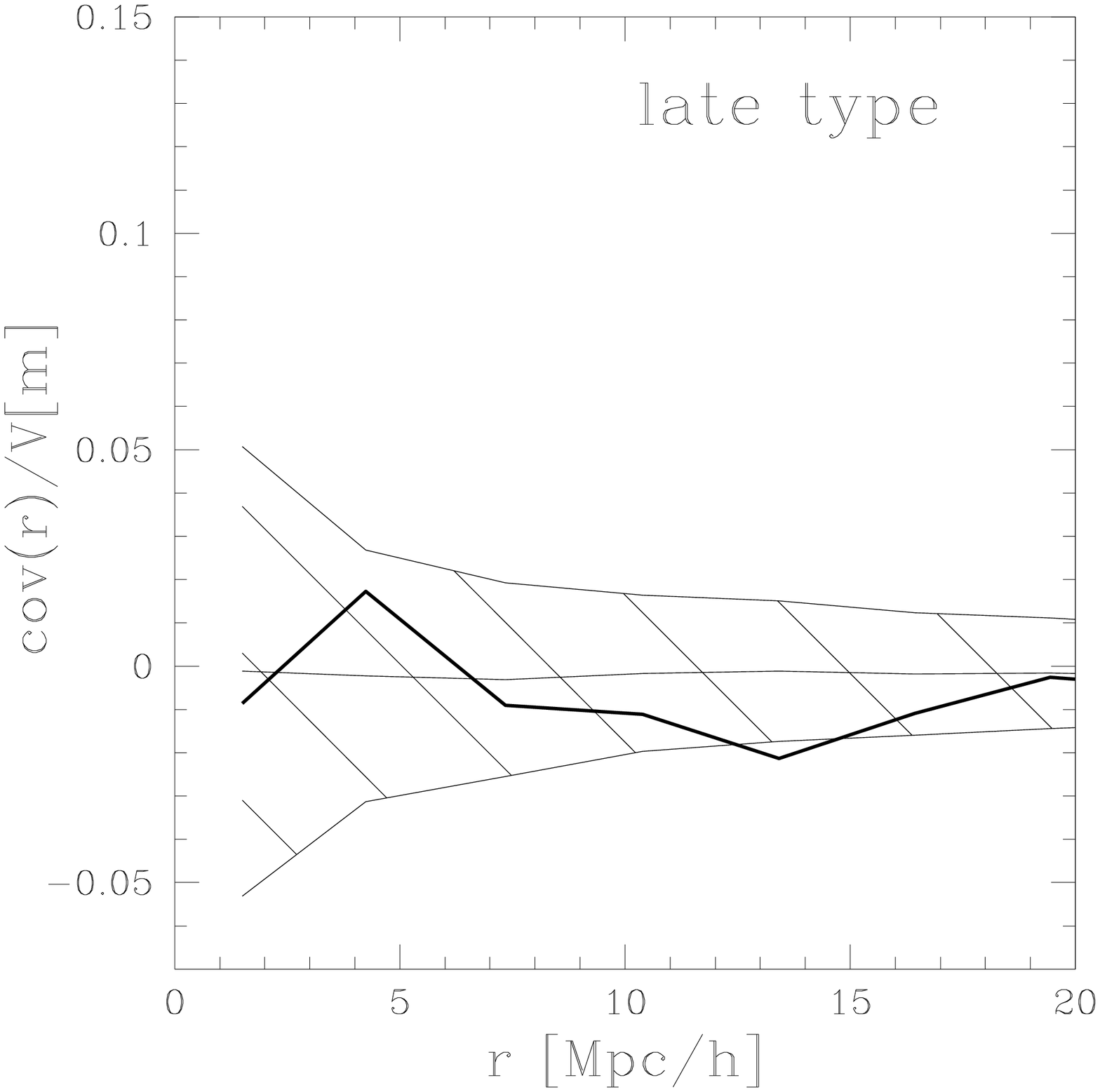}
\caption{\label{fig:ssrs2-markcorr-typedep}    The   mark--correlation
functions  for  a the  early--type and  late--type  galaxies from  the
volume--limited sample of  the SSRS2 with a depth  of $100\hMpc$.  The
shaded  areas  denote  the  1--$\sigma$ errors  for  randomized  marks
estimated from 1000 realizations.}
\end{figure}

As a second  test, we split the 100\hMpc\  volume--limited sample from
the SSRS2 into two subsamples consisting out of early-- and late--type
galaxies each. Using the luminosity  as the (continuous) mark, we look
for  luminosity   segregation  similarly  to   the  investigations  in
Sect.~\ref{sect:ssrs-luminosity-cont}.
From  Fig.~\ref{fig:ssrs2-markcorr-typedep} it  is  evident that  both
subpopulations show luminosity  segregation, but the main contribution
comes from early--type galaxies.  The late--type galaxies show a small
signal in  $k_{mm}$ only.  Clearly, with  this kind of  analysis we do
not pick  up features intrinsic  to the interplay between  early-- and
late--type  galaxies, which  may  add a  further  contribution to  the
observed luminosity segregation (Fig.~\ref{fig:ssrs2-markcorr}).

\subsection{The other way round?}
\label{sect:wayround}

\begin{figure}
\epsscale{0.33}
\plotone{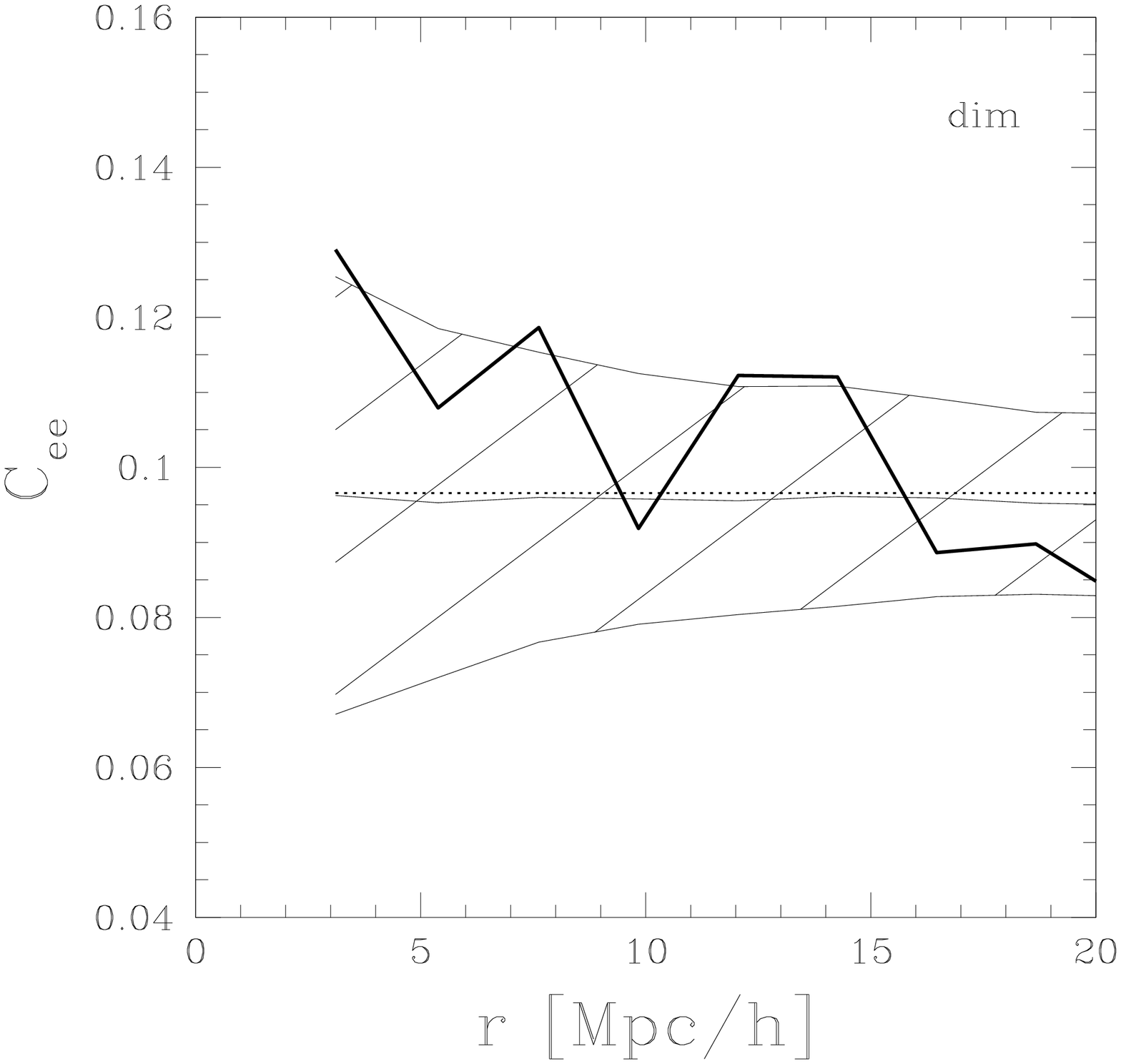}\plotone{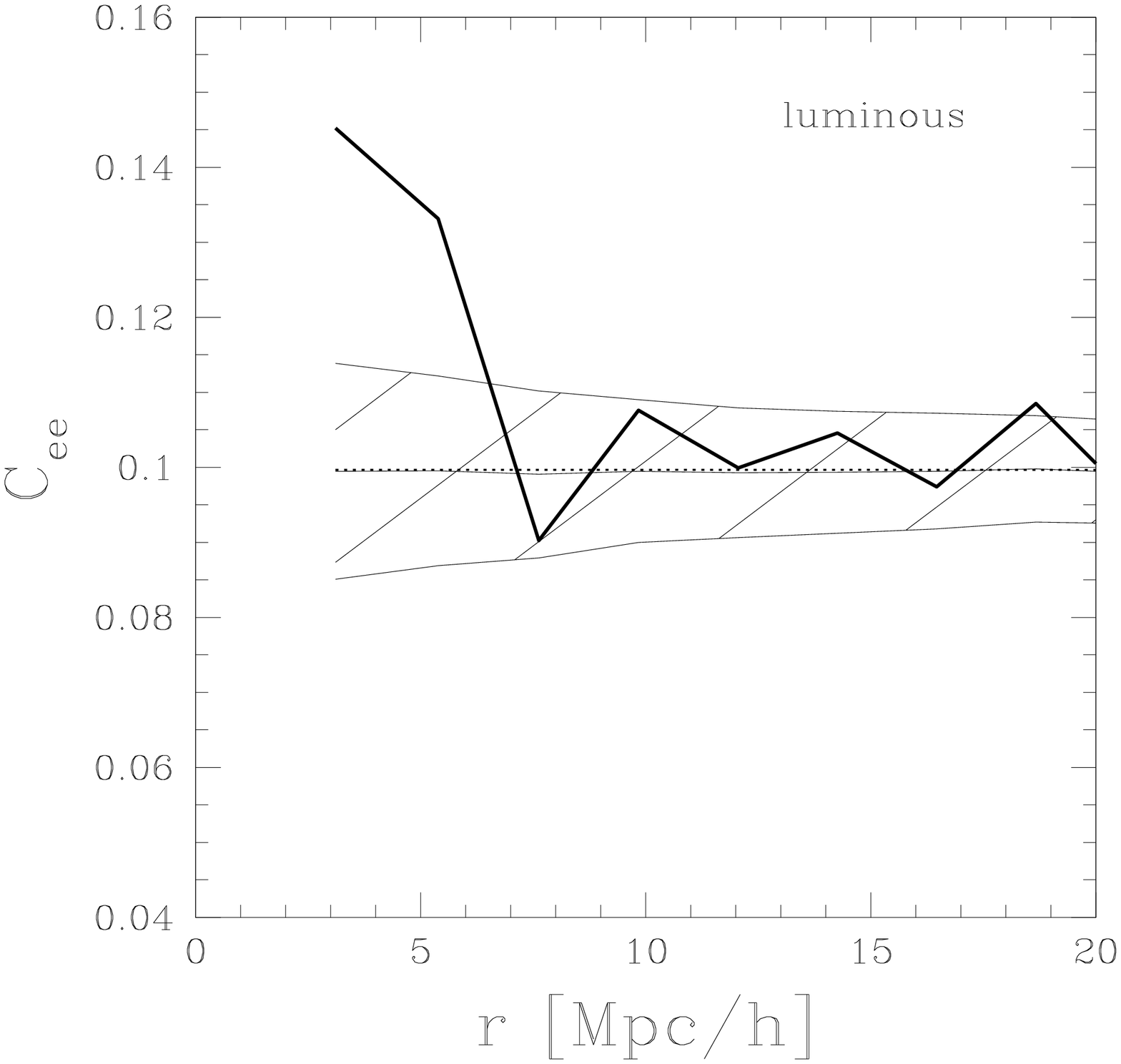}
\caption{\label{fig:morhpology-crosscorr-dimlum}    The    conditional
cross--correlation function $C_{ee}(r)$ of early--type galaxies in the
luminous  and dim  subsamples. The  1--$\sigma$ region  was determined
from 1000 realizations with randomized marks.}
\end{figure}

So far, our  results show clearly, that the  luminosity segregation is
not a pure  effect of the morphology segregation.   To investigate the
opposite  case, where  the  morphology segregation  is  caused by  the
luminosity  segregation,  we  split  the volume--limited  sample  with
100\hMpc\ depth  into two equally sized luminosity  classes, with {\em
  dim} and  {\em luminous} galaxies, respectively.  For  each of these
samples  we  calculate  the  conditional  cross--correlations  between
early-- and late--type  galaxies.  
The  strong  (conditional)  correlations  $C_{ee}(r)$  of  early--type
galaxies on  small scales are now  only visible in the  sample of {\em
  luminous}    galaxies,   confirming    the   trends    reported   by
{}\citet{willmer:southern}.  The conditional anticorrelation indicated
by  the $C_{ll}(r)$  of  the late  type  galaxies on  small scales  is
present in both  subsamples but more pronounced in  the sample of {\em
  luminous}    galaxies     (only    $C_{ee}(r)$    is     shown    in
Fig.~\ref{fig:morhpology-crosscorr-dimlum}).
This test does not allow very strong conclusions, since it is based on
an  ad--hoc division of  the whole  sample.  
A finer division is not  feasible, since very few early--type galaxies
will  populate the  subsamples.  However, our  results strengthen  the
interpretation that  both sorts of mark correlations  are irreducible. 
Neither  is the  luminosity  segregation the  source of  morphological
segregation  nor  is it  the  other  way  around. In  particular,  the
luminous  early--type galaxies  cluster more  strongly than  all other
galaxies.

\section{Summary and Outlook}
\label{sect:summary}

The investigation of luminosity and morphology segregation of galaxies
has been  a scientific task for  many years.  Our results  allow for a
new perspective  and suggest that  both the {\em methodology}  and the
{\em physical interest} should shift slightly.

Methodologically,   we    discussed   luminosity   and   morphological
segregation  in  the  framework   of  marked  point  processes.   This
perspective  provides  us  with  a  unifying view  on  morphology  and
luminosity segregation.   Moreover, the mathematical  theory of marked
point  processes provides  us with  test quantities  and models  to be
compared with the data.
In this  line we discussed the  mark--weighted conditional correlation
functions. These  functions are  not only easy  to estimate,  but also
offer a clear interpretation.  They  may be applied to a single volume
limited sample, a sequence of volume limited samples is not necessary.
As a consequence, they break  the degeneracy between a fractal spatial
structure and  luminosity segregation.  We suggest  that the $k_{mm}$,
$\var$, and $\cov$ functions are  of special interest for a first test
on luminosity segregation.
Since  several  bias--models  assume  scale--dependent bias,  we  need
quantities  like $k_{mm}$,  $\var$, and  $\cov$ which  can  unfold the
scales at which  mass or luminosity segregation is  relevant.  This is
not possible by looking at the amplitude of the two--point correlation
function $\xi_{R}(r)$ alone.  Moreover our method allows for a ``built
in'' significance test, by randomly re--shuffling the marks.
The conditional cross--correlation functions seem to be useful if mark
segregation  has  already  been shown  to  be  present  and is  to  be
understood more closely.  However, they are based on a division of the
whole  sample into  subpopulations, a  division  that has  to be  done
carefully.
The conditional mark--correlation functions are rather flexible.  With
the  peculiar velocities or  the orientations  of galaxies  treated as
marks,  the conditional mark  correlation functions  will allow  for a
fresh  look  at the  pairwise  velocity  dispersion  and on  alignment
effects.
Our methods can be easily extended to higher--order correlations.
In  a forthcoming  work  we  will study  the  mark correlations  using
higher--order       statistics       as       the       $J$--functions
({}\citealt{vanlieshout:j}, {}\citealt{kerscher:regular}).

Concerning  the physical  results, we  were  not only  able to  assess
luminosity segregation as  well as morphological segregation.  Rather,
our perspective allowed us to ask the question: What is the luminosity
and morphological  segregation like?   Our main results  obtained from
the SSRS2 survey are:
\begin{itemize}
\item
The average  luminosity of pairs  of galaxies and the  fluctuations in
the  luminosity on  each  galaxy  is enhanced  for  pairs closer  than
15\hMpc.   Hence, luminosity  correlations  are scale--dependent,  and
they  are significant  even outside  clusters of  galaxies.  On scales
larger than 15\hMpc\ our  results indicate that neither luminosity nor
morphological segregation is present.
\item
The  luminosities of  galaxies in  pairs closer  than 3\hMpc\  show an
increased  covariance  --   close  galaxies  preferably  have  similar
luminosities.
\item 
The  luminosity segregation is  not compatible  with the  random field
model.   Thus,  the  luminosity  does  not trace  an  underlying  {\em
independent} random field.  The luminosity of a galaxy  depends on the
local clustering and on interactions with other galaxies.
\item
There is an interesting feature, a small peak, in $k_{mm}$, $\var$ and
$\cov$ for galaxy pairs  with a separation of approximately $10\hMpc$,
which is currently beyond an explanation.
\item We observe morphological  segregation between early-- and late--
type galaxies for  scales smaller than 10\hMpc. This  effect is mainly
due to highly luminous  galaxies.  Especially the luminous early--type
galaxies  seem to  play an  important  role, both  for luminosity  and
morphology segregation.
\item
The importance  of early--type galaxies for  luminosity segregation is
confirmed by our analysis of  the IRAS samples. These infrared samples
exhibit  a deficit in  early--type galaxies  and consequently  show no
luminosity segregation.
\item 
An  inhomogeneous, scale--invariant  galaxy distribution,  but without
luminosity  segregation,  can  not  account  for the  signal  seen  in
$k_{mm}$,  $\var$, and  $\cov$.  The  lowered correlation  of  the dim
galaxies,  and the enhanced  correlation of  the luminous  galaxies we
found, explains at least in  part why the amplitude of the correlation
function  rises if  deeper, i.e.,  more luminous,  galaxy  samples are
considered.
\item 
With several independent  tests we could show that  it is not possible
to   explain   the    observed   luminosity   segregation   from   the
morphology--density relation alone.
\end{itemize}
Nevertheless, a  couple of  question remain open. \\ 
Concerning the data,  it seems important to confirm  our results using
other  galaxy   surveys.   Also   the  influence  of   redshift  space
distortions and  of galaxy clusters should be  investigated beyond the
simple error--estimates presented in Subsect.~\ref{sect:errors} and
Sect.~\ref{sect:morphology-density}.\\
Our  methods  are  directly  applicable  to  volume--limited  samples,
similar to  the usual way  of assessing luminosity  segregation, where
one needs a  series of volume--limited samples.  Using  models for the
conditional  mark density $\CM_2$  or the  mark--correlation functions
one   may    determine   the   parameters   of    such   models   from
magnitude--limited surveys directly.  Similarly, the influence of mark
segregation on  the two--  and $N$--point correlations  estimated from
magnitude--limited surveys can be estimated.\\
Closely related is the question how strongly the deprojected two-- and
$N$--point correlation functions, determined from 2-dimensional galaxy
catalogs, are  influenced by luminosity segregation.   With models for
the mark--correlations a refined  Limber's equation may be constructed
(see  e.g.,  {}\citealt{gardini:limber}).   Both, the  concerns  about
magnitude--limited surveys and deprojection formulas will be addressed
in future work.

In this  article we focused on clarifying  the mathematical framework,
on  the data--analysis,  and  on the  interpretation  of the  observed
luminosity  and  morphological   segregation.   The  relation  to  the
peak--formalism {}\citep{bardeen:gauss} and other biasing schemes will
be   investigated  in  future   work.  Understanding   the  luminosity
distribution on the galaxies from dynamical models is the major goal.

\acknowledgements  

We  thank Thomas  Buchert,  Niv Drory,  Ulrich  Hopp, Roberto  Saglia,
Dietrich Stoyan,  Alex Szalay, Istvan  Szapudi and Herbert  Wagner for
valuable  discussions and  Alessandro Amici  for kindly  providing the
fractal  point  set used  in  Subsect.~\ref{sect:robust}.   CB and  MK
acknowledge support from  the {\em Sonderforschungsbereich 375 f{\"u}r
Astroteilchenphysik  der DFG}.  MK acknowledges  support from  the NSF
grant AST 9802980.


\appendix

\section{Estimators for mark--correlation functions}
\label{sect:estimators}

In   this   section,   we   discuss  estimators   for   the   weighted
mark--correlation     functions.      For     this    purpose,     let
$\{(\bx_i,m_i)\}_{i=1}^{N}$  denote  the  $N$  empirical  data  points
$\bx_i$ inside  the sample  $\CD$ with their  marks $m_i$.   We prefer
estimators  which  are  unbiased  and  show small  variances.   For  a
detailed      discussion     of     two--point      estimators     see
{}\citep{kerscher:twopoint}.    One  basic   idea   is  to   construct
estimators for $\paverage{f}(r)$ from  a combination of estimators for
the      numerator      and      for      the      denominator      of
eq.~\eqref{eq:kappa-f-separate}. We  first discuss estimators  of this
type, but then introduce a different estimator, which does not use any
boundary conditions. It turns out, that in our case, this estimator is
unbiased and is recommended by its simplicity and low variance.

\subsection{Construction of the estimators}

To calculate the  correlation functions in bins of  width $\Delta$, we
use the indicator function of a set $A$
\begin{equation}
\Beins_A(x) =
\begin{cases}
1 & \text{ if } x\in A\\
0 & \text{ otherwise },
\end{cases}
\end{equation}
and           the            reduced           sample           window
$\CD_{-r}=\{\bx\in\CD|d(\bx,\partial\CD)>r\}$ shrunken by $r$.

Using   these  definitions,   the   ratio--unbiased  minus   estimator
$\phataverage{f}^M(r)$   for  the  weighting   functions  $f(m_1,m_2)$
(compare eq.~\eqref{eq:kappa-f-separate}) is simply
\begin{equation}
\phataverage{f}^M(r) = 
\frac{\sum_{i\ne j=1}^N
\Beins_{\CD_{-r}}(\bx_i)\Beins_{[r, r+\Delta]}(|\bx_i-\bx_j|)\ f(m_1,m_2)}
{\sum_{i\ne j=1}^N
\Beins_{\CD_{-r}}(\bx_i)\Beins_{[r, r+\Delta]}(|\bx_i-\bx_j|)},
\end{equation}
where the  indicator function $\Beins_{\CD_{-r}}(\bx_i)$  assures that
the point $\bx_i$  is further than $r$ from  the boundary (for details
see {}\citealt{kerscher:twopoint}).

In the minus estimator the window is effectively shrun\-ken, resulting
in an  increased variance.  On  the contrary, the  following estimator
uses  all   point  pairs  $\bx_i,\bx_j$,  however   weighted  with  an
geometrical weight $\omega(\bx_i,\bx_j)$.  Such weighting schemes lead
to ratio--unbiased estimators  for the two--point correlation function
(for details see {}\citealt{stoyan:stochgeom}).  The straight--forward
generalization of these concepts results in ratio--unbiased estimators
for $\paverage{f}$:
\begin{equation}
\phataverage{f}^\omega(r) = 
\frac{\sum_{i\ne j=1}^N
\Beins_{[r, r+\Delta]}(|\bx_i-\bx_j|)\ \omega(\bx_i,\bx_j)\ f(m_1,m_2)}
{\sum_{i\ne j=1}^N
\Beins_{[r, r+\Delta]}(|\bx_i-\bx_j|)\ \omega(\bx_i,\bx_j)} .
\end{equation}
Using the  weight
\begin{equation}
\label{eq:setcov}
\omega(\bx_i,\bx_j) 
= \frac{|\CD|}{|\CD\cap\CD_{\bx_i-\bx_j}|},
\end{equation}
we  arrive at  an estimator  $\phataverage{f}^\omega(r)$  suggested by
{}\citet{stoyan:fractals}.  In full analogy  to the estimators for the
two--point  correlation  function   other  weights,  like  the  Ripley
(Rivolo) weight  or the isotropized  version of eq.~\eqref{eq:setcov},
can be used (for details see {}\citet{kerscher:twopoint}).

Instead of estimating $\paverage{f}$  with unbiased estimators for the
numerator and  for the denominator  in Eq.~\eqref{eq:kappa-f-separate}
separately, we suggest to simply use the ratio
\begin{equation}
\phataverage{f}(r) = 
\frac{\sum_{i\ne j=1}^N
\Beins_{[r, r+\Delta]}(|\bx_i-\bx_j|)\ f(m_1,m_2)}
{\sum_{i\ne j=1}^N
\Beins_{[r, r+\Delta]}(|\bx_i-\bx_j|)}.
\end{equation}
This  is   motivated  by  the  observation,   that  $\paverage{f}$  is
calculated  from the  marks under  the  {\em condition}  that the  two
points  are separated  by $r$.   Indeed we  are not  investigating the
spatial distribution  of the  points, but ``divide  spatial two--point
properties out''.   Unfortunately, the unbiasedness  of this estimator
cannot be proven  with the common methods used in  the theory of point
processes,  but it  seems  intuitively clear  that  this estimator  is
unbiased.  In sect.~\ref{sect:compare-estimators} we show this using a
numerical example;  we illustrate furthermore that  this estimator has
preferable   variance   properties   (this   was  also   observed   by
{}\citealt{capobianco:autocovariance}).

\subsection{Comparison of the estimators}
\label{sect:compare-estimators}

\begin{figure}
\epsscale{0.33}
\plotone{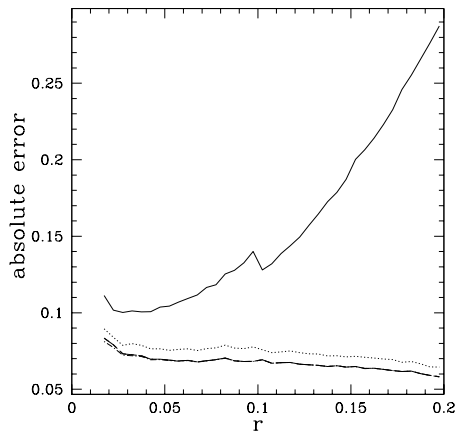}
\caption{ \label{fig:compare-estimators}  The standard error estimated
from 5000 realizations for the $\gamma(r)/V$ is shown: minus estimator
(solid  line),  Rivolo (Ripley)  estimator  (dotted  line), the  other
weighting estimators and the estimator without boundary correction lie
nearly on top of each other (the lowest line).}
\end{figure}

We     use    the    marked     Poisson    process     discussed    in
Sect.~\ref{sect:marked-poisson}   to   numerically   investigate   the
properties  of these estimators  for the  continuous mark--correlation
functions.   The  sample mean  of  the  estimators  coincide with  the
theoretical  mean value  for  all estimators.  Thus, empirically,  all
estimators are unbiased.
In     Fig.~\ref{fig:compare-estimators}      the     variances     of
$\gamma(r)/V$ for the different estimators are shown.
The  variance  of  the  minus  estimator  becomes  unacceptably  large
especially on large scales. The  estimators using a weighting with the
set covariance or  the isotropized set covariance show  the same small
variance,  even smaller  than the  variance of  the estimator  using a
weighting of Rivolo (Ripley)  type.  A detailed inspection shows, that
the  estimator  using  no  boundary  correction  typically  gives  the
smallest variance.
A   qualitatively   similar   behavior   is  found   for   the   other
mark--correlation   functions.    Therefore,   and  for   reasons   of
computational simplicity,  we mainly apply this  estimator. We suggest
to use it for all  mark--weighted correlation functions as the natural
and unbiased choice.

\end{document}